\documentclass[pdflatex,sn-vancouver,Numbered]{sn-jnl}

\usepackage{graphicx}%
\usepackage{multirow}%
\usepackage{amsmath,amssymb,amsfonts}%
\usepackage{amsthm}
\usepackage{mathrsfs}%
\usepackage[title]{appendix}%
\usepackage{xcolor}%
\usepackage{textcomp}%
\usepackage{manyfoot}%
\usepackage{booktabs}%
\usepackage{algorithm}%
\usepackage{algorithmic}%
\usepackage{listings}%

\newtheorem{lemma}{Lemma}

\newtheorem{observation}{Observation}
\usepackage{adjustbox}
\usepackage{diagbox}
\usepackage{bm}

\let\orcidlogo\relax
\usepackage{orcidlink}

\PassOptionsToPackage{authoryear}{natbib}
\usepackage{natbib}
\usepackage{float}
\usepackage{chngcntr}
\usepackage{subcaption}
\usepackage{siunitx}

\usepackage{braket}

\usepackage[most]{tcolorbox}
\newtcolorbox{myhighlight}{
    colback=yellow!30, 
    frame hidden,
    boxrule=0pt,
    sharp corners,
    enhanced,
    breakable 
}

\usepackage{soul}
\soulregister\ref7
\soulregister\eqref7
\soulregister\citep7
\renewcommand{\hl}[1]{#1}

\usepackage{array}
\usepackage{hhline}

\newtheorem{remark}{Remark}%

\newtheorem{definition}{Definition}%

\begin{document}

\title[Article Title]{\hl{Binary integer programming for optimizing ebit cost in distributed quantum circuits with fixed module allocation}}


\author[1]{\fnm{Hyunho} \sur{Cha} \orcidlink{0009-0008-2933-6989}}\email{aiden132435@cml.snu.ac.kr}

\author*[1]{\fnm{Jungwoo} \sur{Lee} \orcidlink{0000-0002-6804-980X}}\email{junglee@snu.ac.kr}

\affil[1]{\orgdiv{NextQuantum and Department of Electrical and Computer Engineering}, \orgname{Seoul National University}, \orgaddress{\city{Seoul}, \postcode{08826}, \country{Republic of Korea}}}



\abstract{\hl{Modular and networked quantum architectures can scale beyond the qubit count of a single device, but executing a circuit across modules requires implementing non-local two-qubit gates using shared entanglement (ebits) and classical communication, making ebit cost a central resource in distributed execution.
The resulting distributed quantum circuit (DQC) problem is combinatorial, motivating prior heuristic approaches such as hypergraph partitioning.
In this work, we decouple module allocation from distribution. For a fixed module allocation (i.e., assignment of each qubit to a specific Quantum Processing Unit), we formulate the remaining distribution layer as an exact binary integer programming (BIP). This yields solver-optimal distributions for the fixed-allocation subproblem and can be used as a post-processing step on top of any existing allocation method. We derive compact BIP formulations for four or more modules and a tighter specialization for three modules.
Across a diverse benchmark suite, BIP post-processing reduces ebit cost by up to 20\% for random circuits and by more than an order of magnitude for some arithmetic circuits.
While the method incurs offline classical overhead, it is amortized when circuits are executed repeatedly.}}




\maketitle

\subsection*{Acknowledgments}
This work is in part supported by the National Research Foundation of Korea (NRF, RS-2024-00451435 (20\%), RS-2024-00413957 (40\%)), Institute of Information \& Communications Technology Planning \& Evaluation (IITP, 2021-0-01059 (40\%)), grant funded by the Ministry of Science and ICT (MSIT), Institute of New Media and Communications (INMAC), and the Brain Korea 21 FOUR program of the Education and Research Program for Future ICT Pioneers.

\section{Introduction}
In recent years, the capacity of quantum computing hardware has steadily increased \citep{arute2019quantum, bravyi2018quantum, neill2018blueprint, kjaergaard2020superconducting}. However, a significant obstacle to realizing the full potential of quantum computing is the limited number of qubits available in a single quantum computer (or module) \citep{terhal2015quantum, fowler2012surface, o2016scalable, bourassa2021blueprint}. To overcome this\vphantom{limitation}, distributing a large quantum computation over a network of modules has emerged as a promising approach \citep{cirac1999distributed, cacciapuoti2019quantum, cuomo2023optimized, yimsiriwattana2004generalized, eisert2000optimal}.

The distributed quantum circuit (DQC) problem involves dividing a quantum circuit across multiple modules. A non-local controlled unitary gate between two modules can be realized \hl{using only local quantum operations and classical communication} if the modules share a Bell state \citep{nielsen2010quantum, sych2009complete, zaman2018counterfactual}, which is referred to as an $\textit{ebit}$ \citep{g2021efficient, andres2019automated, wu2023entanglement, andres2024distributing}.
\hl{Typically, an algorithm used to solve DQC attempts to minimize the ebit cost for circuit execution.}
Its goal is to map the qubits in the circuit to individual modules and determine how to implement the non-local operations between modules in a way that minimizes the number of required ebits.

This paper primarily considers homogeneous networks with $k \geq 3$ modules. In this setting, distributions using a hypergraph partitioning formulation and heuristic solvers proved to be the most effective \citep{wu2023entanglement, andres2024distributing}. However, this approach jointly determines the module allocation and the implementation of non-local gates based on that allocation, even though the former cannot logically depend on the latter. \hl{As our main contribution, we observe that exact binary integer programming (BIP) formulations \citep{papadimitriou1998combinatorial, karp2009reducibility, wolsey2020integer} under a \emph{fixed} module allocation function (i.e., assignment of each qubit to a specific Quantum Processing Unit (QPU)) yield improved ebit costs, demonstrating stability for various types of circuits.} First, we present formulations for $k \geq 4$ modules and then show how it simplifies for $k = 3$ modules. As a result, our method can be combined with any DQC solver by ignoring its distribution steps and adopting only the module allocation it returns. Our approach further reduces ebit costs for various types of circuits, with significant improvements observed in some structured circuits. This highlights the importance of our post-processing for achieving robust results.

\hl{The motivation for using a BIP formulation is that, given a fixed module allocation, selecting migrations (defined in Section~\ref{problem_formulation}) to cover all non-local gates reduces to a discrete subset-selection problem with binary decisions. An exact BIP formulation yields (solver-)optimal solutions for this fixed-module subproblem, enabling a principled post-processing step that can be plugged into any existing method by reusing its module allocation and optimizing only the distribution layer.}

The rest of this paper is organized as follows. Section 2 provides an overview of the background and related works. In Section 3, we present detailed derivations and proofs of our BIP formulations, along with an explanation of the unique characteristics of distributing quantum Fourier transform (QFT) circuits \citep{nielsen2010quantum, coppersmith2002approximate, ruiz2017quantum}. Section 4 presents experimental results, which have been validated on various circuits. Finally, Section 5 concludes the paper and offers a perspective on future work.

\section{Preliminaries}

\hl{At its core, the objective of this work can be summarized as follows: Given a monolithic quantum circuit $C$, how can we optimally distribute its constituent gates across $k$ distinct hardware modules while minimizing the overhead of entanglement resources and non-local operations? In the following sections, we transition into the formal definitions of distributed quantum architectures. The concepts introduced in this section are used throughout and will be required to state our BIP formulation in Section~\ref{sec:section_bip_formulation}.}

\begin{figure}
  \centering
  \begin{subfigure}[c]{0.48\textwidth}
    \centering
    \includegraphics[width=\textwidth]{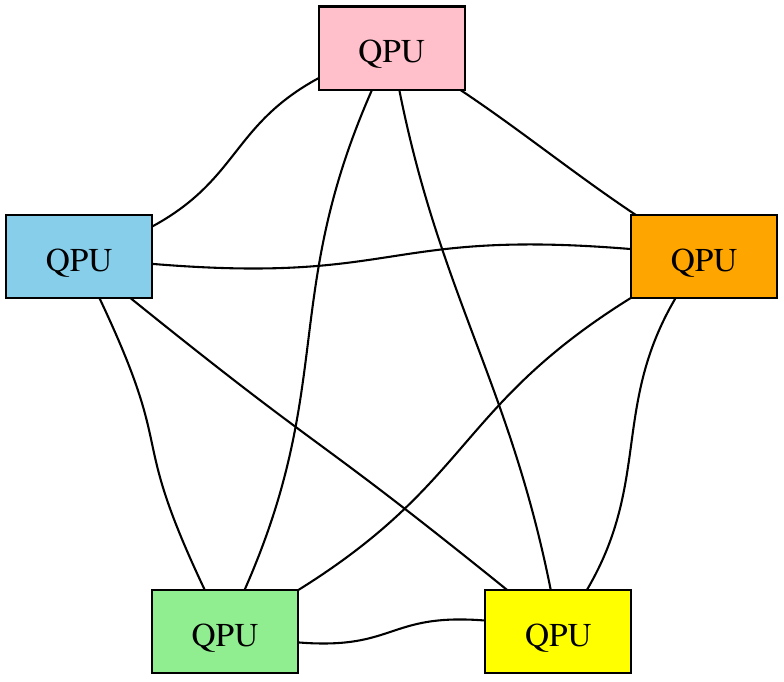}
  \end{subfigure}
  \hfill
  \begin{subfigure}[c]{0.48\textwidth}
    \centering
    \includegraphics[width=\textwidth]{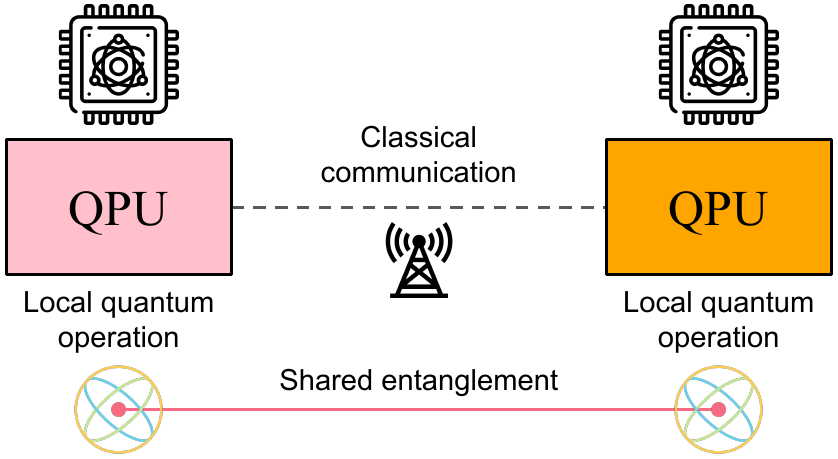}
  \end{subfigure}
  \vspace{1em}
  \caption{\hl{Example network of five QPUs. Every pair of modules shares entanglement resources to facilitate non-local gate operations. This distributed architecture enables large-scale quantum circuits through a combination of local quantum operations and inter-module classical communication.}}
  \label{fig:dqc_network}
\end{figure}

\subsection{Distribution of quantum circuits}

\hl{In the context of distributed quantum computing, a \textit{module} refers to an individual quantum computer that forms part of a larger, interconnected network \citep{andres2024distributing} (represented as a box in Fig.~\ref{fig:dqc_network}). Modern quantum architectures often face scaling limitations where the qubit count of a single device is insufficient for large-scale algorithms. To overcome this, software tools must partition a global circuit into smaller fragments that can be executed across multiple modules, allowing them to function collectively as a single, powerful processor. This distributed execution, however, necessitates sophisticated non-local gate operations and advanced compiler strategies to determine an optimal circuit layout, which amounts to minimizing the entanglement and communication overhead between modules.}

As in previous studies, we consider a quantum circuit to be composed only of unary and binary gates \citep{g2021efficient, andres2019automated, wu2023entanglement, andres2024distributing}. Here, each binary gate is a controlled phase ($CP$) gate \citep{nielsen2010quantum}
\begin{equation*}
CP(\theta) = |0\rangle\langle 0| \otimes \mathbb{I} + |1\rangle\langle 1| \otimes
\begin{pmatrix}
1 & \\
& e^{i\theta}
\end{pmatrix},
\end{equation*}
where $\mathbb{I}$ denotes the single-qubit identity operator. Any circuit can be transformed into an equivalent circuit in this form, a fact that follows from the universality of the set consisting of all unary gates and the CNOT gate \citep{barenco1995elementary}.

Distributing an $n$-qubit quantum circuit involves two steps. First, each qubit must be assigned to a module. Second, non-local gates are identified based on this module allocation, and the optimal method to implement these gates with the least resources is determined. This overall process is referred to as the DQC problem. Typically, a $(k, \epsilon)$ \textit{balanced partition} is used for module allocation, ensuring that each module contains at most $(1 + \epsilon) \frac{n}{k}$ qubits \citep{g2021efficient}. By a reduction from the hypergraph min-cut problem, the DQC problem was shown to be NP-hard \citep{andres2019automated}.
\hl{This complexity implies that finding a globally optimal distribution for large-scale circuits is computationally intractable, traditionally necessitating the use of heuristic approaches such as hypergraph partitioning. However, these heuristics often bundle module allocation and ebit distribution together, potentially missing significant resource savings. By decoupling these steps, we can isolate the ebit distribution layer as a subproblem that can be solved via BIP. This allows our method to refine heuristic-based pipelines, providing a better solution for a given module allocation under which a standard heuristic alone would settle for suboptimal resource costs.}

\begin{figure}
\centering
\includegraphics[width=0.6\columnwidth]{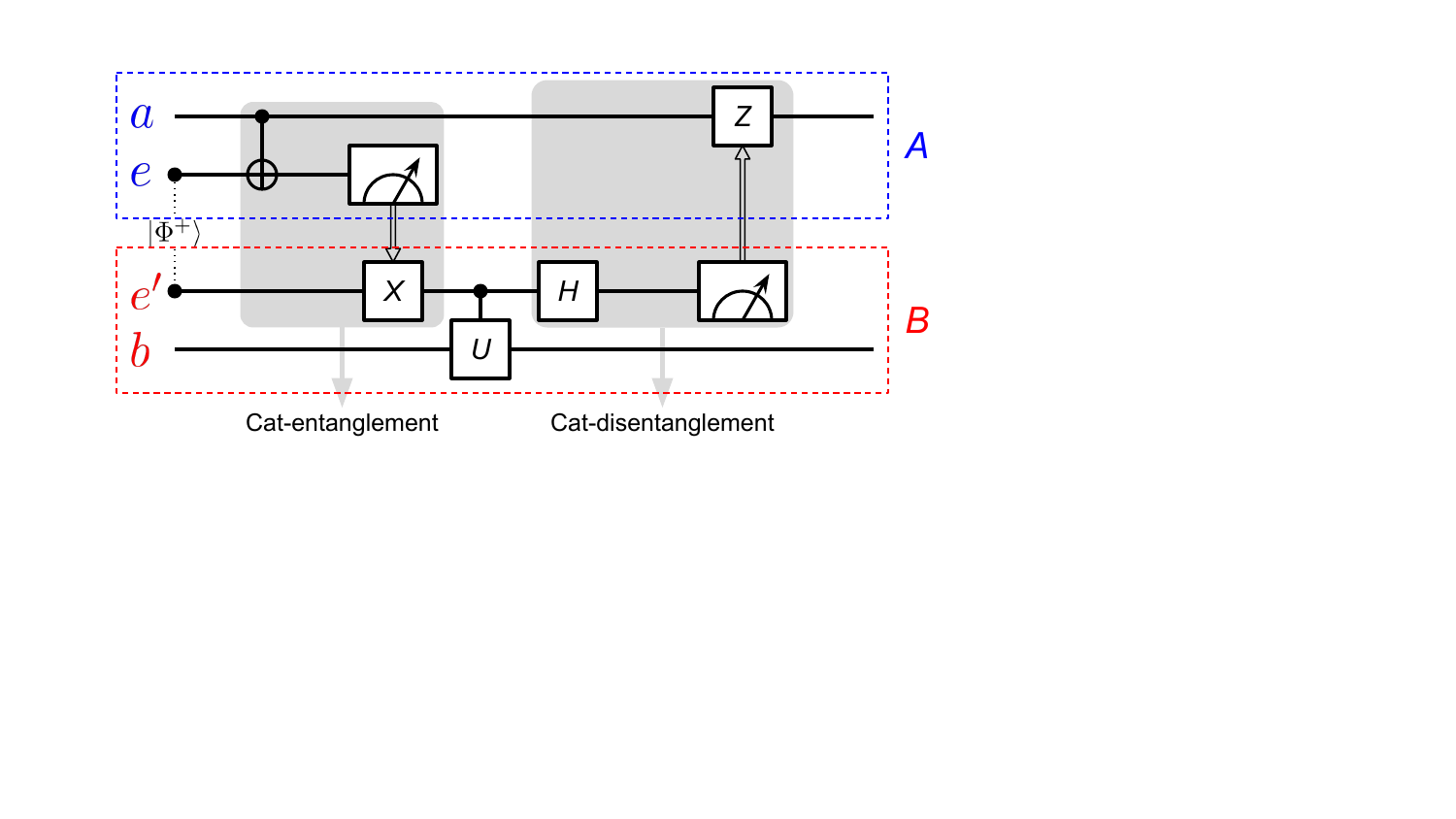}
\vspace{1em}
\caption{\textbf{Implementation of a non-local gate.} Module $A$ contains qubits $(a,e)$ and Module $B$ contains qubits $(e^\prime,b)$. The controlled-$U$ gate $CU_{a\to b}$ with control $a$ and target $b$ can be realized using only local operations and classical communication at the cost of a single ebit $|\Phi^+\rangle_{ee^\prime}$ shared between $A$ and $B$.}
\label{fig:cat_entanglement}
\end{figure}

\subsection{Cat-entanglement and cat-disentanglement}
\label{section:cat_entanglement_intro}

A quantum communication channel enables the creation of the Bell state
\begin{equation*}
|\Phi^+\rangle_{AB} = \frac{|00\rangle_{AB} + |11\rangle_{AB}}{\sqrt{2}}
\end{equation*}
between two modules $A$ and $B$, which is a maximally entangled bipartite state. In the context of DQC, such states are referred to as ebits. Fig.~\ref{fig:cat_entanglement} shows how to implement a non-local controlled unitary gate with a single control qubit $a$ and a single target qubit $b$ at the cost of a single ebit \hl{(for a complete validation of this protocol, see Appendix~\ref{sec:controlled_U_explanation})}. The steps before (after) the controlled unitary gate constitute a \textit{cat-entanglement} (\textit{cat-disentanglement}) process. The former creates a linked copy of $a$ in module $B$, and the latter effectively reverses this process. After cat-disentanglement, operations such as $CP(\theta)_{a\to b}$ can be successfully implemented.
\hl{In practice, the end-to-end fidelity and runtime of this ebit-assisted cat-(dis)entanglement are currently limited by the mid-circuit measurements and real-time feed-forward, but existing demonstrations already mitigate part of this overhead to make non-local operations more robust \citep{carrera2024combining, baumer2025measurement}.}

Suppose we create linked copies of a qubit $q$ at modules other than the one including $q$. This cat-entanglement operation (i) commutes with $CP$ gates acting on $q$ and (ii) generally does not commute with arbitrary unary gates acting on $q$ (see Appendices~\ref{appendix:commute}~and~\ref{appendix:non_commute}). The first property allows us to restrict the creation of a linked copy of $q$ in any module to occur immediately after one of the unary operations on $q$ or at time $0$, assuming that the circuit is compiled exclusively with $CP$ gates and unary gates. The second property implies that in general, all linked copies of $q$ in other modules must be reintegrated into $q$ before a unary gate acts on $q$.

\subsection{Problem formulation}
\label{problem_formulation}

\hl{To formally address the distribution of gates across a modular architecture, we first define the structure of the input quantum circuit $C$. Our goal is to transform this monolithic circuit into a partitioned representation that can be mapped onto $k$ hardware modules. To achieve this, we model the circuit as a collection of \emph{discrete gate events} in space and time, categorized by their qubit dependencies.}

Specifically, we represent a quantum circuit acting on a set of $n$ qubits $Q = \{ q_i \}_{i = 1}^{n}$ as a finite set $C = U \cup B$, where
\begin{equation*}
U = \bigcup_{i=1}^{n} \left\{ (q_i, t) \mid t \in T^{(i)} \subset \mathbb{N} \right\}
\end{equation*}
is a set of unary gates and
\begin{equation*}
B = \bigcup_{1 \leq i < j \leq n} \left\{ (\{q_i, q_j\}, t) \mid t \in T^{(\{i, j\})} \subset \mathbb{N} \right\}
\end{equation*}
is a set of \textit{non-local} binary gates. As is clear from the notation, $(q_i, t)$ represents a unary gate acting on $q_i$ at time $t$ and $(\{q_i, q_j\}, t)$ represents a binary gate acting on $\{ q_i, q_j \}$ at time $t$. Without loss of generality, it is assumed that all $T^{(i)}$ and $T^{(\{i, j\})}$ are mutually exclusive.

In what follows, we assume that there are $k \geq 3$ modules in total, i.e., $P = \{ p_1, p_2, \cdots , p_k \}$, and that we are \textit{given} a module allocation function $\pi : Q \rightarrow P$. This implies that $\pi(q_i) \neq \pi(q_j)$ for all $(\{q_i, q_j\}, t) \in B$.
The concept of \textit{migration} was first defined in Ref. \citep{g2021efficient}.
\hl{Intuitively, a migration does \emph{not} physically move a qubit between modules.
Rather, it denotes a temporary, protocol-level operation in which the quantum information of a logical qubit is made available at another module so that subsequent $CP$ gates that would otherwise be non-local can be executed locally at that module.}
Technically, a migration represents the cat-entanglement operation in Section~\ref{section:cat_entanglement_intro}. It also implicitly assumes that cat-disentanglements are performed whenever necessary.
\begin{definition}[Migration \citep{g2021efficient}]
\label{definition:def_migration}
A migration is a triple $(q, p, t)$, which translates to creating a linked copy of $q$ in module $p$ at time $t$. The set of all candidate migrations is defined as
\begin{equation*}
M \equiv \bigcup_{i = 1}^{n} \bigcup_{p \in P \setminus \{ \pi(q_i) \}} \left\{ (q_i, p, t) \mid t \in \{0\} \cup T^{(i)} \right\}.
\end{equation*}
\end{definition}

\hl{For a non-local $CP$ gate, \emph{home coverage} restricts us to migrating one endpoint qubit to the \emph{home module} of the other endpoint, so that the gate can be executed where one of its qubits already belongs.
In contrast, \emph{joint coverage} allows two migrations executing the gate at a \emph{third module} by migrating \emph{both} endpoint qubits there. This greatly enlarges the combinatorial choices because it couples the selection decisions for the two qubits. Formally, we have the following definitions.}

\begin{definition}[Home coverage \citep{g2021efficient}]
\label{def:coverage}
For a non-local $CP$ gate $g = (\{ q_i, q_j \}, t^*)$, we say that a migration $(q, p, t)$ home-covers $g$ if one of the following conditions holds, thereby enabling the execution of $g$.
\begin{enumerate}
\item \label{item:first} $q = q_i, \quad p = \pi(q_j), \quad t = \max\limits_{\Tilde{t} \in \{0\} \cup T^{(i)}, \, \Tilde{t} \leq t^*} \Tilde{t}$
\item \label{item:second} $q = q_j, \quad p = \pi(q_i), \quad t = \max\limits_{\Tilde{t} \in \{0\} \cup T^{(j)}, \, \Tilde{t} \leq t^*} \Tilde{t}$
\end{enumerate}
\end{definition}

\begin{definition}[Joint coverage \citep{g2021efficient}]
\label{def:joint_coverage}
For a non-local $CP$ gate $g = (\{ q_i, q_j \}, t^*)$, we say that a pair of migrations $\{ (q_{i'}, p, t), (q_{j'}, p, t') \}$ joint-covers $g$ if the following condition holds, thereby enabling the execution of $g$.
\begin{itemize}
\item \label{item:joint_condition} $\{ i', j' \} = \{ i, j \}, \quad p \notin \{ \pi(q_i), \pi(q_j) \},\\
t = \max\limits_{\Tilde{t} \in \{0\} \cup T^{(i')}, \, \Tilde{t} \leq t^*} \Tilde{t}, \qquad t' = \max\limits_{\Tilde{t} \in \{0\} \cup T^{(j')}, \, \Tilde{t} \leq t^*} \Tilde{t}$
\end{itemize}
\end{definition}

\hl{Given $\pi$, the problem of finding the minimum subset of $M$ that covers all gates in $B$ following Definition~\ref{def:coverage} is called \textit{migration selection under home coverage} (MS-HC).
If we further allow joint coverage in Definition~\ref{def:joint_coverage}, then the problem is called \textit{migration selection under general coverage} (MS-GC). We remark that MS-HC can be viewed as a set cover instance with highly restricted structure, and an optimal polynomial-time algorithm for MS-HC given $\pi$ has been identified. On the other hand, joint coverage significantly expands the search space, and MS-GC is conjectured to be NP-hard. A heuristic for MS-GC was proposed, which greedily adds a set of migrations at each iteration based on some criterion \citep{g2021efficient}.}

\hl{Importantly, this work does \emph{not} claim a polynomial-time algorithm for MS-GC.
Instead, our contribution is an exact BIP formulation for MS-GC under a fixed $\pi$, which we use as a post-processing step.
Starting from any candidate allocation $\pi$ (e.g., produced by a heuristic allocator), we solve the corresponding fixed-$\pi$ BIP to obtain an ebit distribution for that allocation.
This improves the ebit cost without changing the module allocation.}

\subsection{Hypergraph partitioning formulation}
\label{sec:martinez_intro}

\begin{figure}
  \centering
  \begin{subfigure}[c]{0.643\textwidth}
    \centering
    \includegraphics[width=\textwidth]{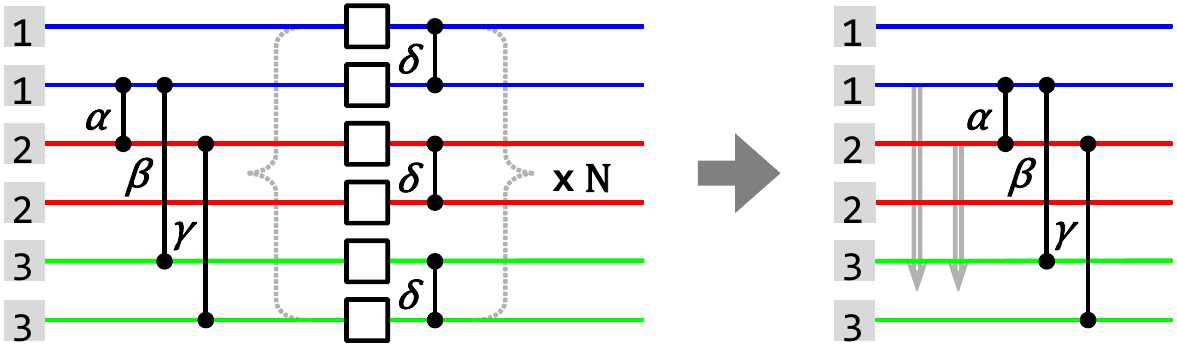}
    \caption{}
    \label{fig:diagram1}
  \end{subfigure}
  \hspace{0.05\textwidth}
  \begin{subfigure}[c]{0.257\textwidth}
    \centering
    \includegraphics[width=\textwidth]{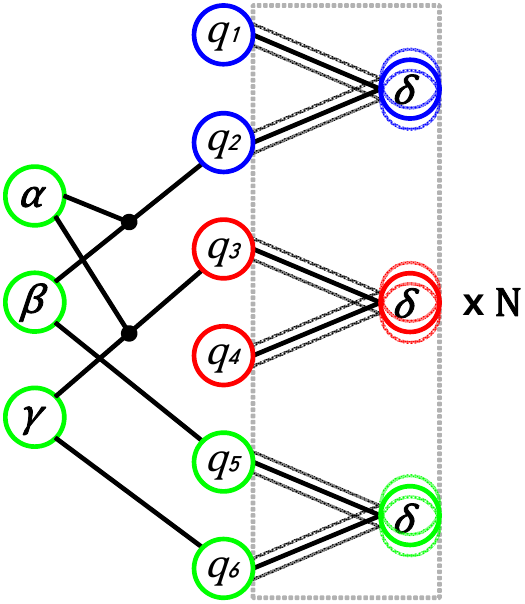}
    \caption{}
    \label{fig:diagram2}
  \end{subfigure}
  \caption{\hl{(a) Example distributed circuit for $k=3$ modules. Wires with the same color belong to the same module. The labels $1,1,2,2,3,3$ on the left are the module indices of the six logical qubits (each module contains two qubits). White squares represent single-qubit gates and the symbols $\alpha,\beta,\gamma,\delta$ denote two-qubit gates (with repeated $\delta$-blocks). Downward arrows indicate migrations (here, to module 3). (b) Hypergraph representation of the same instance, with qubit vertices $q_1,\ldots,q_6$ and gate vertices $\alpha,\beta,\gamma,\delta$. Qubit-vertex colors indicate their home modules, whereas gate-vertex colors indicate the modules where the corresponding gates are executed.}}
  \label{fig:martinez_graph}
\end{figure}

\hl{Our BIP post-processing requires a module allocation $\pi$ as input.
In practice, we can obtain $\pi$ from existing allocation heuristics, and in particular from the hypergraph partitioning formulation of circuit distribution \citep{andres2019automated}, which has achieved the best performance for $k \ge 3$ \citep{wu2023entanglement, andres2024distributing}.
We briefly summarize this construction here for completeness.
Readers interested only in the post-processing formulation may skip to Section~\ref{sec:section_bip_formulation}.}

Given a circuit $C$, the corresponding hypergraph is constructed as follows. First, a node is added for each binary gate and for each qubit. Second, for each qubit $q_i$ and time interval $I = (t, t')$ such that the following conditions are satisfied, a hyperedge that connects $q_i$ and all the binary gates that act on $q_i$ within $I$ is added, if such gate exists:
\begin{itemize}
\item $t, t' \in T^{(i)} \cup \{0, \infty \}, \, t < t'$
\item $\left\{ t'' \mid t'' \in T^{(i)}, \, t < t'' < t' \right\} = \emptyset$
\item $\{ (j, t'') \mid (\{q_i, q_j\}, t'') \in B, \, t < t'' < t' \} \neq \emptyset$
\end{itemize}
Fig.~\ref{fig:martinez_graph} shows an example circuit distributed by this method. The coverage of the gate $\alpha$ is a joint coverage, and the coverages of the gates $\beta$ and $\gamma$ are home coverages.

\section{Algorithmic improvements and examples}
\label{sec:section_bip_formulation}

\hl{The full DQC pipeline consists of two layers: a module allocation step that fixes $\pi:Q\to P$, followed by a distribution step that selects migrations to cover all non-local gates.
The latter step is precisely MS-GC under a fixed $\pi$, and existing approaches rely on heuristics because MS-GC is conjectured to be NP-hard.
Our goal in this section is therefore not to obtain a new allocator, but to provide a post-processing method that, for any given $\pi$, computes the set of migrations for that allocation.
To this end, we formulate fixed-$\pi$ MS-GC as a BIP, which serves as a drop-in improvement layer on top of any module allocation heuristic by optimizing only the migration-selection layer.
In Section~\ref{section:tiny_example}, we include a small illustrative example showing that even on a tiny circuit the hypergraph partitioner can be suboptimal, and that our BIP post-processing reduces the ebit cost.
}

\hl{We begin by describing a BIP formulation of MS-GC for $k \geq 4$. A more efficient formulation for $k = 3$ is described in Section~\ref{sec:bip_formulation_k_3}.}

\subsection{BIP formulation of MS-GC for $k \geq 4$}

For notational simplicity, we define
\begin{equation*}
f(i, t) \equiv \max\left(\left\{t' \in \{0\} \cup T^{(i)} \mid t' \leq t\right\}\right),
\end{equation*}
which denotes the latest time a unary gate is applied to $q_i$ up to time $t$ (or $0$ if none exists). We define the following sets of migrations:
\begin{align}
\label{eq:m1g_m2g_definition}
M^{(1, (\{q_i, q_j\}, t))} & \equiv \{ (q_i, \pi(q_j), f(i, t)), (q_j, \pi(q_i), f(j, t)) \},\nonumber\\
M^{(2, (\{q_i, q_j\}, t))} & \equiv \big\{ \{ (q_i, p, f(i, t)), (q_j, p, f(j, t)) \} \mid p \in P \setminus \{ \pi(q_i), \pi(q_j) \} \big\}.
\end{align}
In simpler terms, $M^{(1, (\{q_i, q_j\}, t))}$ denotes a pair of migrations, where each migration in the pair migrates one of the qubits to the other qubit's module; $M^{(2, (\{q_i, q_j\}, t))}$ denotes a set of pairs of migrations, where each pair in the set comprises migrations that migrate both qubits to a module that contains neither of them. We proceed to define the following unions:
\begin{align}
\label{eq:m1_m2_mtilde_definition}
\Tilde{M}^{(g)} & \equiv M^{(1, g)} \cup M^{(2, g)},\nonumber\\
M^{(1)} & \equiv \bigcup_{g \in B} M^{(1, g)}, \quad M^{(2)} \equiv \bigcup_{g \in B} M^{(2, g)},\\
\Tilde{M} & \equiv M \cup M^{(2)}.\nonumber
\end{align}
\hl{To express the BIP compactly, we fix an arbitrary but consistent ordering of each set and treat it as an \emph{indexed list}.
This allows us to introduce decision variables and constraints using subscripts without repeatedly naming individual migrations.
Specifically, for any set $S$ we employ the notation}
\begin{equation*}
\forall 1 \leq i \leq |S|, \quad S_i \in S \quad \text{and} \quad S.\text{idx}(S_i) = i.
\end{equation*}

\begin{algorithm}
\caption{BIP for MS-GC, $k \geq 4$}
\label{algorithm:bip_k_geq_4}
\begin{algorithmic}[1]
\STATE \textbf{Input:} Unary gates $U$, binary gates $B$, module allocation function $\pi$
\STATE \textbf{Output:} Vector of selected migrations $x$
\STATE \hl{Construct $M$, $M^{(2)}$, and $\tilde M$ using Definition~\ref{definition:def_migration} and Eqs.~\eqref{eq:m1g_m2g_definition}~and~\eqref{eq:m1_m2_mtilde_definition}.}
\STATE \hl{\emph{[Define objective and constraints]}}
\STATE Initialize:
\STATE \quad $v^{\top} \leftarrow \left[\bm{1}_{|M|}^{\top} \bm{0}_{|M^{(2)}|}^{\top}\right]$
\STATE \quad $\Tilde{A} \in \mathbb{B}^{\left(|B| + |M^{(2)}|\right) \times |\Tilde{M}|} \leftarrow 0$
\STATE \quad $b^{\top} \leftarrow \left[\bm{1}_{|B|}^{\top} \bm{0}_{|M^{(2)}|}^{\top}\right]$
\STATE \hl{\emph{[Iterate over migrations and fill coverage/consistency constraints]}}
\FOR{$j = 1$ to $|\Tilde{M}|$}
\FOR{$i = 1$ to $|B|$}
\IF{$\Tilde{M}_j \in \Tilde{M}^{(B_i)}$}
\STATE $\Tilde{A}_{ij} \leftarrow 1$
\ENDIF
\ENDFOR
\FOR{$i = 1$ to $|M^{(2)}|$}
\IF{$\Tilde{M}_j \in M^{(2)}_i$}
\STATE $\Tilde{A}_{(|B| + i)j} \leftarrow 1$
\ELSIF{$\Tilde{M}_j = M^{(2)}_i$}
\STATE $\Tilde{A}_{(|B| + i)j} \leftarrow -2$
\ENDIF
\ENDFOR
\ENDFOR
\STATE \hl{\emph{[Use an off-the-shelf solver]}}
\STATE Solve \hl{the following} BIP for $x \in \mathbb{B}^{|\Tilde{M}|}$:
\STATE \label{step:minimize_step} \quad Minimize $v^{\top} x$ subject to $\Tilde{A} x \geq b$.
\STATE \textbf{return} $x$
\end{algorithmic}
\end{algorithm}

Consider the binary matrix
\begin{equation*}
A \in \{0, 1\}^{|B| \times |\Tilde{M}|},
\end{equation*}
where
\begin{equation}
\label{equation:def_A_first}
A_{ij} =
\begin{cases} 
1 & \text{if } \Tilde{M}_j \in \Tilde{M}^{(B_i)}\\
0 & \text{otherwise}
\end{cases}\text{.}
\end{equation}
Each row of $A$ contains $2 + (k - 2) = k$ non-zero elements. Let $x \in \{0, 1\}^{|\Tilde{M}|}$ represent the set of all selected migrations and pairs of migrations, which is a subset of $\Tilde{M}$. To ensure that $x$ is a valid representation, we impose the constraint that if a pair of migrations is selected, then each migration in the pair must also be selected. For each pair of migrations $\textbf{m} \in M^{(2)}$, we add a constraint to the vector $x$ as $e^{(\textbf{m})\top} x \geq 0$, where
\begin{equation*}
e^{(\textbf{m})} \equiv e_{\Tilde{M}.\text{idx}(\textbf{m}_1)} + e_{\Tilde{M}.\text{idx}(\textbf{m}_2)} - 2 e_{\Tilde{M}.\text{idx}(\textbf{m})}
\end{equation*}
and $e_i$ denotes the $i$-th standard unit vector in $|\Tilde{M}|$ dimensions.

Let $\bm{0}_d$ and $\bm{1}_d$ denote the all-zeros vector and the all-ones vector in $d$ dimensions, respectively. Then the objective is defined as
\begin{align}
\label{equation:formulation_1}
\text{minimize} \quad & \left[\bm{1}_{|M|}^{\top} \bm{0}_{|M^{(2)}|}^{\top}\right] x \nonumber \\
\text{subject to} \quad & \Tilde{A} x \geq \left[\bm{1}_{|B|}^{\top} \bm{0}_{|M^{(2)}|}^{\top}\right]^{\top},
\end{align}
where
\begin{equation*}
\Tilde{A} \equiv \left[A^{\top} e^{\big(M^{(2)}_{1}\big)} e^{\big(M^{(2)}_{2}\big)} \cdots e^{\Big(M^{(2)}_{|M^{(2)}|}\Big)}\right]^{\top}
\end{equation*}
and $A$ is defined as in Eq.~\eqref{equation:def_A_first} (see Algorithm~\ref{algorithm:bip_k_geq_4}).
\hl{Algorithm~\ref{algorithm:bip_k_geq_4} obtains the minimizer in Step~\ref{step:minimize_step} by passing the BIP instance to a standard mixed-integer solver (we use Gurobi \citep{gurobi}), which searches over $x\in\mathbb{B}^{|\tilde{M}|}$.
Therefore, the returned $x$ specifies an MS-GC solution for the given $\pi$, and its objective value equals the ebit cost achievable with that solution.}

\hl{To summarize, we enforce that every gate is covered by selecting at least one admissible covering option (either a home coverage or a joint coverage), while guaranteeing logical consistency by preventing the selection of a joint coverage unless its constituent migrations are also selected.
The objective in Eq.~\eqref{equation:formulation_1} then counts the number of selected \emph{single} migrations (i.e., the ebit cost), so minimizing it yields the minimum-cost feasible cover under general coverage for the fixed allocation $\pi$.
Formally, we explain the validity of this formulation through the following observation.}

\begin{table}
\caption{Truth table for Eqs.~\eqref{equation:linear_constraint_1}~and~\eqref{equation:linear_constraint_2}.}
\label{table:truth_table}
\centering
\begin{tabular}{|c|c|c|c|c|}
\hline
$x_{\Tilde{M}.\text{idx}(\textbf{m}_1)}$ & $x_{\Tilde{M}.\text{idx}(\textbf{m}_2)}$ & $x_{\Tilde{M}.\text{idx}(\textbf{m})}$ & Eq.~\eqref{equation:linear_constraint_1} & Eq.~\eqref{equation:linear_constraint_2} \\
\hline
0 & 0 & 0 & T & T \\
0 & 0 & 1 & F & F \\
0 & 1 & 0 & T & T \\
0 & 1 & 1 & F & F \\
1 & 0 & 0 & T & T \\
1 & 0 & 1 & F & F \\
1 & 1 & 0 & T & T \\
1 & 1 & 1 & T & T \\
\hline
\end{tabular}
\end{table}

\begin{observation}
\label{observation:optimality_1}
A solution of Eq.~\eqref{equation:formulation_1} is an optimal solution of MS-GC.
\end{observation}

\begin{proof}
We show that it suffices to relax the constraint
\begin{equation}
\label{equation:constraint_equality}
x_{\Tilde{M}.\text{idx}(\textbf{m})} = x_{\Tilde{M}.\text{idx}(\textbf{m}_1)} x_{\Tilde{M}.\text{idx}(\textbf{m}_2)}
\end{equation}
to
\begin{equation}
\label{equation:constraint_inequality_version}
x_{\Tilde{M}.\text{idx}(\textbf{m})} \leq x_{\Tilde{M}.\text{idx}(\textbf{m}_1)} x_{\Tilde{M}.\text{idx}(\textbf{m}_2)}.
\end{equation}
Since $x$ is a binary vector, Eq.~\eqref{equation:constraint_inequality_version} is equivalent to
\begin{align}
\label{equation:linear_constraint_1}
x_{\Tilde{M}.\text{idx}(\textbf{m}_1)} - x_{\Tilde{M}.\text{idx}(\textbf{m})} & \geq 0, \nonumber \\
x_{\Tilde{M}.\text{idx}(\textbf{m}_2)} - x_{\Tilde{M}.\text{idx}(\textbf{m})} & \geq 0,
\end{align}
and these two constraints can be merged into a single constraint as
\begin{equation}
\label{equation:linear_constraint_2}
x_{\Tilde{M}.\text{idx}(\textbf{m}_1)} + x_{\Tilde{M}.\text{idx}(\textbf{m}_2)} - 2 x_{\Tilde{M}.\text{idx}(\textbf{m})} \geq 0,
\end{equation}
reducing the number of constraints by half. The equivalence of Eqs.~\eqref{equation:linear_constraint_1} and~\eqref{equation:linear_constraint_2} is verified in Table~\ref{table:truth_table}.

The only combination that violates Eq.~\eqref{equation:constraint_equality} while making Eq.~\eqref{equation:linear_constraint_2} true is
\begin{equation*}
\left(x_{\Tilde{M}.\text{idx}(\textbf{m}_1)}, x_{\Tilde{M}.\text{idx}(\textbf{m}_2)}, x_{\Tilde{M}.\text{idx}(\textbf{m})}\right) = (1, 1, 0).
\end{equation*}
However, it is easy to see that this does not disturb the formulation. Let $\hat{x}$ be an optimal solution of Eq.~\eqref{equation:formulation_1} and assume that
\begin{equation*}
\hat{x}_{\Tilde{M}.\text{idx}(\textbf{m}_1)} = \hat{x}_{\Tilde{M}.\text{idx}(\textbf{m}_2)} = 1, \quad \hat{x}_{\Tilde{M}.\text{idx}(\textbf{m})} = 0
\end{equation*}
for some $\textbf{m} \in M^{(2)}$. Then $\hat{x}^{(\textbf{m})} \equiv \hat{x} + e_{\Tilde{M}.\text{idx}(\textbf{m})}$ is also an optimal solution because
\begin{equation*}
\left[\bm{1}_{|M|}^{\top} \bm{0}_{|M^{(2)}|}^{\top}\right] \hat{x}^{(\textbf{m})} = \left[\bm{1}_{|M|}^{\top} \bm{0}_{|M^{(2)}|}^{\top}\right] \hat{x}
\end{equation*}
and
\begin{align*}
\Tilde{A} \hat{x}^{(\textbf{m})} & = \Tilde{A} \hat{x} + \Tilde{A} e_{\Tilde{M}.\text{idx}(\textbf{m})}\\
& \geq \Tilde{A} \hat{x} = \left[\bm{1}_{|B|}^{\top} \bm{0}_{|M^{(2)}|}^{\top}\right]^{\top},
\end{align*}
i.e., $\hat{x}^{(\textbf{m})}$ is also an optimal solution of Eq.~\eqref{equation:formulation_1}, which satisfies Eq.~\eqref{equation:constraint_equality}.
\end{proof}

\begin{algorithm}
\caption{BIP for MS-GC, $k = 3$}
\label{algorithm:bip_k_eq_3}
\begin{algorithmic}[1]
\STATE \textbf{Input:} Unary gates $U$, binary gates $B$, module allocation function $\pi$
\STATE \textbf{Output:} Vector of selected migrations $x$
\STATE \hl{Construct $M^{(1,g)}$, $M^{(2,g)}$, and $M^{(1)}$ using Eqs.~\eqref{eq:m1g_m2g_definition}~and~\eqref{eq:m1_m2_mtilde_definition}.}
\STATE \hl{\emph{[Define objective and constraints]}}
\STATE Initialize:
\STATE \quad $v^{\top} \leftarrow \bm{1}_{|M^{(1)}|}^{\top}$
\STATE \quad $A \in \mathbb{B}^{|B| \times |M^{(1)}|} \leftarrow 0$
\STATE \quad $b \leftarrow 2 \cdot \bm{1}_{|B|}$
\STATE \hl{\emph{[Iterate over migrations and fill coverage constraints]}}
\FOR{$i = 1$ to $|B|$}
\FOR{$j = 1$ to $|M^{(1)}|$}
\IF{$M_j^{(1)} \in M^{(1, B_i)}$}
\STATE $A_{ij} \leftarrow 2$
\ELSIF{$M_j^{(1)} \in M^{(2, B_i)}_1$}
\STATE $A_{ij} \leftarrow 1$
\ENDIF
\ENDFOR
\ENDFOR
\STATE \hl{\emph{[Use an off-the-shelf solver]}}
\STATE Solve \hl{the following} BIP for $x \in \mathbb{B}^{|M^{(1)}|}$:
\STATE \quad Minimize $v^{\top} x$ subject to $A x \geq b$.
\STATE \textbf{return} $x$
\end{algorithmic}
\end{algorithm}

\subsection{BIP formulation of MS-GC for $k = 3$}
\label{sec:bip_formulation_k_3}

If $k = 3$, we observe that the number of BIP variables and constraints can be further reduced. Consider the matrix
\begin{equation*}
A \in \mathbb{B}^{|B| \times |M^{(1)}|},
\end{equation*}
where
\begin{equation}
\label{equation:def_A_second}
A_{ij} =
\begin{cases} 
2 & \text{if } M_j^{(1)} \in M^{(1, B_i)}\\
1 & \text{if } M_j^{(1)} \in M^{(2, B_i)}_1\\
0 & \text{otherwise}
\end{cases}
\end{equation}
(see Algorithm~\ref{algorithm:bip_k_eq_3}). To avoid confusion, notice that
\begin{equation*}
M^{(2, g)}_1 = \{ (q_i, p, f(i, t)), (q_j, p, f(j, t)) \},
\end{equation*}
where $g = (\{ q_i, q_j \}, t)$ and $p \in P \setminus \{ \pi(q_i), \pi(q_j) \}$ is unique.
Before we describe the objective function, we remark that the optimization is \hl{with respect to} $x \in \mathbb{B}^{|M^{(1)}|}$ instead of $x \in \mathbb{B}^{|M|}$. We justify this reduction of variables through the following lemma.

\begin{lemma}
\label{lemma:k_3_prune_migrations}
If $k = 3$, the power set $2^{M^{(1)}}$ includes an optimal choice of migrations.
\end{lemma}

\begin{proof}
We prove by contradiction. Let $P = \{p_1, p_2, p_3\}$ and $\pi(q_i) = p_1$. Suppose $m = (q_i, p_2, t) \notin M^{(1)}$ for some $m \in M$ is strictly necessary. By the definition of $M^{(1)}$, there is no gate $(\{q_i, q_j\}, t')$ such that $\pi(q_j) = p_2$ and $f(i, t') = t$. This means that the non-local binary gates that $m$ can cover (by itself or in a pair with another migration) must be in the form of $(\{q_i, q_j\}, t')$, where $\pi(q_j) = p_3$ and $f(i, t') = t$ (and in fact, $m$ cannot cover one of these gates by itself because it migrates $q_i$ in $p_1$ to $p_2$ while the gate acts on a qubit in $p_3$; it must be paired with a migration that sends $q_j$ to $p_2$). Since we assumed that $m$ is strictly necessary, there must exist at least one such gate. Meanwhile, notice that those gates are all covered by the \textit{single} migration $m' = (q_i, p_3, t) \in M^{(1)}$. Therefore, replacing $m$ with $m'$ does not increase the number of selected migrations without introducing any uncovered gates, which implies that any migration $m \notin M^{(1)}$ is not strictly necessary; each of them can be replaced by some $m' \in M^{(1)}$.
\end{proof}

\begin{figure}
\centering
\includegraphics[width=0.5\columnwidth]{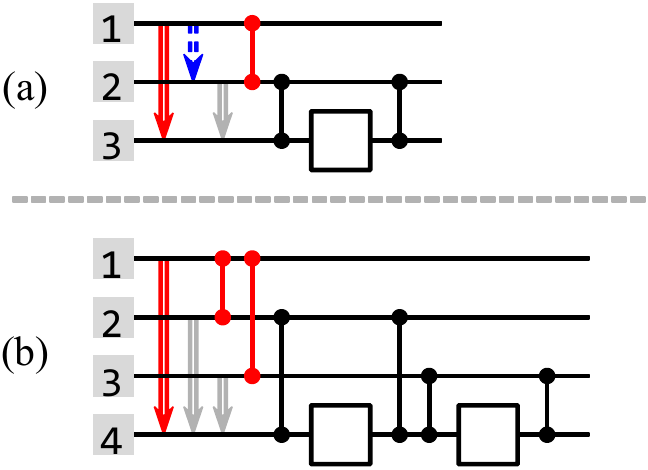}
\vspace{1em}
\caption{
\hl{Illustration of why $k=3$ admits a reduced formulation (Lemma~\ref{lemma:k_3_prune_migrations}) while $k\ge4$ does not.
Gray boxes show module indices (each module contains a single qubit), and arrows denote migrations.
(a) $k=3$: Removing the red migration makes the red gate uncovered, but it can still be covered by the single blue migration because the blue migration (which is an element of $M^{(1)}$) moves the same qubit to the other endpoint's home module. This reflects that migrations outside $M^{(1)}$ can be replaced without increasing cost.
(b) $k=4$: The circuit includes an additional set of gates, and removing the red migration leaves the red gates uncovered in a way that cannot be repaired by any single migration.
This contrast explains why we can restrict migrations to $M^{(1)}$ for $k=3$, but not for $k\ge4$.}
}
\label{fig:prune_some_migrations}
\end{figure}

Fig.~\ref{fig:prune_some_migrations} provides an example circuit to visualize Lemma~\ref{lemma:k_3_prune_migrations}, as well as a circuit with $k = 4$ to which the lemma does not apply. \hl{The contrast stems from the fact that when $k=3$ there is only \emph{one} ``third'' module besides the two endpoint-modules of any migration, whereas for $k\ge4$ there are \emph{multiple} possible third modules.}

The objective is defined as
\begin{align}
\label{equation:formulation_2}
\text{minimize} \quad & \bm{1}_{|M^{(1)}|}^{\top} x \nonumber \\
\text{subject to} \quad & A x \geq 2 \cdot \bm{1}_{|B|},
\end{align}
where $A$ is defined as in Eq.~\eqref{equation:def_A_second}. We explain the validity of this formulation through the following observation.
\begin{observation}
\label{observation:optimality_2}
A solution of Eq.~\eqref{equation:formulation_2} is an optimal solution of MS-GC for $k = 3$.
\end{observation}
\begin{proof}
By the definition of $A$, it is clear that $A_{i,:} x \geq 2$ if and only if a non-empty subset of $\left\{M^{(1, B_i)}_1, M^{(1, B_i)}_2, M^{(2, B_i)}_1\right\}$ is selected, where $A_{i,:}$ denotes the $i$-th row of $A$.
\end{proof}

Note that this is not a valid formulation for $k \geq 4$. For a gate $g = (\{q_i, q_j\}, t)$, consider two modules $p, p' \in P \setminus \{ \pi(q_i),\pi(q_j) \}$ such that $p \neq p'$. We cannot distinguish the selection of the pair
\begin{equation*}
\{ (q_i, p, f(i, t)), (q_j, p, f(j, t)) \}
\end{equation*}
which covers $g$
from the pair
\begin{equation*}
\{ (q_i, p, f(i, t)), (q_j, p', f(j, t)) \}
\end{equation*}
which does not cover $g$. In general, it is impossible for four positive ``rewards'' $r_1$, $r_2$, $r_3$, and $r_4$ to satisfy the conditions where $r_1 + r_2$ and $r_3 + r_4$ are greater than or equal to a threshold $\text{th}$, while $r_1 + r_3$, $r_1 + r_4$, $r_2 + r_3$, and $r_2 + r_4$ are all less than $\text{th}$.

\begin{remark}
The BIP formulations in Eqs.~\eqref{equation:formulation_1}~and~\eqref{equation:formulation_2} are naturally generalized to heterogeneous settings, where the communication cost between each pair of modules is not constant. We simply replace $\left[\bm{1}_{|M|}^{\top} \bm{0}_{|M^{(2)}|}^{\top}\right]$ and $\bm{1}_{|M^{(1)}|}^{\top}$ with the vectors that represent the cost function.
\end{remark}

\subsection{\hl{Illustrative example}}
\label{section:tiny_example}

\begin{figure}
  \centering
  \includegraphics[width=0.9\textwidth]{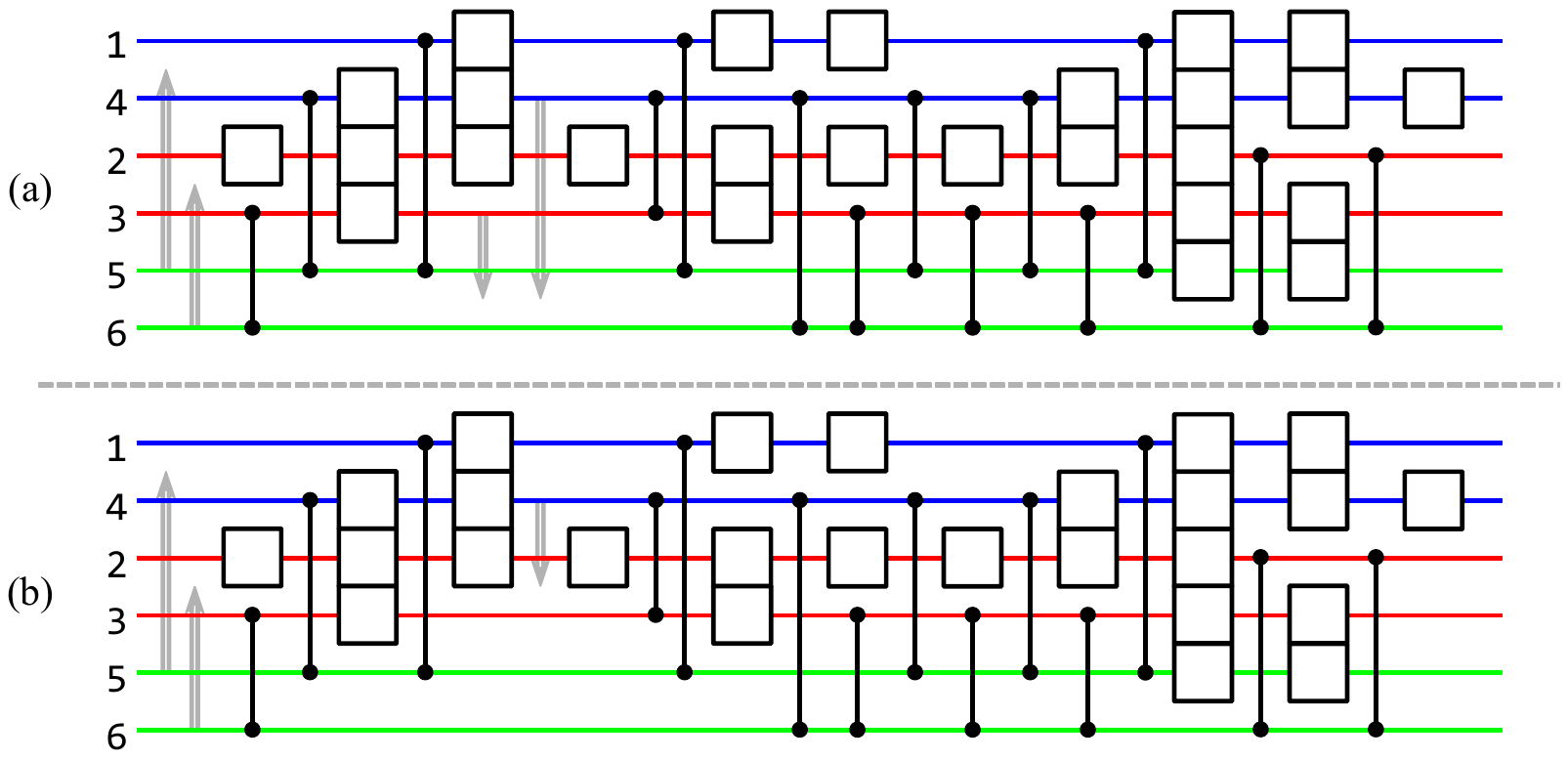}
  \vspace{1em}
  \caption{(a) Distribution found by a hypergraph partitioner and (b) BIP post-processing. Numbers denote qubit indices and wires of the same color are assigned to the same module. \hl{Arrows indicate migrations. The ebit cost drops from $4$ in (a) to $3$ in (b). The fact that the hypergraph partitioner is suboptimal even on this tiny instance motivates our BIP post-processing step.}}
  \label{fig:circuit_text}
\end{figure}

\hl{Selecting the module allocation together with the corresponding migrations and ebit distribution is combinatorially difficult.
Even for small circuits, the hypergraph partitioner can return suboptimal distributions, and our BIP post-processing can reduce the ebit cost.
For example, Fig.~\ref{fig:circuit_text} compares the distribution produced by the hypergraph partitioning approach in Section~\ref{sec:martinez_intro} with the distribution after BIP post-processing: the ebit cost drops from $4$ to $3$.}

\subsection{\hl{Problem size of the BIP formulations}}
Assuming that there are at least $O(n)$ gates in $C$, the number of binary variables is
\begin{align*}
|M| + |M^{(2)}| & \leq (k - 1)(|U| + n) + (k - 2)|B|\\
& = O(k |C|)
\end{align*}
for Eq.~\eqref{equation:formulation_1} and
\begin{equation*}
|M^{(1)}| \leq 2|B|
\end{equation*}
for Eq.~\eqref{equation:formulation_2}. Also, the number of constraints is
\begin{equation*}
|B| + |M^{(2)}| \leq (k - 1)|B|
\end{equation*}
for Eq.~\eqref{equation:formulation_1} and $|B|$ for Eq.~\eqref{equation:formulation_2}.

\subsection{Partitioning QFT}
\label{sec:partitioning_qft}

\hl{In the previous sections, we focused on the migration-selection layer.
That is, for a given module allocation $\pi$, we showed that MS-GC can be formulated as a BIP.
We also illustrated that the distribution produced by hypergraph partitioning \citep{andres2019automated} can be suboptimal.

In this section, we turn to the module allocation layer itself and discuss a structured case where $\pi$ can be chosen in a principled way.
We focus on QFT because it is a core subroutine used in many quantum algorithms, so improving its distributed implementation is valuable in its own right.
Moreover, assuming a \emph{balanced distribution}, QFT exhibits additional structure that makes identifying an optimal $\pi$ tractable in the (restricted) MS-HC setting.}

\begin{figure}
  \centering
  \includegraphics[width=\columnwidth]{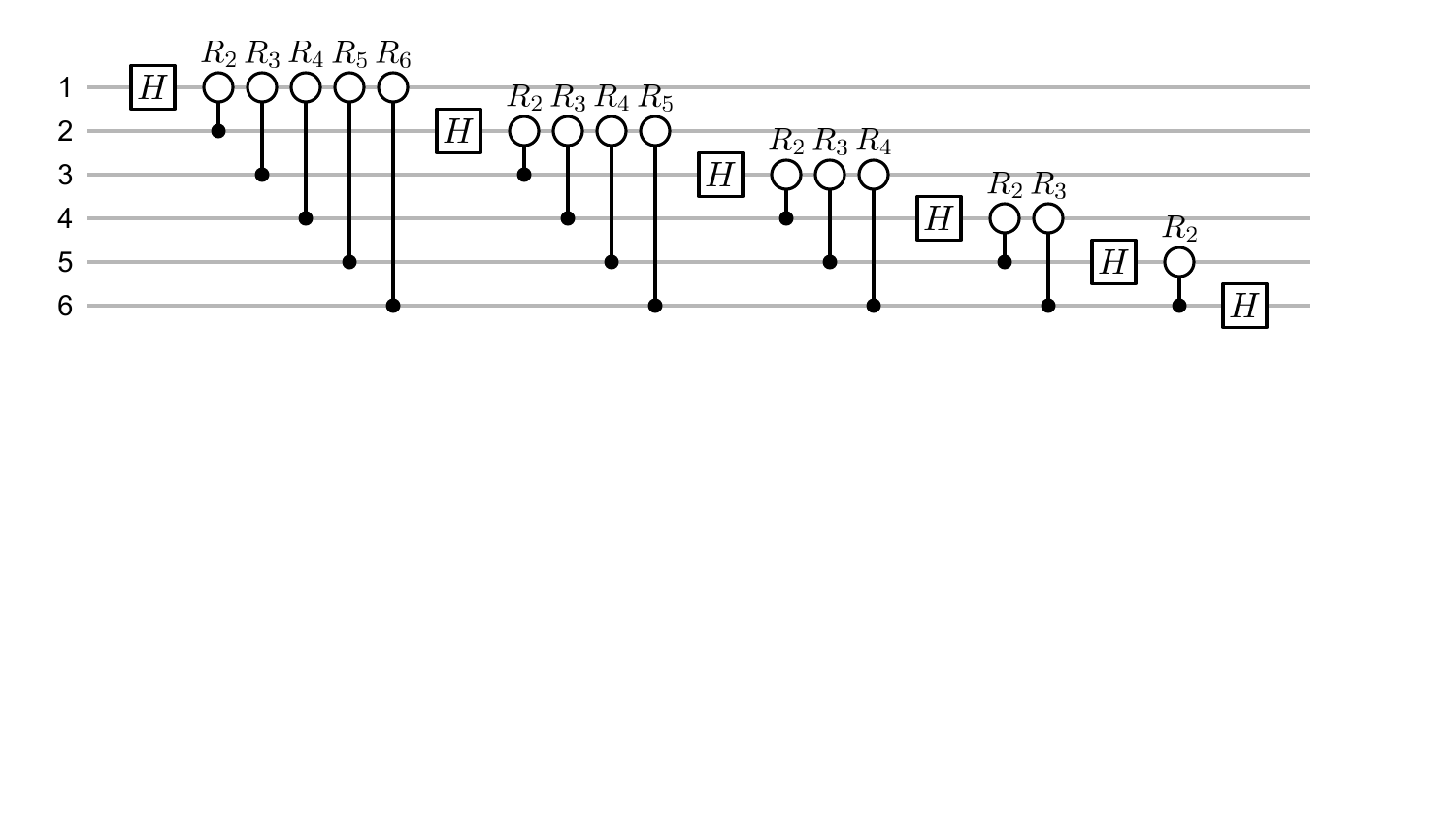}
  \caption{\hl{A monolithic $6$-qubit} QFT circuit (SWAP gates omitted). After fixing a module allocation, we remove gates that are local to each module and optimize only the remaining non-local operations.}
  \label{fig:qft}
\end{figure}

The basic structure of QFT before distribution is shown in Fig.~\ref{fig:qft}. It consists of the Hadamard gates $H$ and $CP$ gates
\begin{equation*}
CP(2 \pi / 2^k) = |0\rangle\langle 0| \otimes \mathbb{I} + |1\rangle\langle 1| \otimes R_k.
\end{equation*}
Note that we can omit the final SWAP gates and interpret the qubits in reverse order \citep{ruiz2017quantum, weinstein2001implementation}. In the subsequent diagrams, we replace each $H$/$CP(2 \pi / 2^k)$ gate with a general unary/$CP$ gate symbol.

\begin{definition}
A $(k, m)$-balanced distribution of QFT is a distribution of the QFT circuit for $n = km$ qubits, where $k$ and $m$ denote the number of modules and the number of qubits per module, respectively.
\end{definition}

It turns out that we can derive an optimal solution for any $(k, m)$-balanced distribution of QFT, assuming MS-HC. \hl{The resulting ebit cost coincides with that derived independently by \citep{kaur2025optimized}. However, we strengthen their result by establishing that this cost attains a matching lower bound, thereby proving the optimality of the scheme in the MS-HC setting. Furthermore, the module allocation introduced in the proof will also be reused in the MS-GC setting.}

\begin{remark}
\hl{While we omit the final \textnormal{SWAP} gates in our QFT distribution, \citep{kaur2025optimized} achieves the same ebit cost even when these operations are explicitly included.
They accomplish this by employing qubit routing, but such techniques fall outside the scope of our specific model (i.e., MS-GC under fixed $\pi$).
More importantly, as noted in \citep{neumann2020imperfect}, these final \textnormal{SWAP} gates can be omitted without introducing overhead in downstream processing.
This is because the \textnormal{SWAP} gates merely permute the logical qubits, and their effect can be compensated for through simple qubit reindexing.}
\end{remark}

\hl{In QFT, each qubit has exactly one unary gate and then interacts via $CP$ with every other qubit, so under MS-HC the natural migration time is immediately after that unary gate.
In a $(k,m)$-balanced allocation, migrating a qubit into a module can cover at most $m$ gates between that qubit and the $m$ qubits in the destination module, giving a tight lower bound on the number of migrations. Formally, we have the following lemma.}

\begin{lemma}
\label{lemma:qft_optimal}
The minimum ebit cost for a $(k, m)$-balanced distribution of the QFT circuit for $n = km$ qubits with MS-HC is $m \binom{k}{2}$.
\end{lemma}
\begin{proof}
We provide a simple constructive proof. The QFT circuit for $n = km$ qubits can be fully described by the sets
\begin{equation*}
U = \left\{ \left(q_i, t_{q_i}\right) \right\}_{i = 1}^n
\end{equation*}
and
\begin{equation*}
\Tilde{B} = \left\{ \left(\{q_i, q_j\}, t_{\{q_i, q_j\}}\right) \right\}_{1 \leq i < j \leq n},
\end{equation*}
because the circuit includes only one unary gate per qubit and one binary gate per pair of qubits. That is, $t_{q_i}$ and $t_{\{q_i, q_j\}}$ are uniquely defined. Assume that the qubits and gates are arranged in the conventional order as shown in Fig.~\ref{fig:qft} and let $\pi$ be a balanced module allocator function, i.e.,
\begin{equation*}
|\{ i | \pi(q_i) = p_{\kappa} \}| = m, \quad \forall \kappa \in \{ 1, 2, \cdots , k \}.
\end{equation*}
Consider a migration $(q, p, t)$ where $\pi(q) \neq p$, and let $S$ be the set of non-local binary gates that can be covered by this migration. Since we assume MS-HC, $S$ is a subset of
\begin{align*}
\Big\{ \left(\{q_i, q_j\}, t_{\{q_i, q_j\}}\right) \mid (q, p) \in \{ (q_i, \pi(q_j)), (q_j, \pi(q_i)) \} \Big\},
\end{align*}
which is equal to
\begin{equation*}
\left\{ \left(\{q, q'\}, t_{\{q, q'\}}\right) \mid \pi(q') = p \right\},
\end{equation*}
so $|S| \leq m$. Meanwhile, the total number of non-local binary gates in the distributed circuit is
\begin{align*}
\# \text{ binary} - \# \text{ local binary} & = \binom{n}{2} - k\binom{m}{2}\\
= & \frac{km(km - 1)}{2} - k\frac{m(m - 1)}{2}\\
= & \frac{k(k - 1)m^2}{2}.
\end{align*}
This gives a lower bound on the ebit cost:
\begin{equation}
\label{equation:ebit_cost_lower_bound}
\frac{k(k - 1)m^2}{2m} = \frac{k(k - 1)m}{2}\text{.}
\end{equation}
Now take $\pi = \pi^*$, where the \textit{canonical partition} $\pi^*$ is defined as
\begin{equation}
\label{equation:canonical_partition}
\pi^*(q_i) = p_{\lceil i / m \rceil}.
\end{equation}
Then the set of all non-local binary gates is
\begin{align}
\label{equation:non_local_shrink}
B = \bigcup_{i = 1}^{(k - 1)m} \bigcup_{\kappa = \lceil i / m \rceil + 1}^{k} \Big\{ \left(\{q_i, q_j\}, t_{\{q_i, q_j\}}\right) \mid \lceil j / m \rceil = \kappa \Big\}
\end{align}
and the circuit to distribute is $C = U \cup B$. But each $\left(\{q_i, q_j\}, t_{\{q_i, q_j\}} \right)$ in Eq.~\eqref{equation:non_local_shrink} is covered by
\begin{equation*}
\left(q_i, p_{\lceil j / m \rceil}, f\left(i, t_{\{q_i, q_j\}}\right) = t_{q_i} \right),
\end{equation*}
and it follows that the set
\begin{equation*}
\left\{ \left(q_i, p_{\kappa}, t_{q_i}\right) \mid \lceil i / m \rceil < \kappa \leq k \right\}
\end{equation*}
covers all gates in $B$ with ebit cost $k(k - 1)m / 2$, which achieves its lower bound Eq.~\eqref{equation:ebit_cost_lower_bound}.
\end{proof}

\begin{figure}[H]
  \centering
  \includegraphics[width=\textwidth]{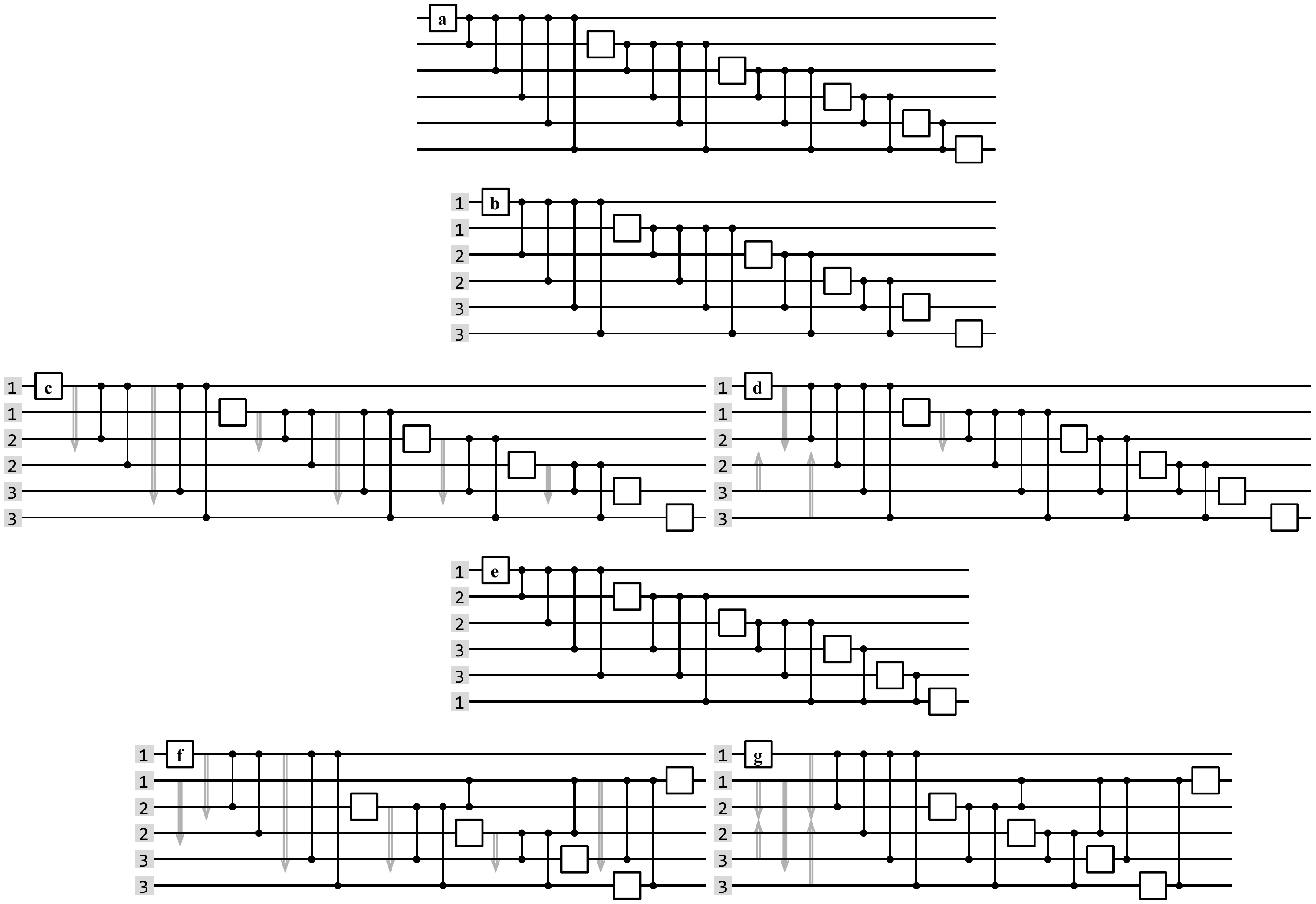}
  \vspace{1em}
  \caption{(a) A $6$-qubit QFT circuit. (b–d) Circuit $C$, MS-HC, and MS-GC for module allocation `$112233$'. (e–g) Circuit $C$, MS-HC, and MS-GC for module allocation `$122331$'. Both (c) and (f) involve $6$ migrations, which is optimal for MS-HC. However, (d) involves $4$ migrations while (g) involves $5$ migrations for MS-GC.}
  \label{fig:qft_2by2}
\end{figure}

\begin{table}[h]
\centering
\begin{tabular}{|c|c|c|c|}
\hline
$\pi$ & ebit cost & $\pi$ & ebit cost \\
\hline
112233 & 4 & 112323 & 5 \\
\hline
112332 & 5 & 121233 & 5 \\
\hline
121323 & 6 & 121332 & 6 \\
\hline
122133 & 5 & 123123 & 6 \\
\hline
123132 & 6 & 122313 & 6 \\
\hline
123213 & 6 & 123312 & 6 \\
\hline
122331 & 5 & 123231 & 6 \\
\hline
123321 & 6 & $-$ & $-$ \\
\hline
\end{tabular}
\caption{Module allocations and ebit costs found by BIP for $(3, 2)$-balanced distributions of QFT.}
\label{table:partition_msgc}
\end{table}

\begin{figure}[H]
    \centering
    \begin{subfigure}{0.6\columnwidth} 
        \centering
        \includegraphics[width=\linewidth]{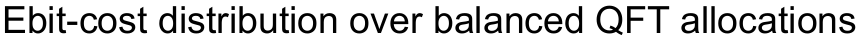}
    \end{subfigure}
    \begin{subfigure}{0.4\columnwidth}
        \includegraphics[width=\linewidth]{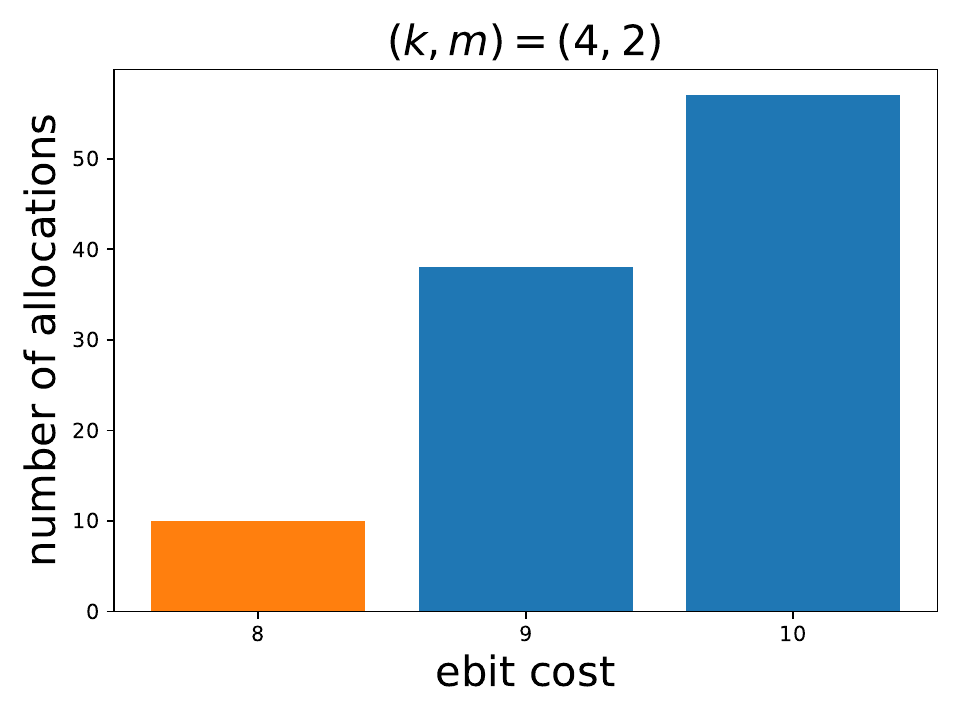}
    \end{subfigure}
    \begin{subfigure}{0.4\columnwidth}
        \includegraphics[width=\linewidth]{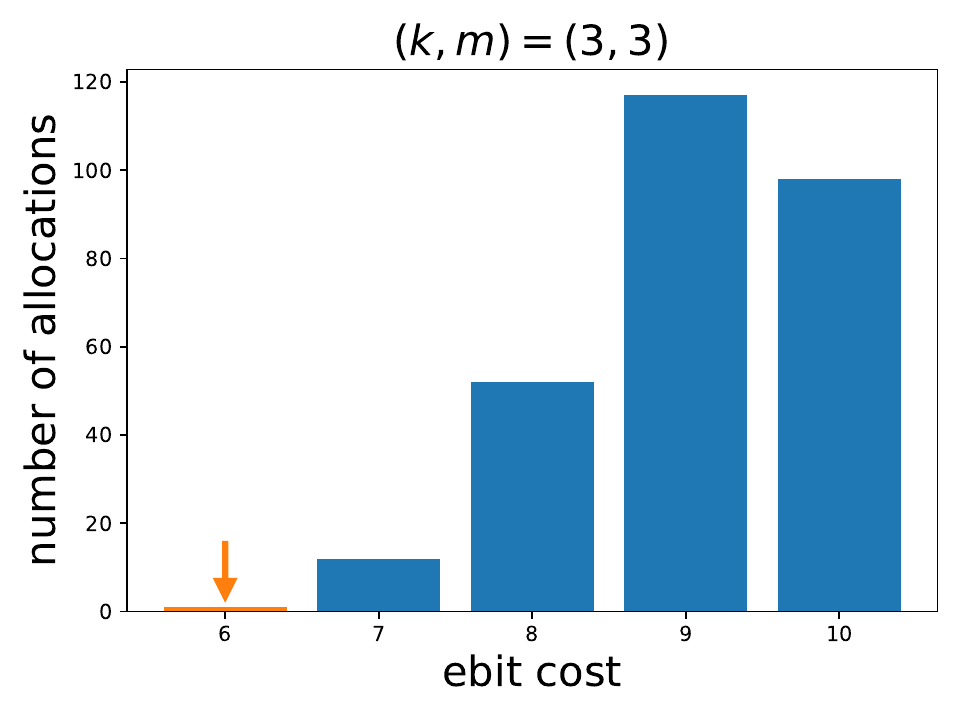}
    \end{subfigure}
    \caption{\hl{Histograms of MS-GC ebit cost for $(k,m)$-balanced distributions of QFT, computed by enumerating all balanced module allocations for the indicated $(k,m)$ and solving the associated BIP. The bar including the canonical partition $\pi^*$ is highlighted in orange. In both cases, $\pi^*$ attains the minimum MS-GC ebit cost (after BIP) among all balanced allocations, supporting our use of $\pi^*$ as the default module allocation for MS-GC.}}
    \label{fig:qft_42_33}
\end{figure}

Note that $\pi^*$ in Lemma~\ref{lemma:qft_optimal} is not the only optimal module allocation for MS-HC (see Fig.~\ref{fig:qft_2by2}). However, for $(k, m) = (3, 2)$, it is highly likely that $\pi^*$ \textit{is} the only optimal module allocation for MS-GC (see Table~\ref{table:partition_msgc}). We also ran exhaustive experiments to find the optimal allocations among all balanced allocations for $(k, m) = (4, 2)$ and $(k, m) = (3, 3)$ (see Fig.~\ref{fig:qft_42_33}). In both cases, the canonical partitions $\pi^*$ were optimal for MS-GC (assuming that the BIP solver found optimal solutions for these small examples). That is, distributing a QFT circuit by solving the BIP associated with $\pi^*$ is expected to yield the lowest ebit cost we can hope for. Indeed, we observe that this heuristic proves effective (see Section~\ref{sec:qft_experiment_section}).

\begin{remark}
\hl{One might be tempted to extend this idea beyond QFT by (i) choosing $\pi$ to be optimal for MS-HC and (ii) applying our BIP post-processing to solve MS-GC under that fixed $\pi$.
In general circuits, however, Step (i) is itself a nontrivial combinatorial problem.
Unlike QFT, we do not currently have a generic way to certify an optimal module allocation $\pi$ even under MS-HC, and moreover an allocation that is optimal for MS-HC need not be optimal for
MS-GC.
That said, there may be additional circuit families with exploitable structure where an allocation strategy can be characterized analytically or with lightweight combinatorial methods.
Natural candidates include state-preparation circuits for highly symmetric target states (e.g., GHZ- and W-type constructions), as well as phase-estimation-type circuits, which contain (inverse) QFT as a subroutine.
Identifying such structured classes and deriving allocation rules or provable bounds for them (and then combining these with our BIP post-processing) is an interesting direction for future work.}
\end{remark}

\section{Experimental results}
\label{sec:experiments}

\begin{table}
\centering
\begin{tabular}{|p{3cm}| >{\centering\arraybackslash}p{4cm} | >{\centering\arraybackslash}p{4cm} |}
\hline
\multirow{2}{*}{\textbf{Experiment}} & \multicolumn{2}{c|}{\textbf{Ebit cost reduction (\%)}} \\ \hhline{|~|--}
& \textbf{Max.} & \textbf{Min.} \\ \hline\hline
$CZ$ fraction-10\% & 1.05 & 0.00 \\ 
\hline
$CZ$ fraction-30\% & 1.56 & 0.00 \\
\hline
$CZ$ fraction-50\% & 3.33 & 0.63 \\
\hline
$CZ$ fraction-70\% & 6.77 & 0.48 \\
\hline
$CZ$ fraction-90\% & 19.6 & 2.99 \\
\hline
Quantum Volume & 0.86 & 0.00 \\
\hline
QFT & 23.1 & 0.00 \\
\hline
DraperQFTAdder & 30.4 & 0.00 \\
\hline
RGQFTMultiplier & 95.8 & 0.00 \\
\hline
AND & 84.4 & 0.00 \\
\hline
InnerProduct & 85.7 & 0.00 \\
\hline
\end{tabular}
\caption{\hl{Summary of the ebit-cost savings achieved by BIP post-processing on top of the hypergraph-partitioning baseline (HP). For each benchmark family, we report the maximum and minimum percentage reduction in total ebit cost over all tested configurations.}}
\label{table:results_compact}
\end{table}

The same algorithm and configuration described in Ref. \citep{andres2024distributing} were used for hypergraph partitioning. For the BIP solver, we used Gurobi 11.0.0 with a Named-User Academic License \citep{gurobi}. It is based on the \textit{branch-and-bound} paradigm for discrete optimization problems \citep{clausen1999branch, land2009automatic}.
\hl{Table~\ref{table:results_compact} outlines the types of circuits considered in our experiments and summarizes the results by reporting the \emph{maximum} and \emph{minimum} ebit-cost reduction (in \%) across all tested configurations for each family. Full results for every parameter setting are deferred to Appendix~\ref{sec:all_histograms_appendix}. Notably, the minimum reduction is nonnegative in all cases, i.e., the BIP post-processing never produces a worse ebit cost than the HP baseline in any run.}

\begin{remark}
\hl{Our BIP post-processing is performed during offline distribution and therefore adds classical runtime compared to a baseline solver. However, since the goal is to reduce execution-time ebit consumption, this overhead is often amortized when a distributed circuit is run many times. In such cases, the one-time post-processing overhead may be outweighed by the repeated savings in ebit cost during execution.}
\hl{Moreover, our method reduces MS-GC to a BIP and then delegates optimization to a standard solver (Gurobi).
In the current implementation, the largest instances solved in our environment are around $n=50$, $k=8$.
The current scalability ceiling is primarily driven by solver capability on these instances.
We plan to extend this regime by exploiting circuit-level local regularities to eliminate redundant variables and constraints and thereby accelerate the optimization, while leaving improvements to general-purpose BIP technology outside the scope of this manuscript.}
\end{remark}

\hl{In what follows, we describe the experimental circuit families and their structures in detail.
These benchmarks are largely divided into two categories: families of random circuits and families of structured arithmetic circuits.
For each family we provide a representative circuit diagram.}

\clearpage
\subsection{Random circuits}

\subsubsection{$CZ$ fraction circuits}

\begin{algorithm}[h]
\caption{$CZ$ fraction}
\label{algorithm:cz_fraction}
\begin{algorithmic}[1]
\STATE \textbf{Input:} $n, d, p$
\STATE \textbf{Output:} A random $n$-qubit circuit
\STATE Initialize:
\STATE \quad $U \leftarrow \emptyset$
\STATE \quad $B \leftarrow \emptyset$
\FOR{$l = 1$ to $d$}
\FOR{$i = 1$ to $n$}
\STATE With probability $1 - p$, add a unary gate acting on $q_i$ to $U$.
\ENDFOR
\STATE Randomly pair the qubits to which no unary gate was applied and add a $CZ$ gate to $B$ for each pair.
\ENDFOR
\STATE \textbf{return} $C = U \cup B$
\end{algorithmic}
\end{algorithm}

\begin{figure}[h]
\centering
\includegraphics[width=0.4\columnwidth]{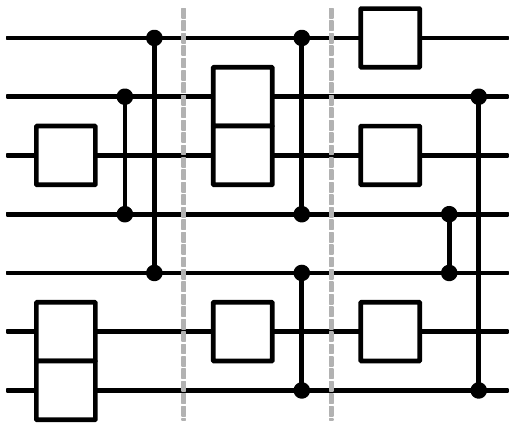}
\vspace{1em}
\caption{An example $7$-qubit $3$-layer $CZ$ fraction circuit with $p = 4 / 7$.}
\label{fig:cz_fraction_example}
\end{figure}

Given parameters $(n, d, p)$, an $n$-qubit $CZ$ fraction circuit is generated as described in Algorithm~\ref{algorithm:cz_fraction} \citep{g2021efficient}. Fig.~\ref{fig:cz_fraction_example} shows an example $CZ$ fraction circuit.
\hl{It is observed from Table~\ref{table:results_compact} that as the $CZ$ fraction increases from 10\% to 90\%, the maximum ebit saving achieved by the BIP post-processing grows.}
Ebit costs for distributing random $CZ$ fraction circuits are shown in Fig.~\ref{fig:many_figures} in Appendix~\ref{section:cz_fraction_appendix}.

\clearpage

\subsubsection{Quantum volume circuits}

\begin{algorithm}[h]
\caption{Quantum volume}
\label{algorithm:quantum_volume}
\begin{algorithmic}[1]
\STATE \textbf{Input:} $n, d$
\STATE \textbf{Output:} A random $n$-qubit circuit
\FOR{$l = 1$ to $d$}
\STATE Split $Q$ into $\lfloor n / 2 \rfloor$ pairs $\{\{q_i, q'_i\}\}_{1 \leq i \leq \lfloor n / 2 \rfloor}$.
\FOR{$i = 1$ to $\lfloor n / 2 \rfloor$}
\STATE Generate a random unitary from $\text{SU}(4)$ according to the Haar measure and apply it to $\{q_i, q'_i\}$.
\ENDFOR
\ENDFOR
\STATE Transpile the resulting circuit to unary gates $U$ and $CP$ gates $B$.
\STATE \textbf{return} $C = U \cup B$
\end{algorithmic}
\end{algorithm}

\begin{figure}[h]
\centering
\includegraphics[width=0.4\columnwidth]{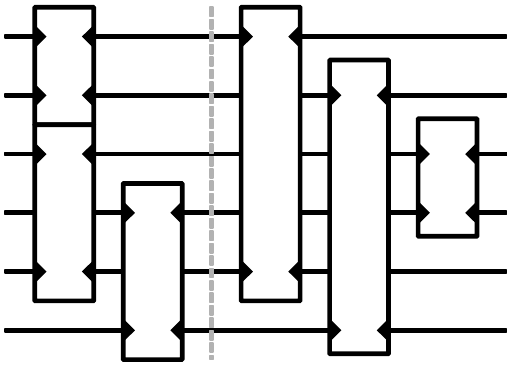}
\vspace{1em}
\caption{An example $6$-qubit $2$-layer quantum volume circuit before transpilation.}
\label{fig:volume_example}
\end{figure}

Given parameters $(n, d)$, an $n$-qubit quantum volume circuit is generated as described in Algorithm~\ref{algorithm:quantum_volume} \citep{cross2019validating}. Fig.~\ref{fig:volume_example} (right) shows an example quantum volume circuit. Ebit costs for distributing random quantum volume circuits are shown in Fig.~\ref{fig:many_figures_volume} in Appendix~\ref{section:quantum_volume_appendix}.

\subsection{Arithmetic circuits}
\subsubsection{QFT circuits}
\label{sec:qft_experiment_section}

Ebit costs for distributing QFT circuits are shown in Fig.~\ref{fig:qft_result_figure} in Appendix~\ref{section:qft_appendix}.
We remark that replacing $\pi$ (found by a hypergraph partitioner) with $\pi^*$ defined in Eq.~\eqref{equation:canonical_partition} improved the ebit cost in every single run of the BIP solver. In this regard, we conjecture that for the distribution of QFT circuits, $\pi^*$ is not only the optimal module allocation function for MS-HC but also for MS-GC.

\clearpage

\subsubsection{DraperQFTAdder}

\begin{figure}[h]
    \centering
    \begin{subfigure}{0.80344\textwidth}
        \centering
        \includegraphics[width=\textwidth]{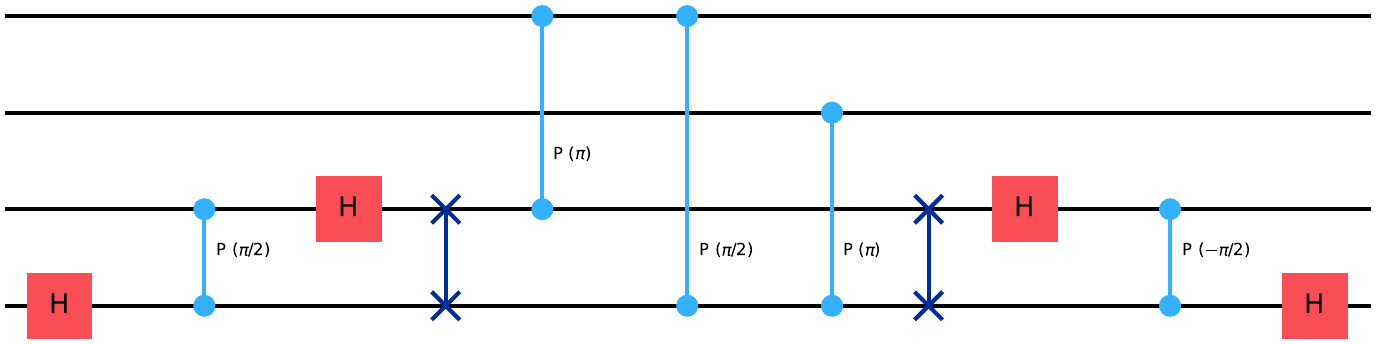}
        \caption{}
    \end{subfigure}
    \vspace{0.5em}
    \vspace{0.5em}
    \begin{tikzpicture}
        \draw[dashed] (0,0) -- (\linewidth,0);
    \end{tikzpicture}
    \begin{minipage}[c]{0.27\textwidth}
        \begin{subfigure}{\textwidth}
            \centering
            \includegraphics[width=\textwidth]{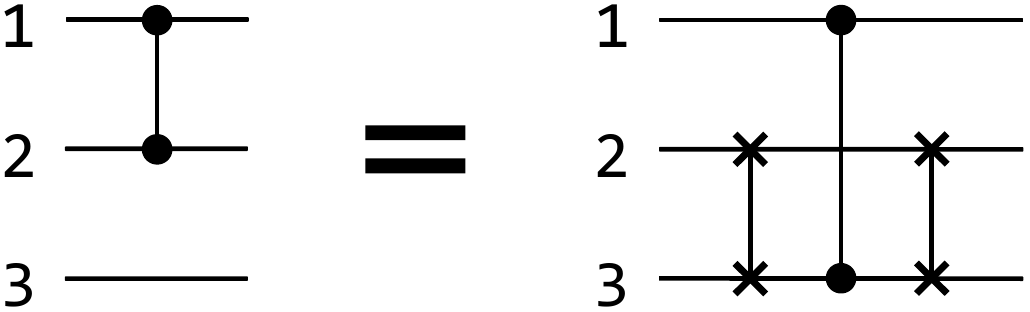}
            \caption{}
        \end{subfigure}
    \end{minipage}
    \begin{minipage}[c]{0.02\textwidth}
        \centering
        \begin{tikzpicture}
            \draw[dashed] (0,0) -- (0,-5);
        \end{tikzpicture}
    \end{minipage}
    \begin{minipage}[c]{0.68992\textwidth}
        \begin{subfigure}{\textwidth}
            \centering
            \includegraphics[width=\textwidth]{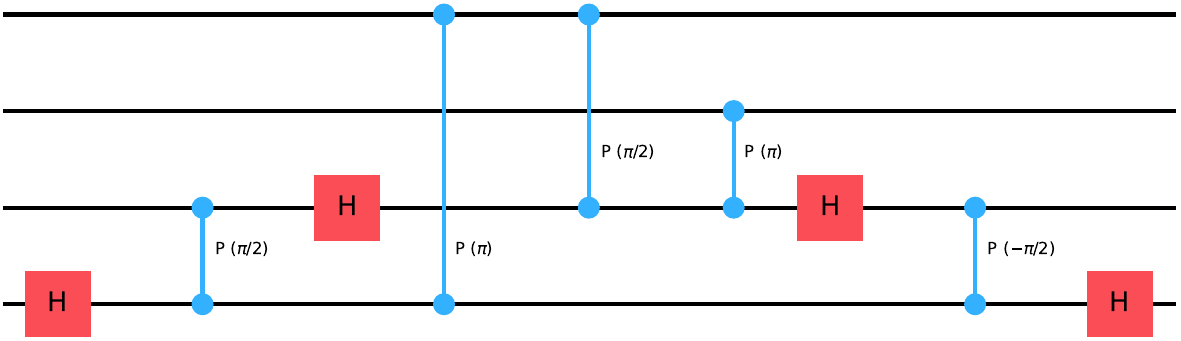}
            \caption{}
        \end{subfigure}
    \end{minipage}
    \vspace{0.5em}
    \caption{(a) Example DraperQFTAdder circuit that performs in-place addition (modulo $2^2$) on two $2$-qubit registers. (b) Circuit representation of Eq.~\eqref{equation:swap_identity}. (c) Equivalent circuit obtained by applying Eq.~\eqref{equation:swap_identity}.}
    \label{fig:draper_explain}
\end{figure}

The DraperQFTAdder is a quantum circuit that leverages QFT to perform addition \citep{draper2000addition}. Before distribution, we perform a minor optimization on this circuit. Consider the identity
\begin{equation}
\label{equation:swap_identity}
CP_{12}(\theta) = \text{SWAP}_{23}CP_{13}(\theta)\text{SWAP}_{23},
\end{equation}
where the subscripts denote qubit indices (see Appendix~\ref{section:draper_appendix}). The right-hand side appears many times in DraperQFTAdder circuits (Fig.~\ref{fig:draper_explain}). Since SWAP gates require binary gates to implement, we replace the right-hand side of Eq.~\eqref{equation:swap_identity} with the left-hand side. Ebit costs for distributing DraperQFTAdder circuits are shown in Fig.~\ref{fig:many_figures_draper} in Appendix~\ref{section:draper_appendix}.

\clearpage

\subsubsection{RGQFTMultiplier}

\begin{figure}[h]
\centering
\includegraphics[width=0.4\textwidth]{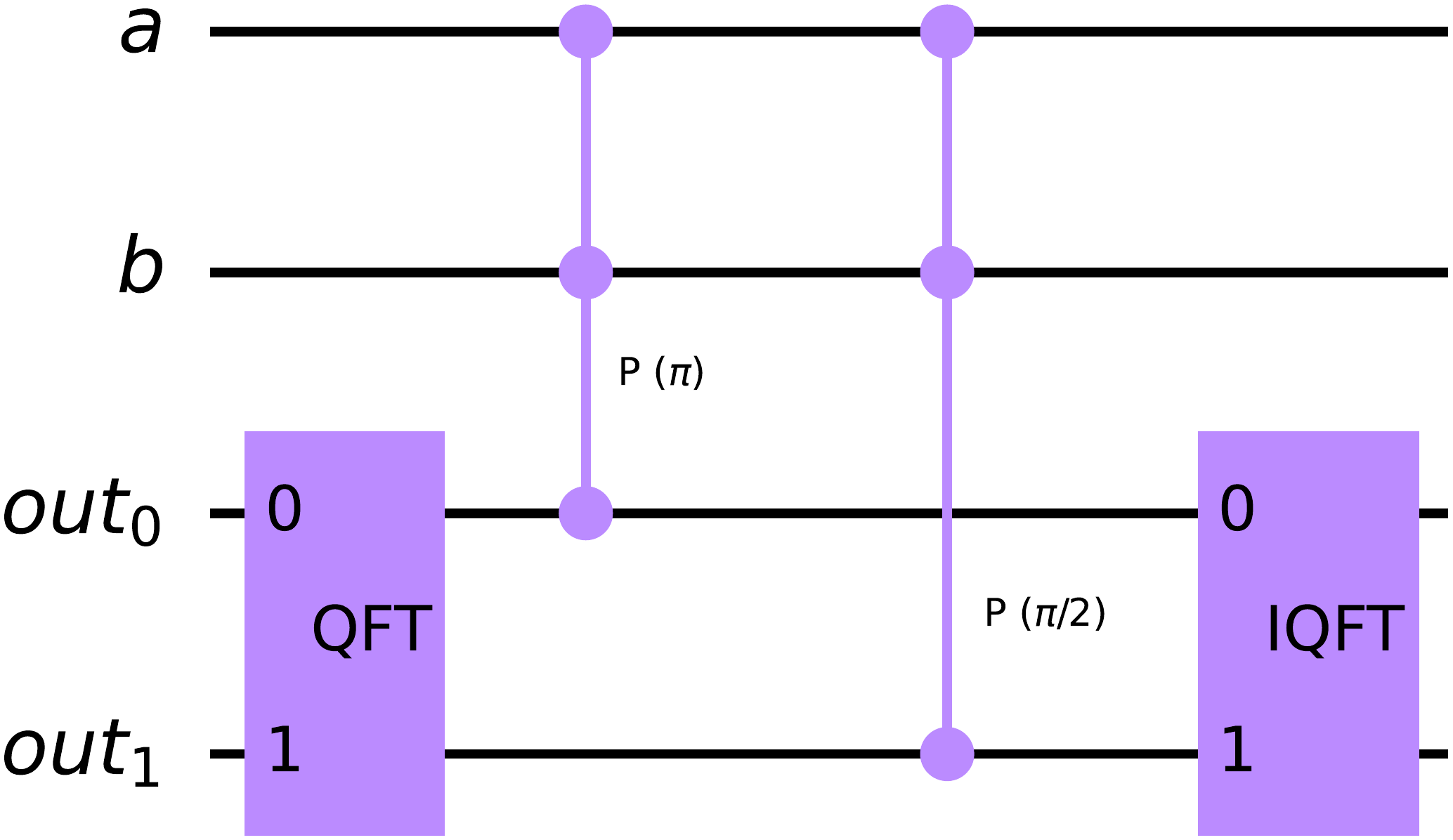}
\vspace{1em}
\caption{Example RGQFTMultiplier circuit that computes the product of two bits ($n' = 1$).}
\label{fig:rgqft_example_figure}
\end{figure}

The RGQFTMultiplier is a quantum circuit that leverages QFT to perform multiplication \citep{ruiz2017quantum}. It stores the product of two $n'$-bit inputs out-of-place. By default, the output register has $2 n'$ qubits, resulting in a total of $4 n'$ qubits (Fig.~\ref{fig:rgqft_example_figure}).

Ebit costs for distributing RGQFTMultiplier circuits are shown in Fig.~\ref{fig:rgqft_results} in Appendix~\ref{section:rgqftmultiplier_appendix}.
Surprisingly, the hypergraph partitioner alone returns extremely inefficient distributions in many cases, and BIP post-processing reduces the ebit costs significantly.

\subsection{Boolean logic circuits}
The hypergraph partitioning approach performs well for random circuits but sometimes fails to yield good solutions for circuits with fixed structures, as observed in the case of RGQFTMultiplier. Unfortunately, we have yet to identify the reasons behind this failure.

The BIP post-processing based on a given module allocation function (found by any means) enhances stability. We present two additional examples of Boolean logic circuits: the AND circuit and the InnerProduct circuit. Even for small circuit size parameters, the impact of BIP post-processing is significant.

\clearpage

\subsubsection{AND circuits}

\begin{figure}[h]
    \centering
    \begin{subfigure}[c]{0.32\columnwidth}
        \includegraphics[width=\linewidth]{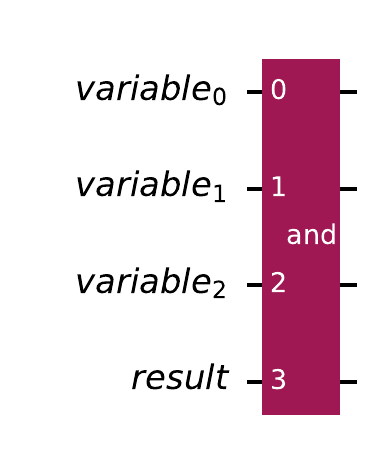}
        \caption{}
    \end{subfigure}
    \hspace{2em}
    \begin{subfigure}[c]{0.4\columnwidth}
        \includegraphics[width=\linewidth]{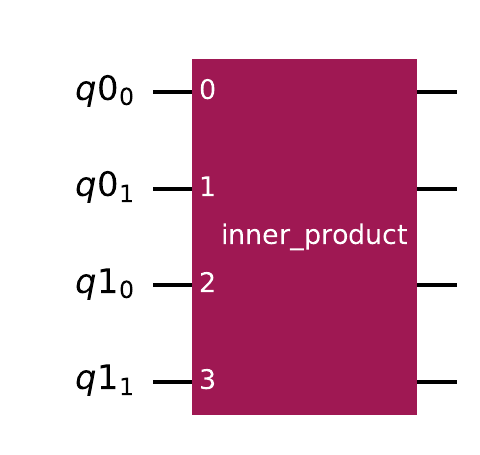}
        \caption{}
    \end{subfigure}
    \caption{(a) Example logical AND operation on 3 qubits. (b) Example $4$-qubit Boolean function that computes the inner product of two $2$-qubit vectors over $F_2$.}
    \label{fig:and_inner_examples}
\end{figure}

The AND circuit implements the logical AND operation on a number of qubits, i.e., it is a multi-controlled-$X$ gate (Fig.~\ref{fig:and_inner_examples}a). Ebit costs for distributing AND circuits are shown in Fig.~\ref{fig:and_plots} in Appendix~\ref{section:and_appendix}.

\subsubsection{InnerProduct circuits}

InnerProduct is a 2$n$-qubit Boolean function that computes the inner product of two $n$-qubit vectors over $F_2$ (Fig.~\ref{fig:and_inner_examples}b):
\begin{equation*}
\mathcal{IP}_{2n} (x_1, \cdots , x_n, y_1, \cdots , y_n) = (-1)^{x \cdot y}.
\end{equation*}
Ebit costs for distributing InnerProduct circuits are shown in Fig.~\ref{fig:inner_plots} in Appendix~\ref{section:innerproduct_appendix}.


\section{Conclusion and outlook}
\hl{We proposed exact BIP formulations for the DQC problem. Given a fixed module allocation function, we optimize the remaining migration-selection layer to minimize ebit cost. This decoupling turns the fixed-allocation subproblem into a step that can be used as post-processing on top of any allocation method, including hypergraph partitioning. We further derived a tighter specialization for the three-module case, leveraging the fact that it admits a reduced migration search space.

Our experiments indicate that BIP post-processing is a reliable choice for distributions obtained from heuristic allocators. Concretely, we observed reductions of up to 20\% for random circuits, and in several structured arithmetic benchmarks the reduction exceeded an order of magnitude. We also found that the magnitude of improvement depends on circuit structure. For instance, circuits with higher fractions of non-local gates tend to admit larger savings.

The BIP is performed offline during compilation and therefore introduces additional classical runtime, but this overhead can be amortized when the same distributed circuit is executed arbitrarily many times. In this regime, even moderate reductions in ebit cost translate directly into recurring savings in entanglement consumption and classical communication overhead at execution time.

Several directions can further improve scalability and broaden applicability.
First, many practical circuits exhibit local regularities (e.g., repeated gate motifs or restricted interaction patterns). Exploiting such structure to eliminate redundant variables and constraints could substantially speed up the optimization.
Second, while we highlighted quantum Fourier transform as a structured case where allocation admits principled choices, there may be other structured families where the optimal allocation can be nicely characterized. Combining such allocation rules with BIP post-processing is a promising avenue.
Finally, it would be valuable to extend integer-programming models beyond the cat-entanglement-based protocol, and to incorporate additional system constraints such as restricted inter-module connectivity or heterogeneous link costs \citep{ferrari2023modular}.}

\clearpage

\begin{appendices}
\renewcommand{\thefigure}{\arabic{figure}}
\setcounter{figure}{19}

\section{\hl{Implementing non-local controlled unitary gate using a shared Bell pair}}
\label{sec:controlled_U_explanation}

This section details the underlying mechanics of Fig.~\ref{fig:cat_entanglement}.
Let $a,e,e',b$ be single-qubit registers. Let $U$ be an arbitrary single-qubit unitary acting on $b$.
We assume the input on $ab$ is an arbitrary pure state
\begin{equation}
\label{eq:input_ab}
\ket{\psi}_{ab}=\alpha\ket{00}+\beta\ket{01}+\gamma\ket{10}+\delta\ket{11},
\end{equation}
and $ee'$ share the Bell pair
\begin{equation*}
\ket{\Phi^+}_{ee'}=\frac{\ket{00}+\ket{11}}{\sqrt2}.
\end{equation*}
We show that the following procedure implements controlled-$U$ with control $a$ and target $b$:
\begin{enumerate}
\item Apply $\mathrm{CNOT}$ with control $a$ and target $e$.
\item Measure $e$ in the computational basis, obtaining outcome $m\in\{0,1\}$.
      If $m=1$, apply $X$ on $e'$.
\item Apply controlled-$U$ with control $e'$ and target $b$.
\item Apply $H$ on $e'$, measure $e'$ in the computational basis, obtaining outcome $n\in\{0,1\}$.
      If $n=1$, apply $Z$ on $a$.
\end{enumerate}
The register order is permuted as needed to maintain readability.

\textbf{Step 0 (initial state).}
The joint initial state on $a,e,e',b$ is
\begin{equation*}
\ket{\Psi_0}
=\ket{\psi}_{ab}\otimes \ket{\Phi^+}_{ee'}
=\frac{1}{\sqrt2}\Big(\ket{\psi}_{ab}\ket{00}_{ee'}+\ket{\psi}_{ab}\ket{11}_{ee'}\Big).
\end{equation*}

\textbf{Step 1 (apply $\mathrm{CNOT}_{a\to e}$).}
Applying $\mathrm{CNOT}$ with control $a$ and target $e$ flips $e$ iff $a=1$.
A direct expansion gives
\begin{align*}
\ket{\Psi_1}
=\frac{1}{\sqrt2}\Big(
&(\alpha\ket{00}+\beta\ket{01})_{ab}\ket{00}_{ee'}
+(\gamma\ket{10}+\delta\ket{11})_{ab}\ket{10}_{ee'}\\
+&(\alpha\ket{00}+\beta\ket{01})_{ab}\ket{11}_{ee'}
+(\gamma\ket{10}+\delta\ket{11})_{ab}\ket{01}_{ee'}
\Big).
\end{align*}
Grouping by the computational basis of $e$ yields
\begin{align*}
\ket{\Psi_1}
=\frac{1}{\sqrt2}\Big(
&\ket{0}_e\Big[(\alpha\ket{00}+\beta\ket{01})_{ab}\ket{0}_{e'}
+(\gamma\ket{10}+\delta\ket{11})_{ab}\ket{1}_{e'}\Big]\\
+&\ket{1}_e\Big[(\gamma\ket{10}+\delta\ket{11})_{ab}\ket{0}_{e'}
+(\alpha\ket{00}+\beta\ket{01})_{ab}\ket{1}_{e'}\Big]
\Big).
\end{align*}

\textbf{Step 2 (measure $e$ and conditionally correct $e'$).}
Measure $e$ in the computational basis, obtaining outcome $m\in\{0,1\}$.
If $m=0$, the (unnormalized) post-measurement state on $a,e',b$ is
\begin{equation*}
\left|\Psi_2^{(0)}\right\rangle
=(\alpha\ket{00}+\beta\ket{01})_{ab}\ket{0}_{e'}
+(\gamma\ket{10}+\delta\ket{11})_{ab}\ket{1}_{e'}.
\end{equation*}
If $m=1$, the (unnormalized) post-measurement state on $a,e',b$ is
\begin{equation*}
\left|\Psi_2^{(1)}\right\rangle
=(\gamma\ket{10}+\delta\ket{11})_{ab}\ket{0}_{e'}
+(\alpha\ket{00}+\beta\ket{01})_{ab}\ket{1}_{e'}.
\end{equation*}
Applying $X$ on $e'$ when $m=1$ swaps $\ket{0}_{e'}\leftrightarrow \ket{1}_{e'}$, hence
\begin{equation*}
X_{e'}\left|\Psi_2^{(1)}\right\rangle
=(\alpha\ket{00}+\beta\ket{01})_{ab}\ket{0}_{e'}
+(\gamma\ket{10}+\delta\ket{11})_{ab}\ket{1}_{e'}.
\end{equation*}
Therefore, after the classical feed-forward correction, \emph{independently of $m$} the
(unnormalized) state can be written as
\begin{align}
\label{eq:psi2_clean}
\ket{\Psi_2}
&=(\alpha\ket{00}+\beta\ket{01})_{ab}\ket{0}_{e'}
+(\gamma\ket{10}+\delta\ket{11})_{ab}\ket{1}_{e'} \nonumber\\
&=\ket{0}_a\ket{0}_{e'}(\alpha\ket{0}+\beta\ket{1})_b
+\ket{1}_a\ket{1}_{e'}(\gamma\ket{0}+\delta\ket{1})_b.
\end{align}
In particular, $e'$ is perfectly correlated with $a$ in the computational basis.

\textbf{Step 3 (apply controlled-$U$ with control $e'$ and target $b$).}
Applying $CU_{e'\to b}$ to Eq.~\eqref{eq:psi2_clean} yields
\begin{equation*}
\ket{\Psi_3}
=\ket{0}_a\ket{0}_{e'}(\alpha\ket{0}+\beta\ket{1})_b
+\ket{1}_a\ket{1}_{e'}\,U(\gamma\ket{0}+\delta\ket{1})_b.
\end{equation*}

\textbf{Step 4 (apply $H$ on $e'$ and measure $e'$).}
Using $H\ket{0}=(\ket{0}+\ket{1})/\sqrt2$ and $H\ket{1}=(\ket{0}-\ket{1})/\sqrt2$,
we obtain
\begin{align*}
\ket{\Psi_4}
=\frac{1}{\sqrt2}\Big[
&\ket{0}_{e'}\Big(\ket{0}_a(\alpha\ket{0}+\beta\ket{1})_b
+\ket{1}_a\,U(\gamma\ket{0}+\delta\ket{1})_b\Big)\nonumber\\
+&\ket{1}_{e'}\Big(\ket{0}_a(\alpha\ket{0}+\beta\ket{1})_b
-\ket{1}_a\,U(\gamma\ket{0}+\delta\ket{1})_b\Big)
\Big].
\end{align*}
Now measure $e'$ in the computational basis, obtaining outcome $n\in\{0,1\}$.
Conditioned on $n$, the (unnormalized) post-measurement state on $ab$ is:
\begin{align}
\label{eq:psi5_n0}
n=0:\quad \left|\Psi_5^{(0)}\right\rangle
&=\ket{0}_a(\alpha\ket{0}+\beta\ket{1})_b
+\ket{1}_a\,U(\gamma\ket{0}+\delta\ket{1})_b,\\
\label{eq:psi5_n1}
n=1:\quad \left|\Psi_5^{(1)}\right\rangle
&=\ket{0}_a(\alpha\ket{0}+\beta\ket{1})_b
-\ket{1}_a\,U(\gamma\ket{0}+\delta\ket{1})_b.
\end{align}

\textbf{Step 5 (apply $Z$ on $a$ if $n=1$).}
If $n=1$, apply $Z$ on $a$. Since $Z\ket{0}=\ket{0}$ and $Z\ket{1}=-\ket{1}$,
this removes the relative minus sign in Eq.~\eqref{eq:psi5_n1}, giving
\begin{equation}
\label{eq:psi_final}
\ket{\Psi_{\mathrm{final}}}
=\ket{0}_a(\alpha\ket{0}+\beta\ket{1})_b
+\ket{1}_a\,U(\gamma\ket{0}+\delta\ket{1})_b,
\end{equation}
independently of the measurement outcomes $m,n$.

\textbf{Identify $CU_{a\to b}$.}
Writing the original input Eq.~\eqref{eq:input_ab} as
\begin{equation*}
\ket{\psi}_{ab}
=\ket{0}_a(\alpha\ket{0}+\beta\ket{1})_b
+\ket{1}_a(\gamma\ket{0}+\delta\ket{1})_b,
\end{equation*}
we see that Eq.~\eqref{eq:psi_final} is exactly the result of applying controlled-$U$
with control $a$ and target $b$, i.e.,
\begin{equation*}
\ket{\Psi_{\mathrm{final}}}
=CU_{a\to b}\ket{\psi}_{ab}.
\end{equation*}
Hence the protocol implements the desired non-local controlled-$U$ gate.

\section{Cat-entanglements commute with $CP$ gates}
\label{appendix:commute}
\setcounter{figure}{13}
\begin{figure}[h]
  \centering
  \includegraphics[width=\columnwidth]{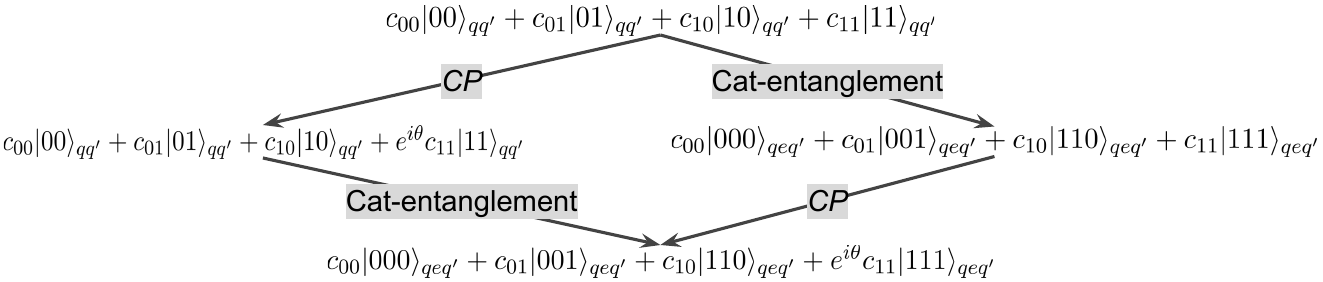}
  \vspace{1em}
  \caption{A linked copy $e$ of $q$ is created at the module containing $q'$. The order of cat-entanglement and $CP$ does not matter.}
  \label{fig:commute_general}
\end{figure}

\clearpage

\section{Cat-entanglements do not commute with unary gates}
\label{appendix:non_commute}
\begin{figure}[h]
  \centering
  \includegraphics[width=\columnwidth]{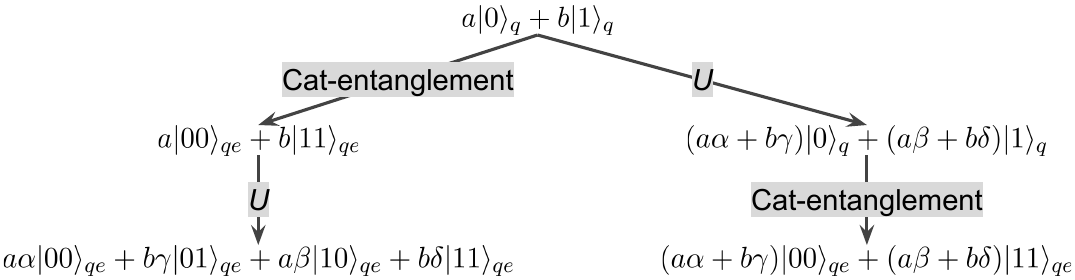}
  \vspace{1em}
  \caption{A linked copy $e$ of $q$ is created at the module containing $q'$. The order of cat-entanglement and $U$ matters.}
  \label{fig:non_commute}
\end{figure}

\section{Additional experimental results}
\label{sec:all_histograms_appendix}

\hl{This section provides the full ebit-cost plots for all benchmark families in Section~\ref{sec:experiments}.
Additionally, the execution time for each typical BIP instance is shown as a line plot overlaid on the corresponding bars.}

\clearpage

\subsection{$CZ$ fraction circuits}
\label{section:cz_fraction_appendix}

\begin{figure}[h]
    \centering
    \begin{subfigure}{0.15\textwidth}
        \includegraphics[width=\linewidth]{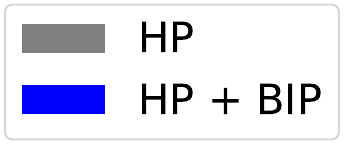}
    \end{subfigure}
    \\
    \begin{subfigure}{0.25\textwidth}
        \includegraphics[width=\linewidth]{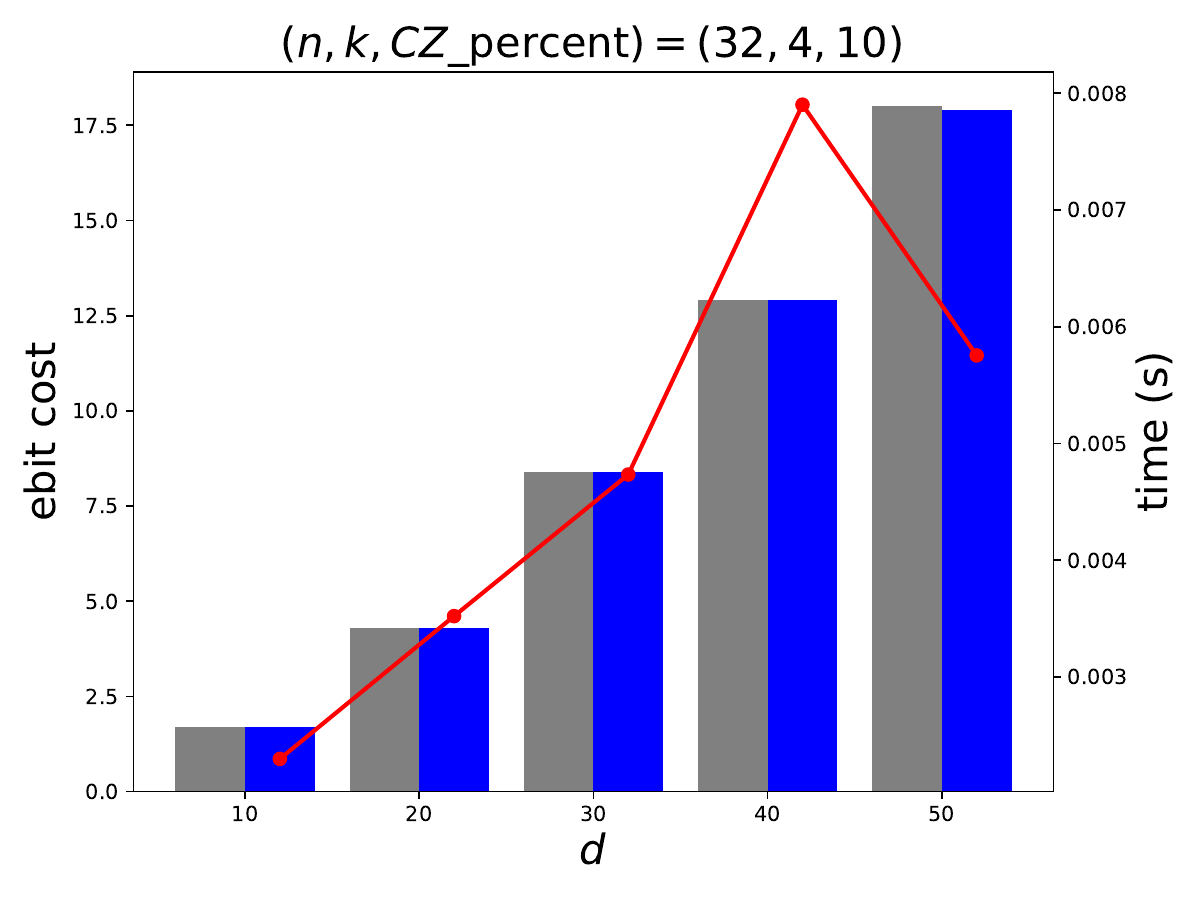}
    \end{subfigure}
    \begin{subfigure}{0.25\textwidth}
        \includegraphics[width=\linewidth]{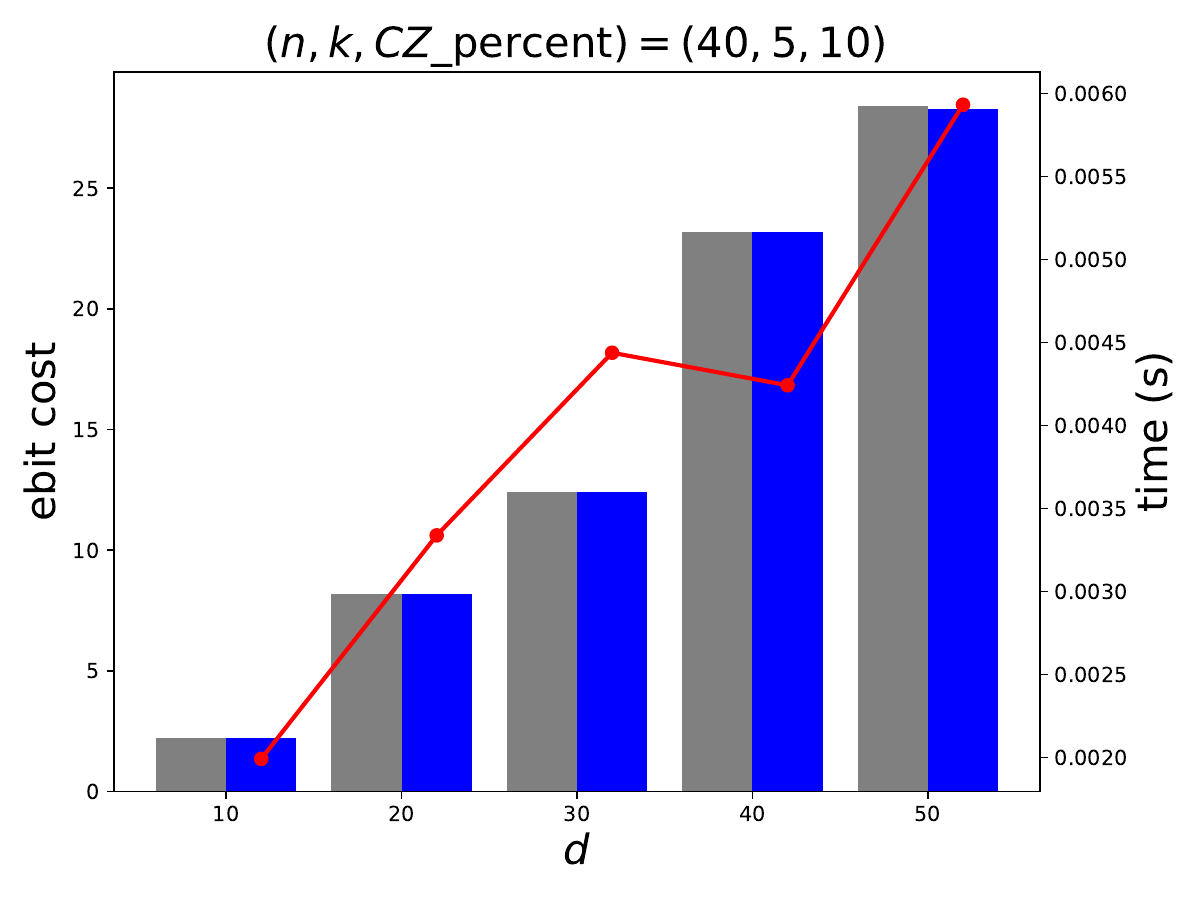}
    \end{subfigure}
    \begin{subfigure}{0.25\textwidth}
        \includegraphics[width=\linewidth]{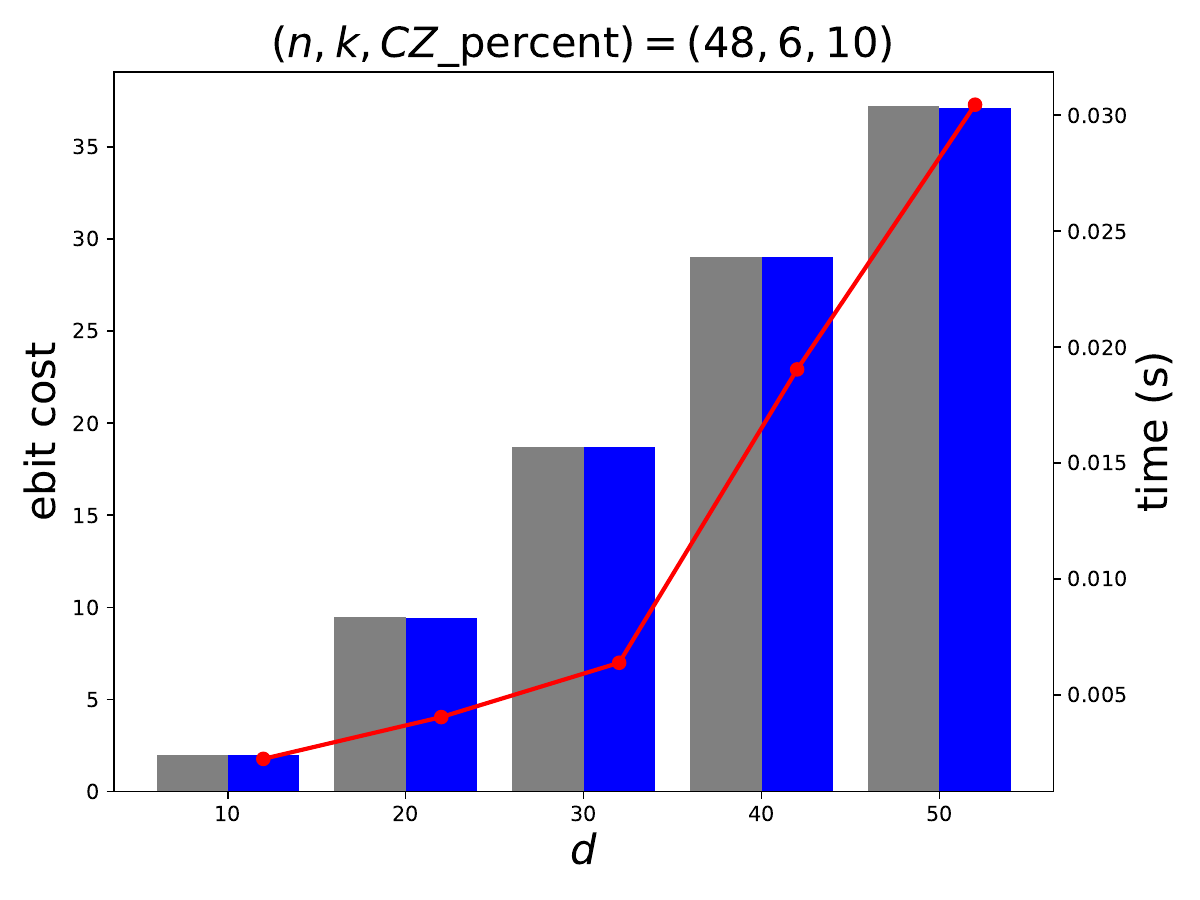}
    \end{subfigure}
    \begin{subfigure}{0.25\textwidth}
        \includegraphics[width=\linewidth]{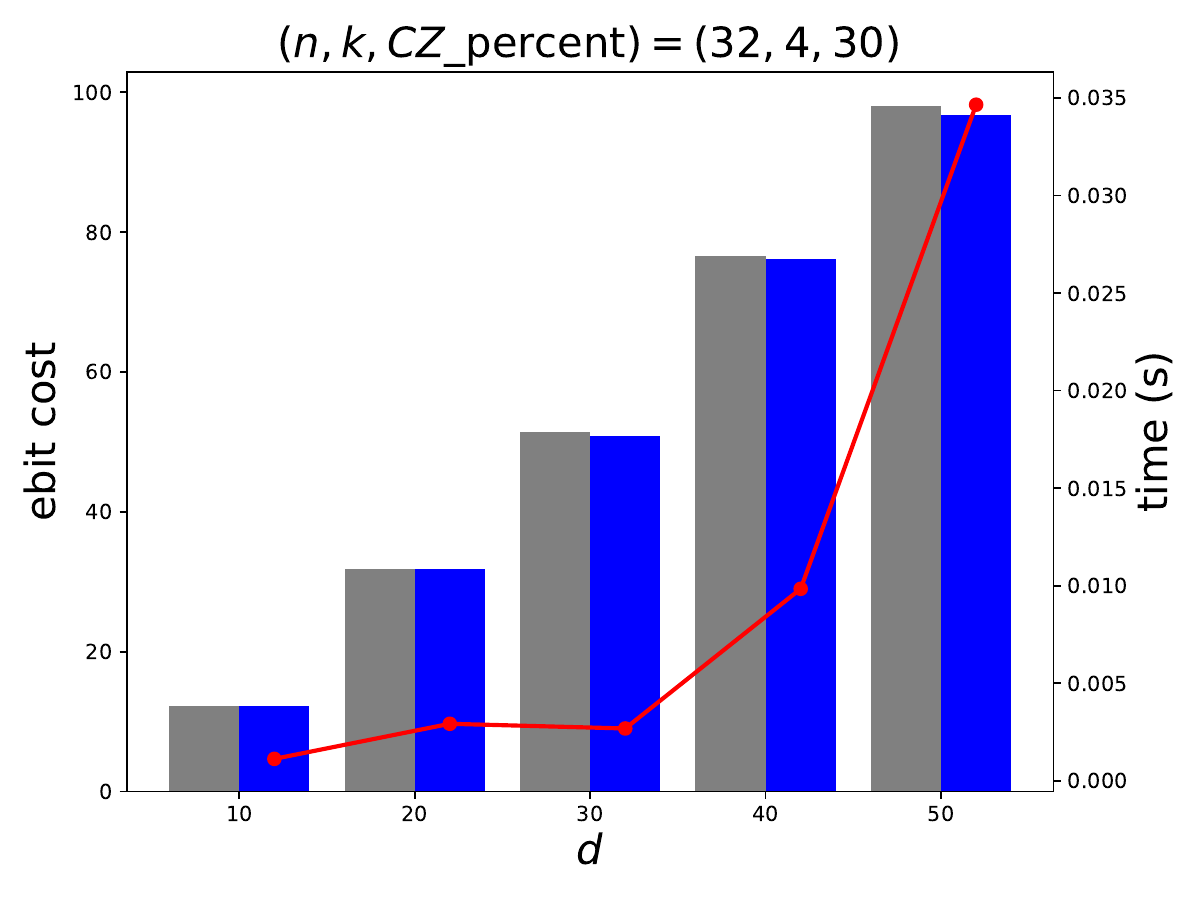}
    \end{subfigure}
    \begin{subfigure}{0.25\textwidth}
        \includegraphics[width=\linewidth]{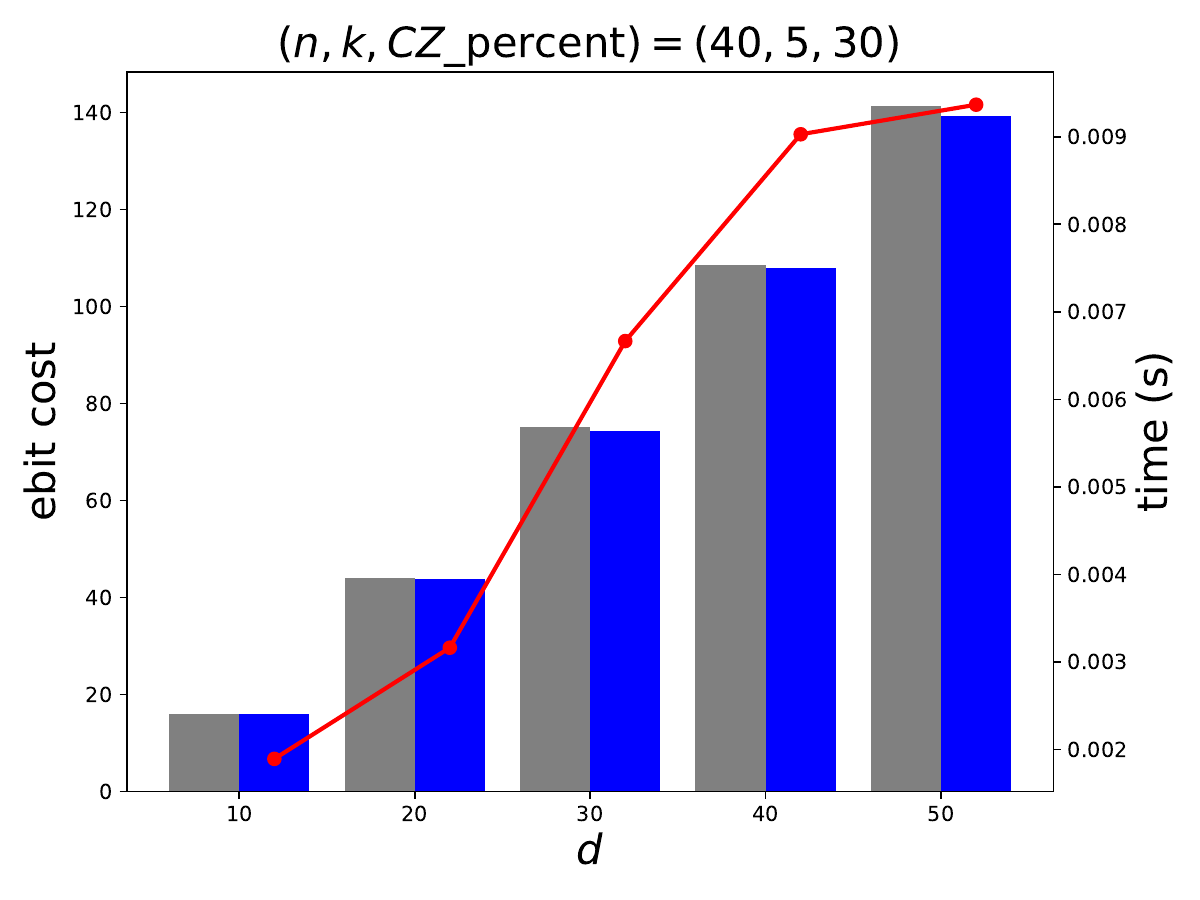}
    \end{subfigure}
    \begin{subfigure}{0.25\textwidth}
        \includegraphics[width=\linewidth]{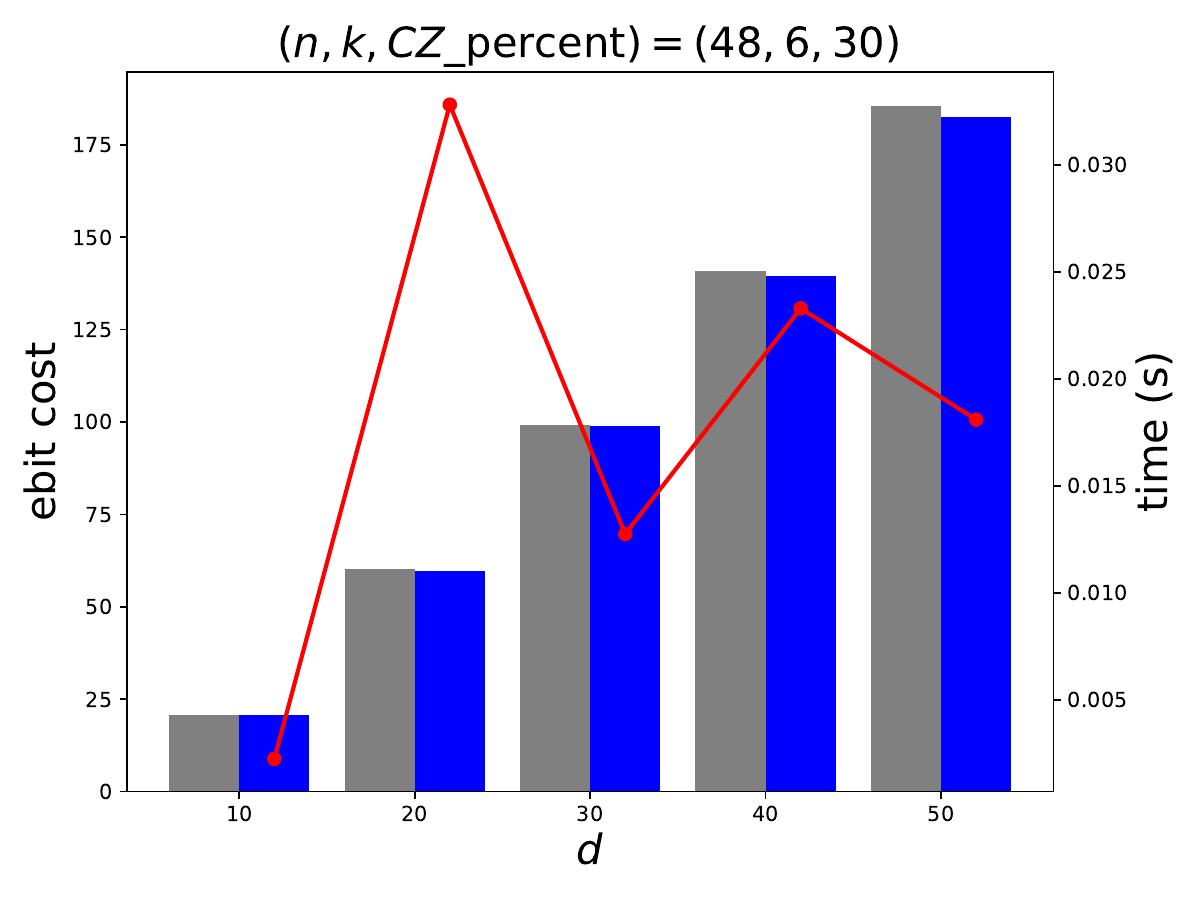}
    \end{subfigure}
    \begin{subfigure}{0.25\textwidth}
        \includegraphics[width=\linewidth]{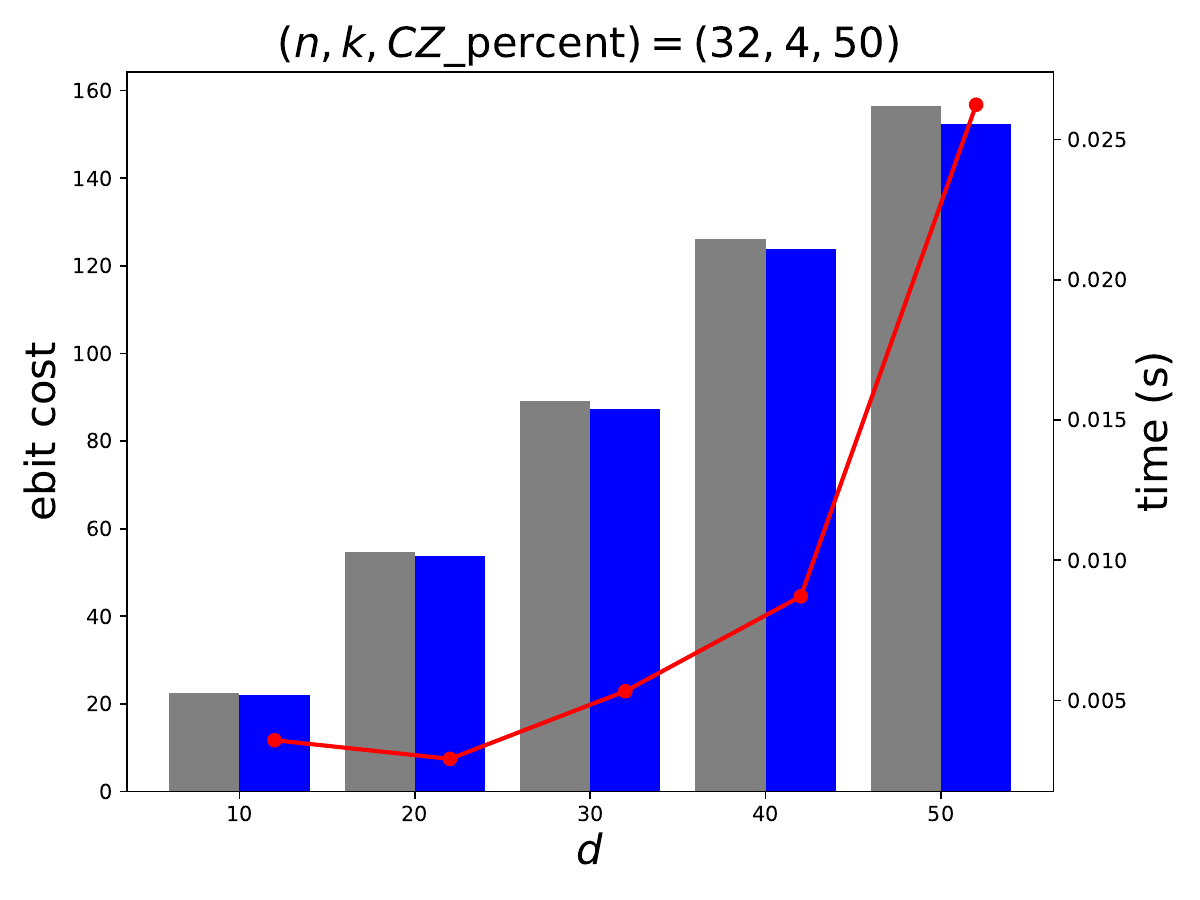}
    \end{subfigure}
    \begin{subfigure}{0.25\textwidth}
        \includegraphics[width=\linewidth]{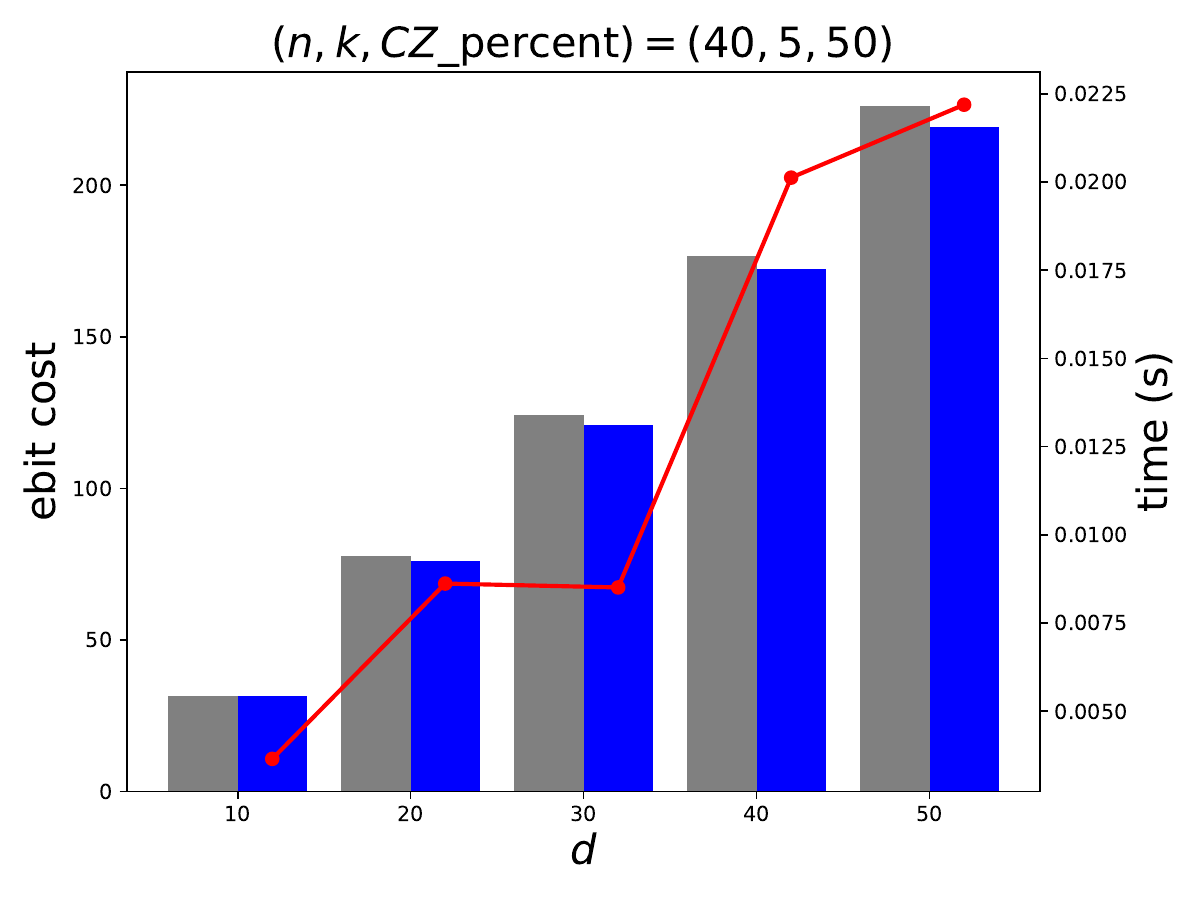}
    \end{subfigure}
    \begin{subfigure}{0.25\textwidth}
        \includegraphics[width=\linewidth]{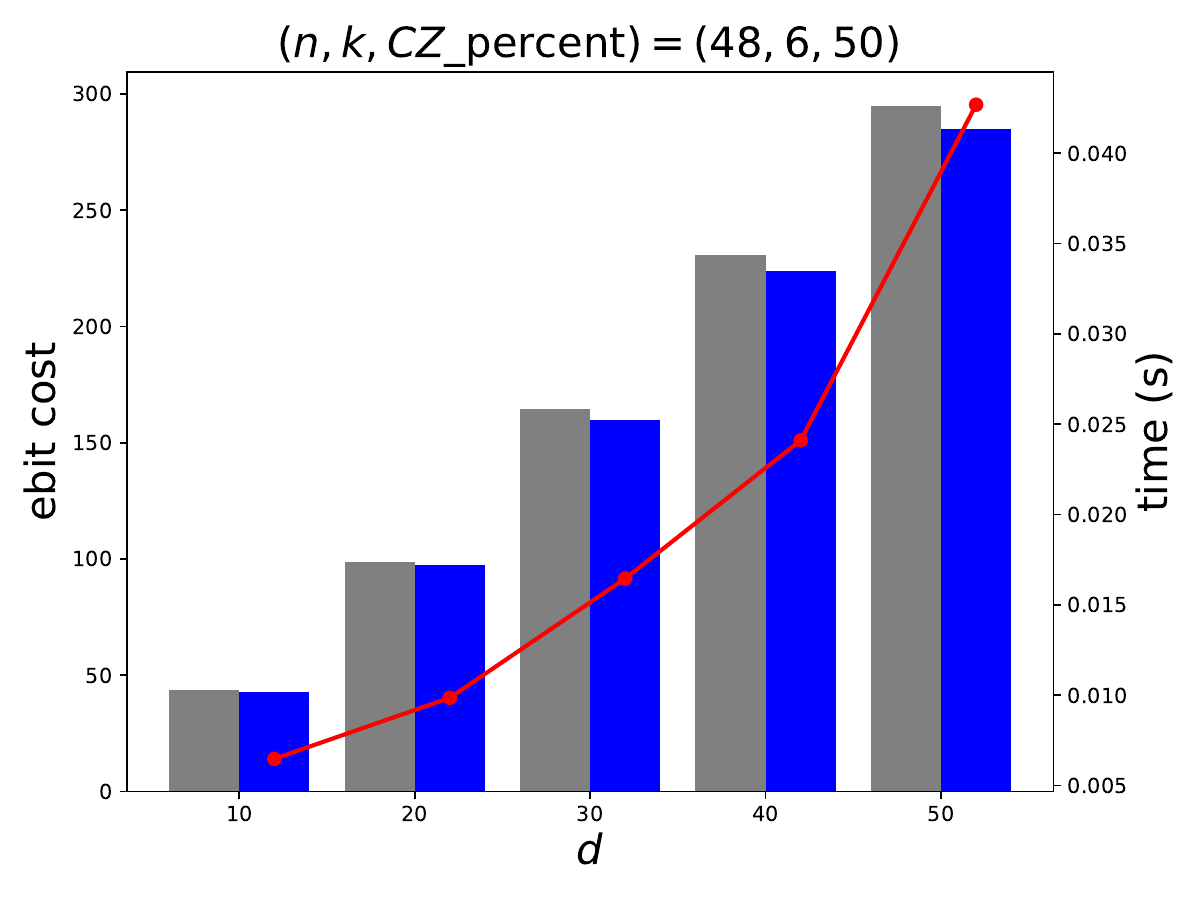}
    \end{subfigure}
    \begin{subfigure}{0.25\textwidth}
        \includegraphics[width=\linewidth]{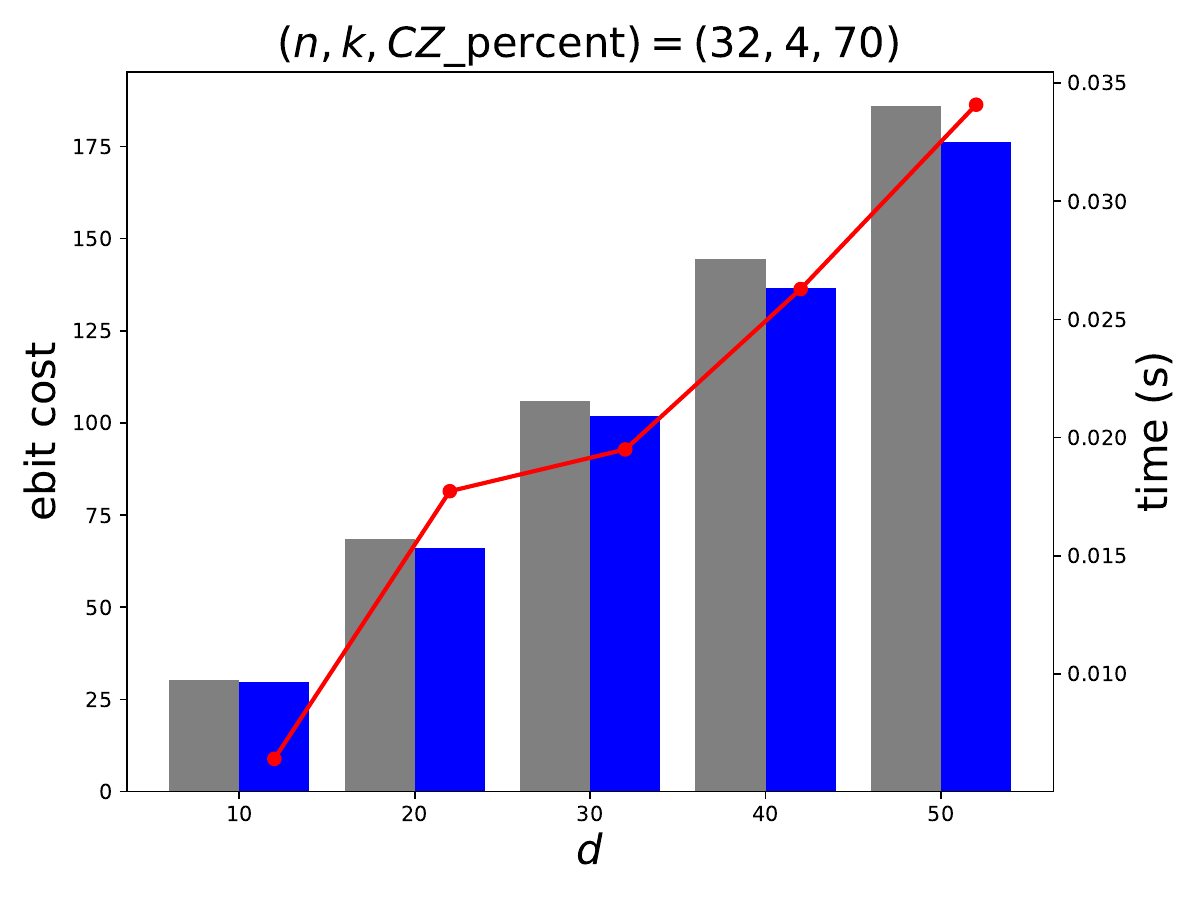}
    \end{subfigure}
    \begin{subfigure}{0.25\textwidth}
        \includegraphics[width=\linewidth]{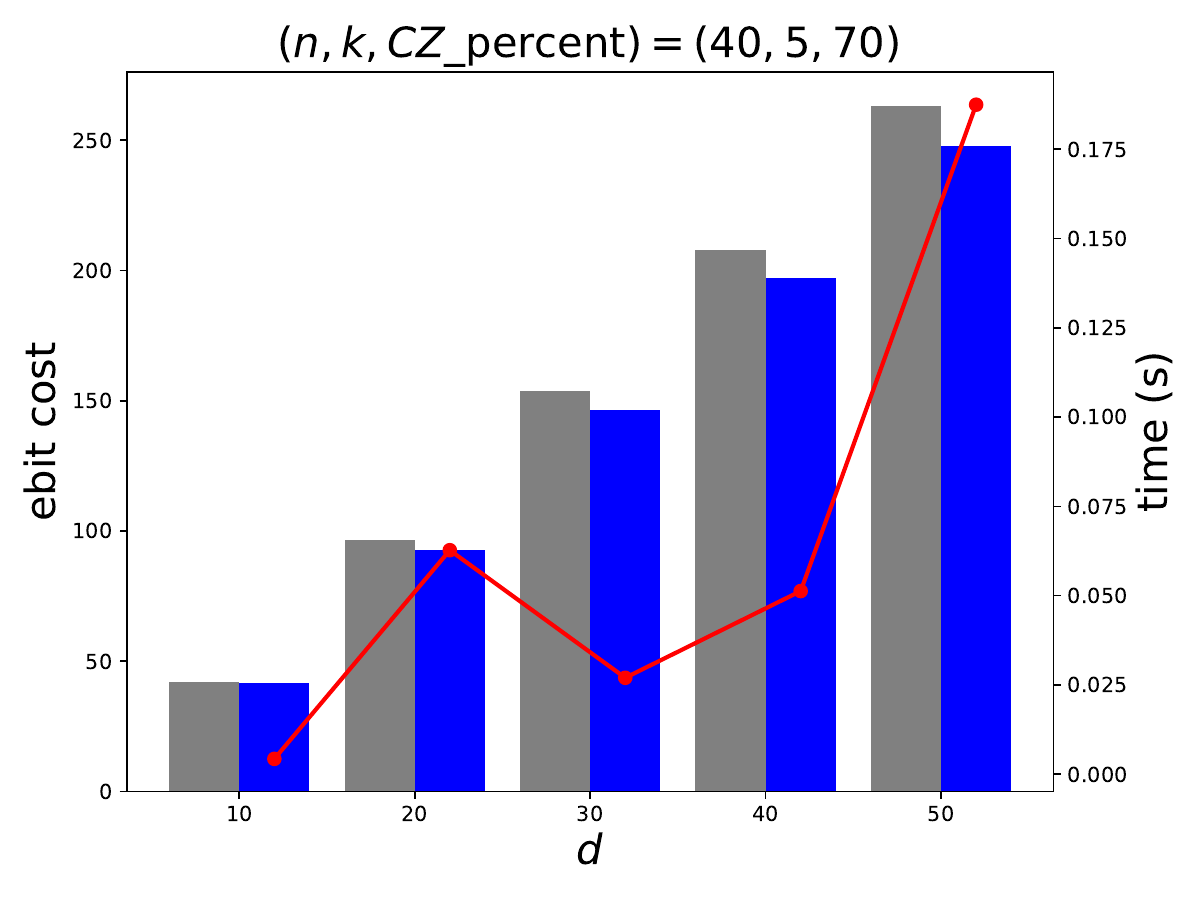}
    \end{subfigure}
    \begin{subfigure}{0.25\textwidth}
        \includegraphics[width=\linewidth]{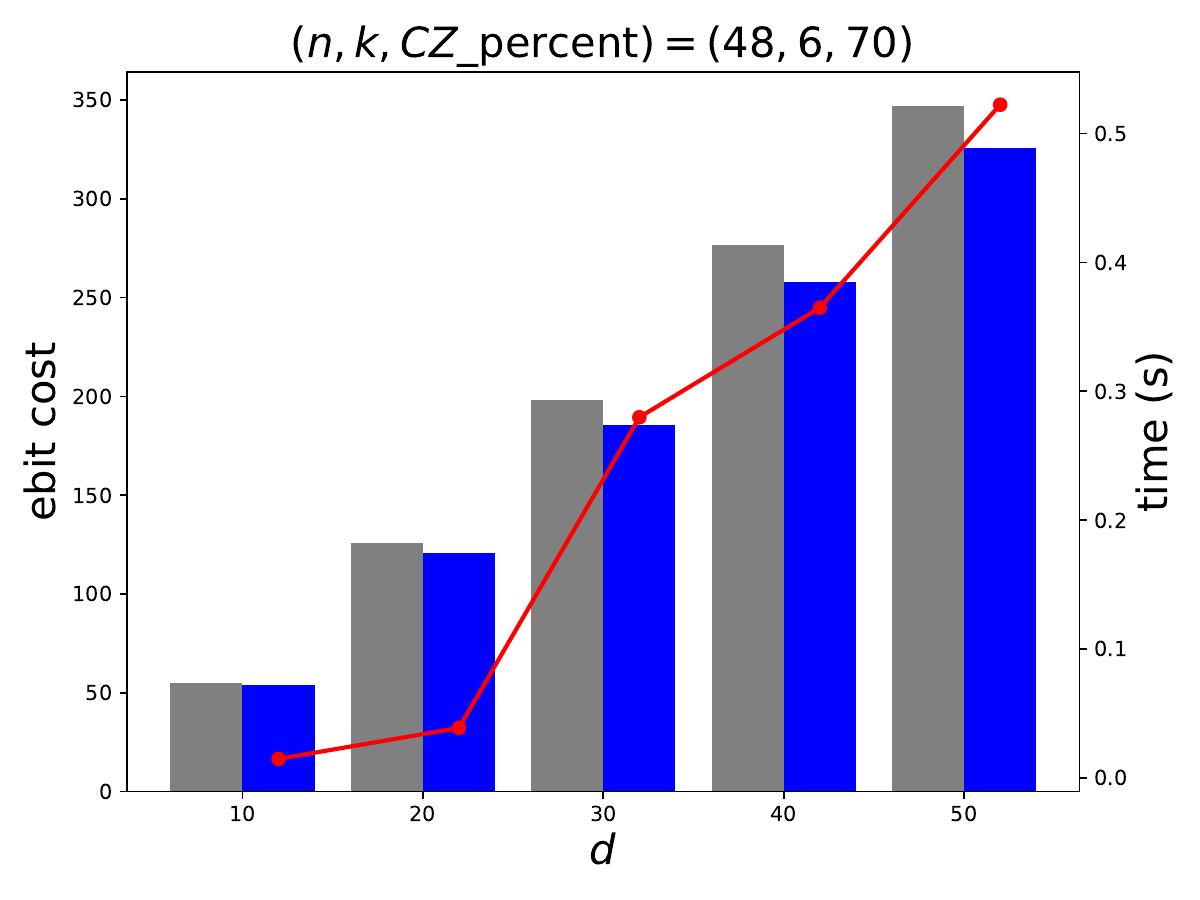}
    \end{subfigure}
    \begin{subfigure}{0.25\textwidth}
        \includegraphics[width=\linewidth]{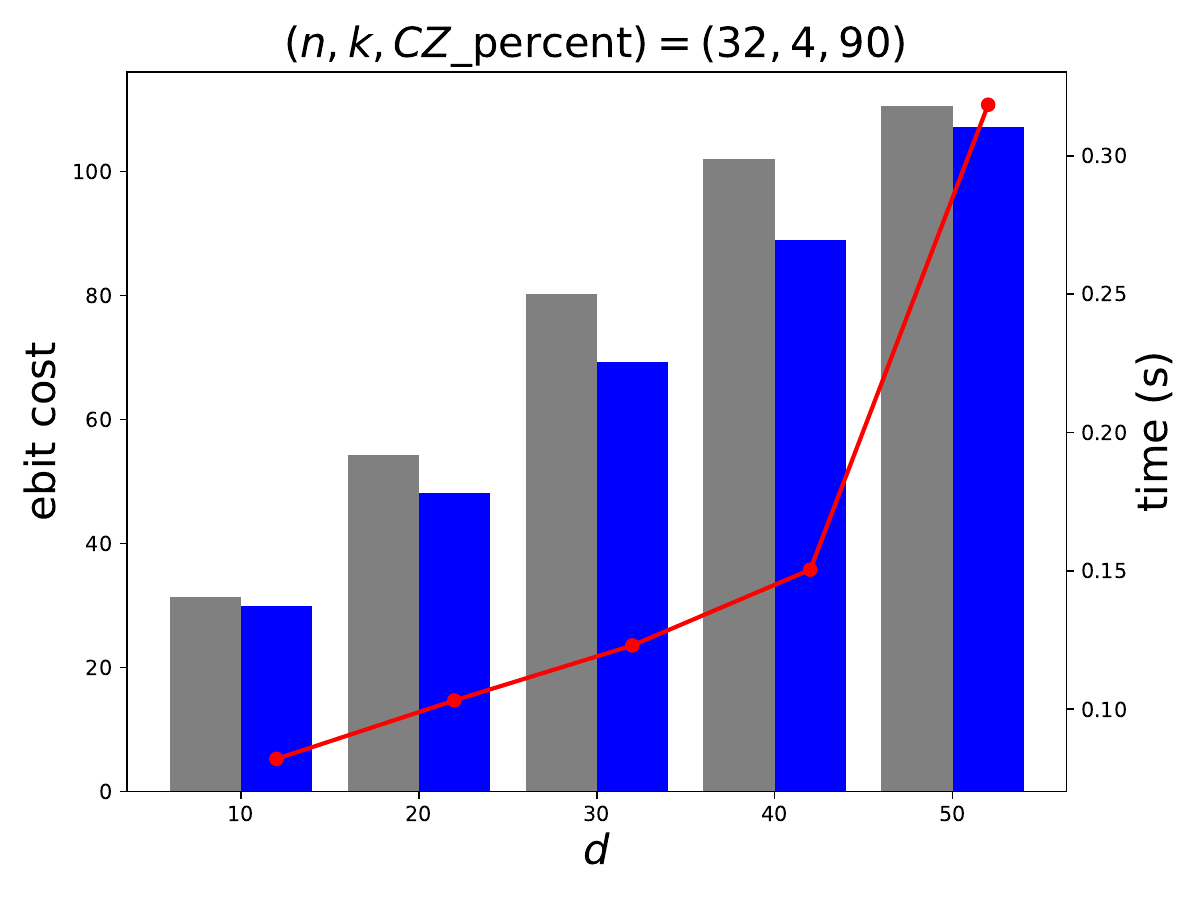}
    \end{subfigure}
    \begin{subfigure}{0.25\textwidth}
        \includegraphics[width=\linewidth]{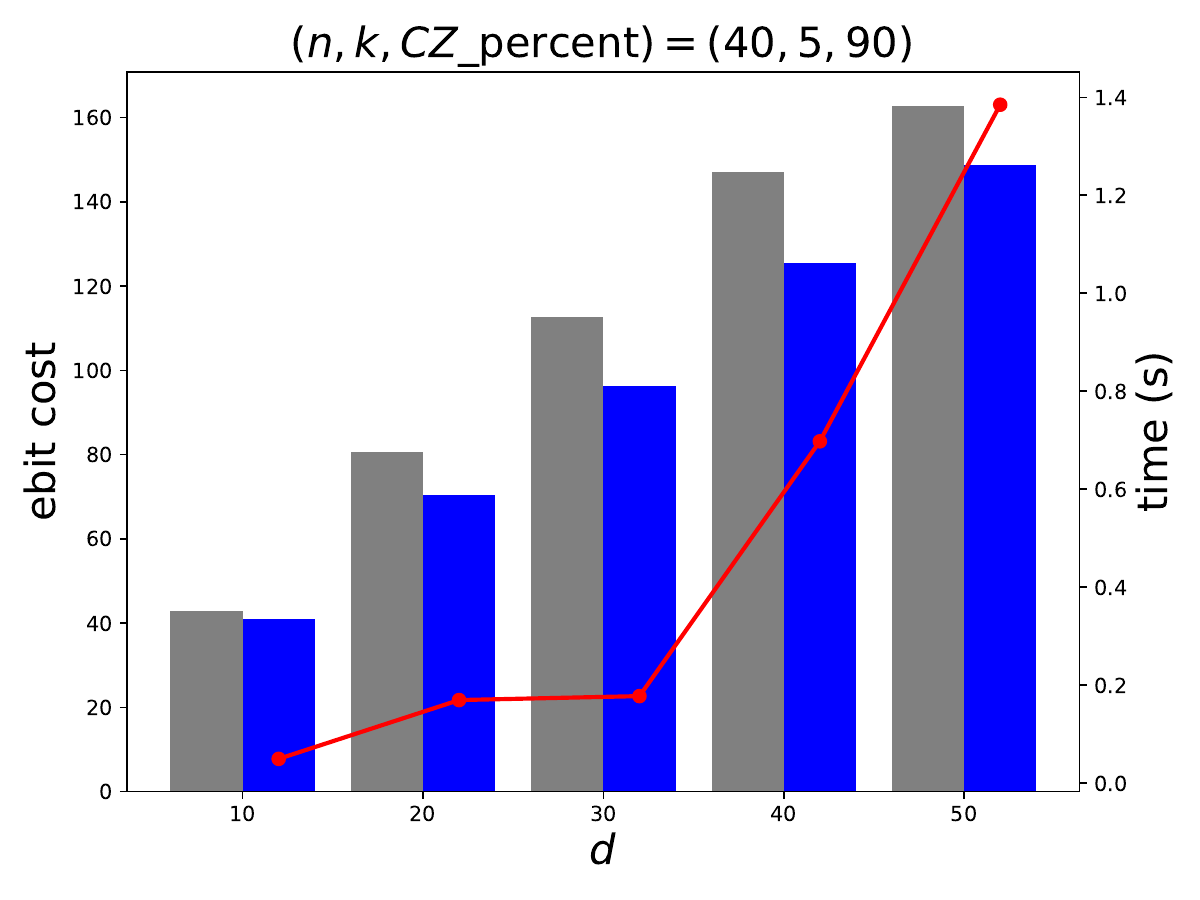}
    \end{subfigure}
    \begin{subfigure}{0.25\textwidth}
        \includegraphics[width=\linewidth]{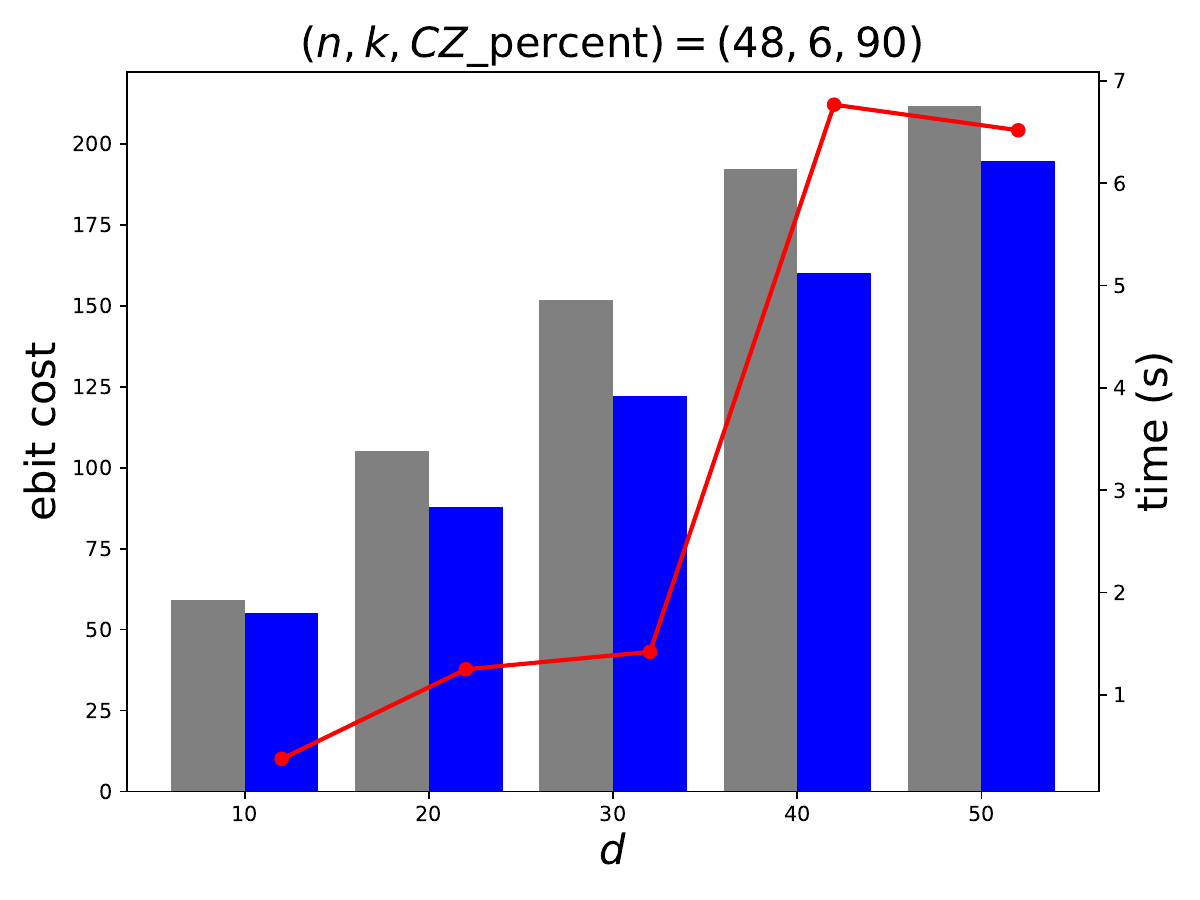}
    \end{subfigure}

    \caption{BIP improvements observed in $CZ$ fraction circuits. \hl{Here, $n$ is the number of qubits, $d$ is the circuit depth, and $k$ is the number of modules.} The reduction in ebit cost generally increases with higher $CZ$ fractions, since they correspond to more difficult DQC instances.}
    \label{fig:many_figures}
\end{figure}

\clearpage

\subsection{Quantum Volume circuits}
\label{section:quantum_volume_appendix}

\begin{figure}[h]
    \centering
    \begin{subfigure}{0.15\textwidth}
        \includegraphics[width=\linewidth]{plots/legend_basic.pdf}
    \end{subfigure}
    \\
    \begin{subfigure}{0.25\textwidth}
        \includegraphics[width=\linewidth]{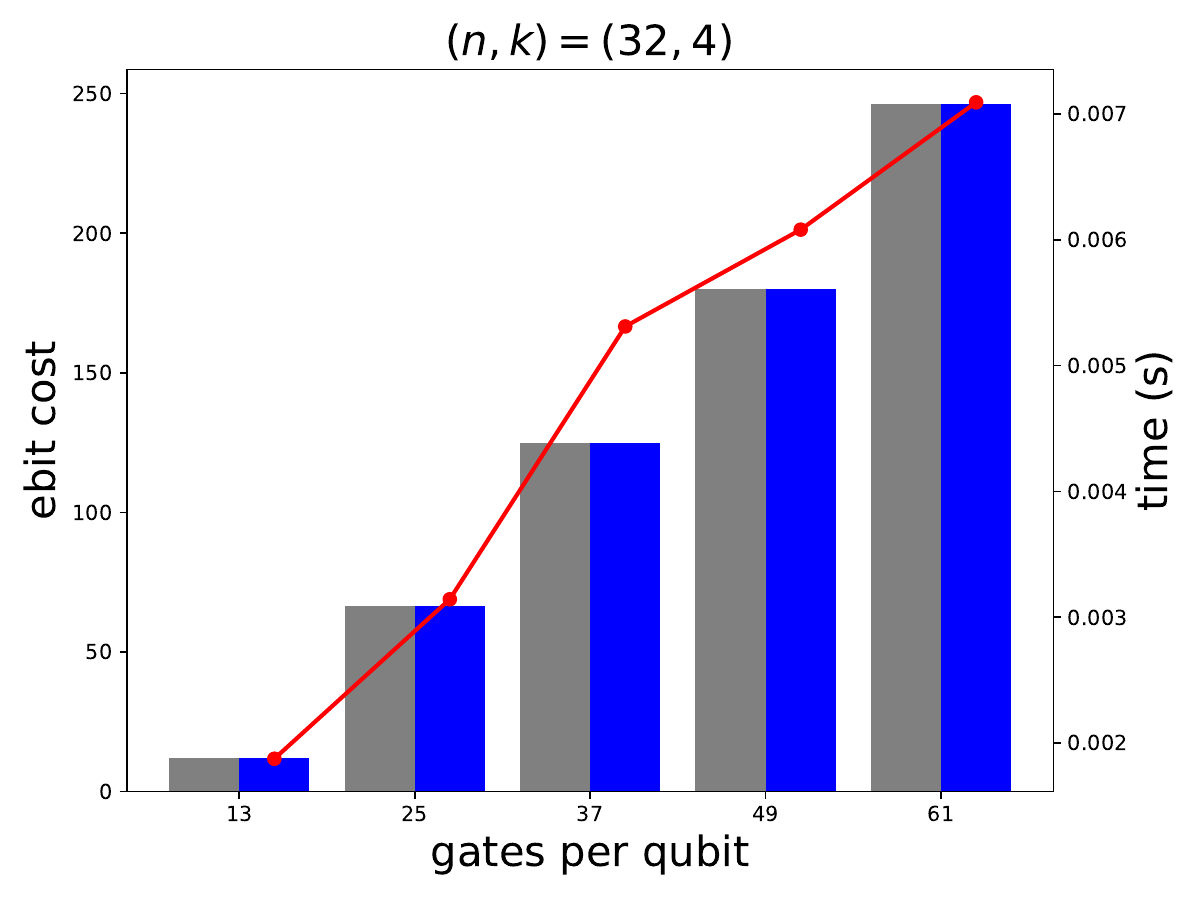}
    \end{subfigure}
    \begin{subfigure}{0.25\textwidth}
        \includegraphics[width=\linewidth]{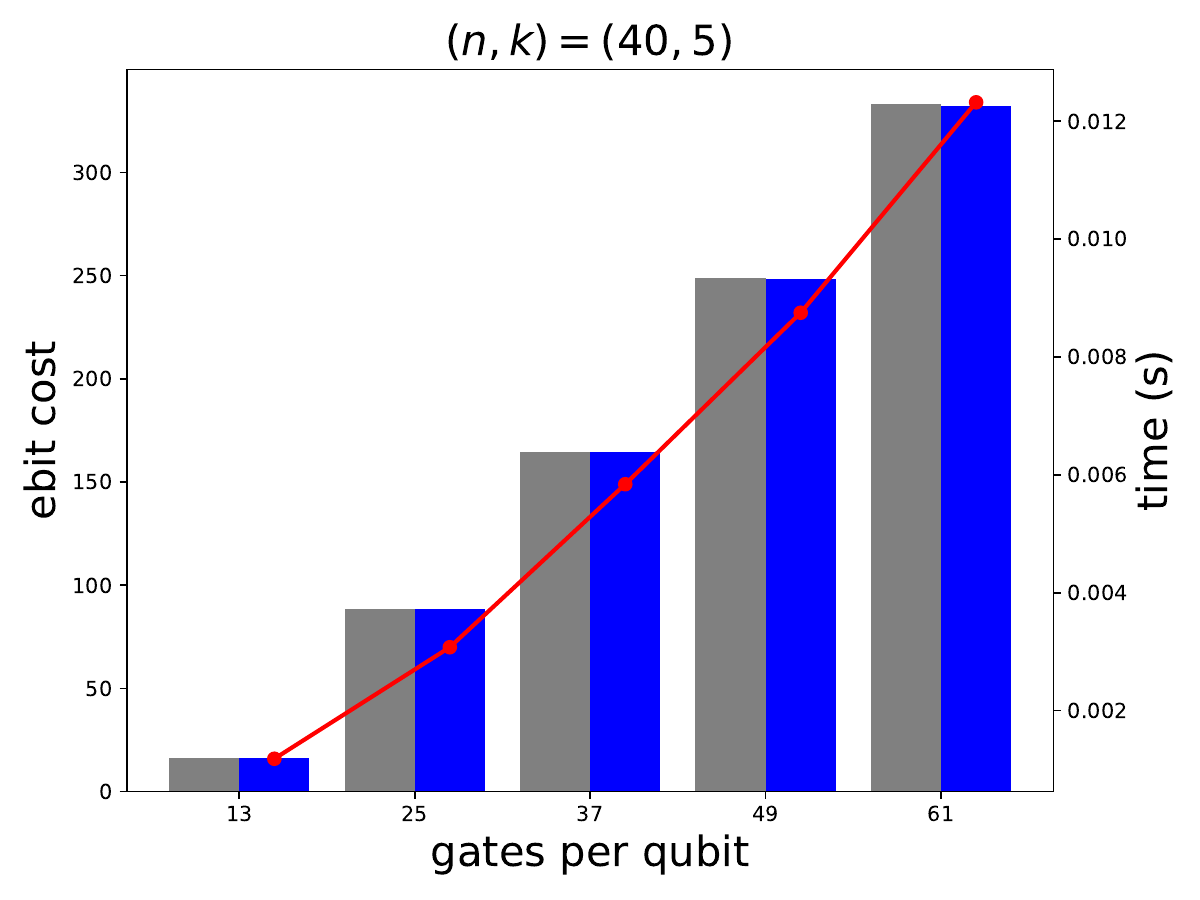}
    \end{subfigure}
    \begin{subfigure}{0.25\textwidth}
        \includegraphics[width=\linewidth]{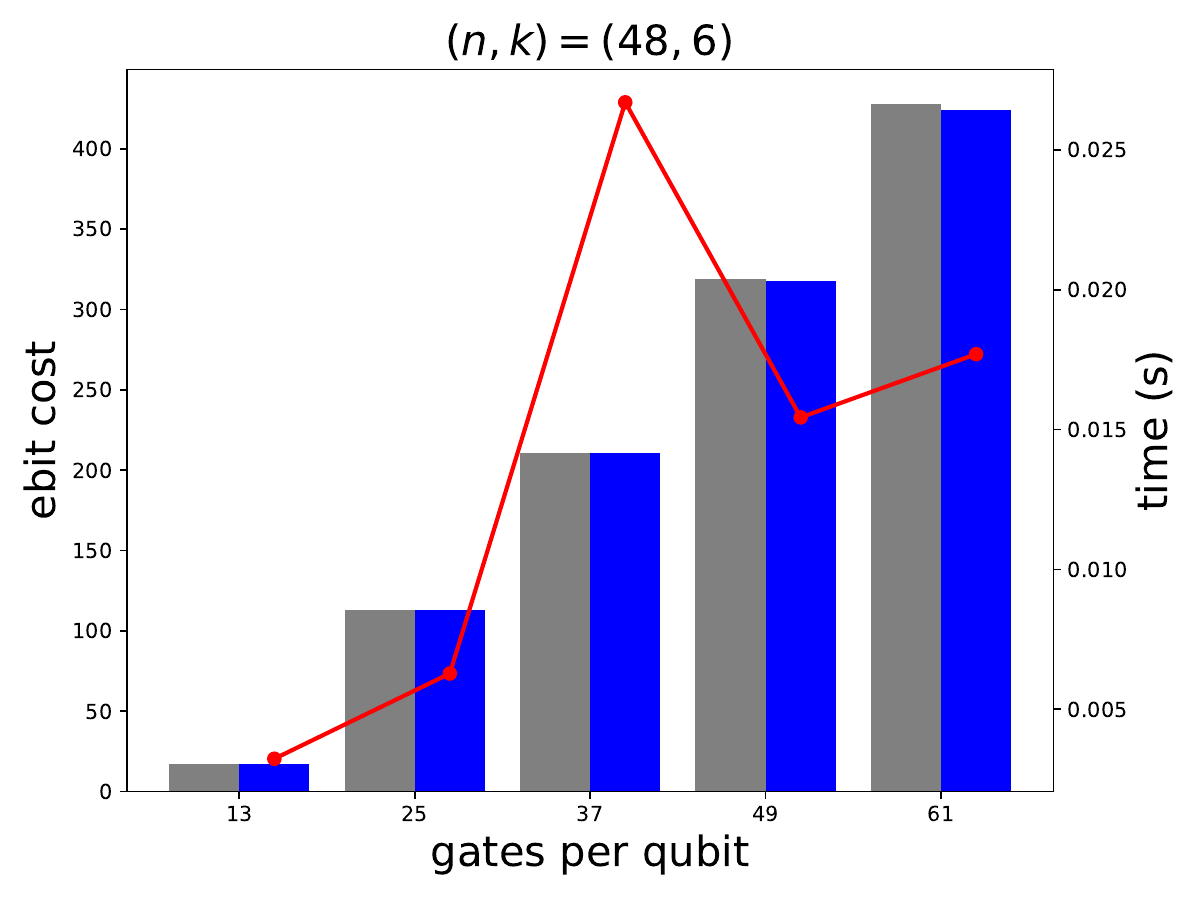}
    \end{subfigure}

    \caption{BIP improvements observed in quantum volume circuits. \hl{Here, $n$ is the number of qubits and $k$ is the number of modules.} The reduction in ebit cost is relatively marginal for these circuits. This is because as mentioned in \citep{andres2024distributing}, rewriting a given quantum volume circuit such that all non-unary gates are $CP$ gates makes the fraction of binary gates quite low. Similar to the results for circuits with low $CZ\_\text{percent}$ shown in Fig.~\ref{fig:many_figures}, little improvement is expected.}
    \label{fig:many_figures_volume}
\end{figure}

\subsection{QFT circuits}
\label{section:qft_appendix}

\begin{figure}[h]
    \centering
    \begin{subfigure}{0.15\textwidth}
        \includegraphics[width=\linewidth]{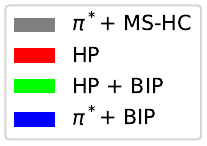}
    \end{subfigure}
    \\
    \begin{subfigure}{0.25\textwidth}
        \includegraphics[width=\linewidth]{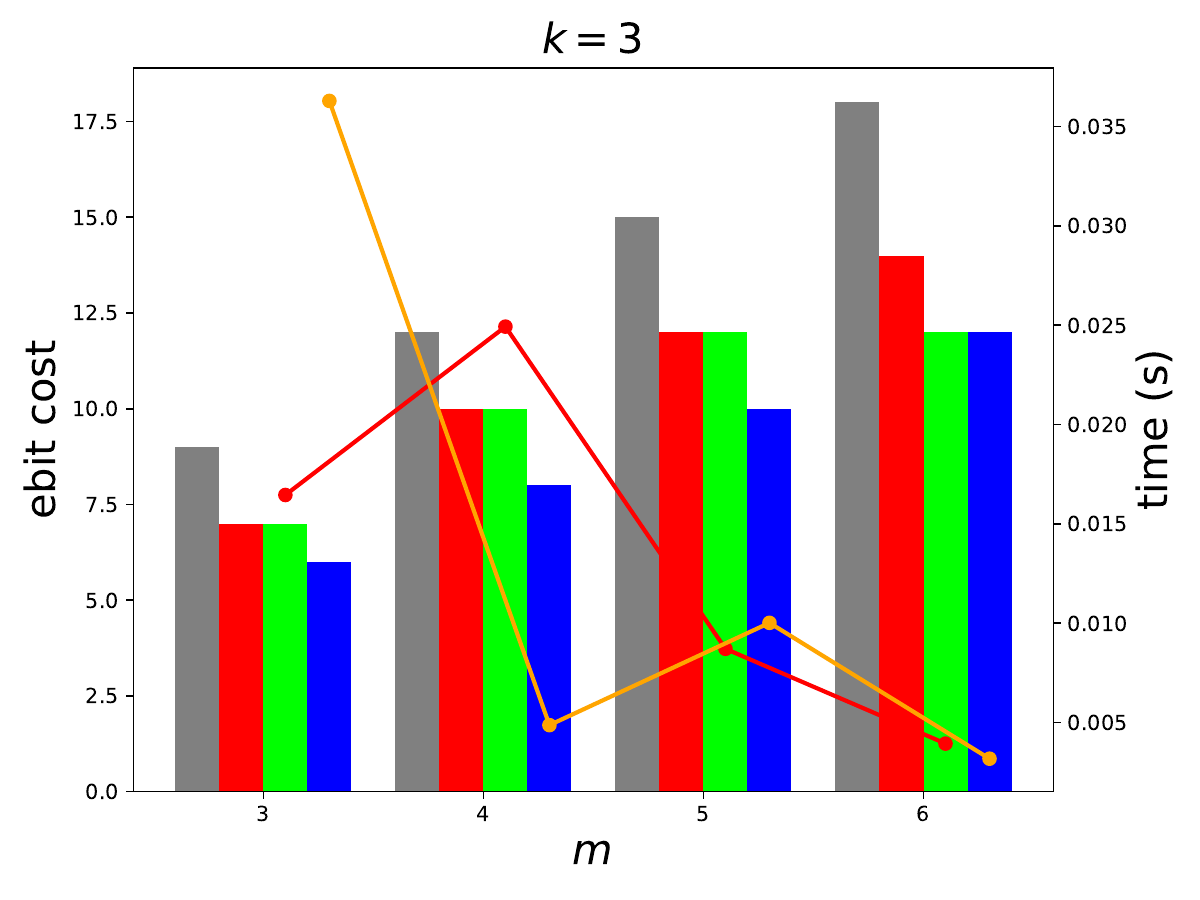}
    \end{subfigure}
    \begin{subfigure}{0.25\textwidth}
        \includegraphics[width=\linewidth]{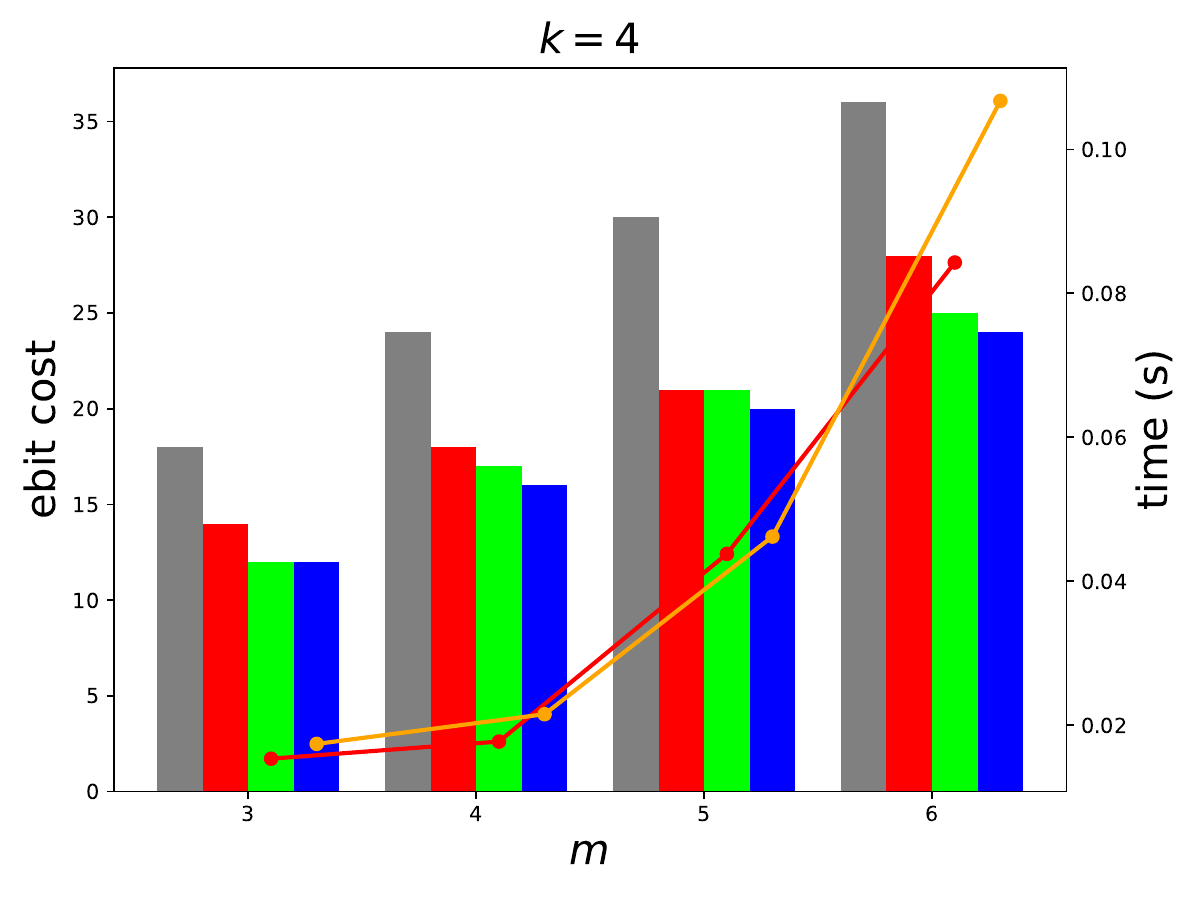}
    \end{subfigure}
    \begin{subfigure}{0.25\textwidth}
        \includegraphics[width=\linewidth]{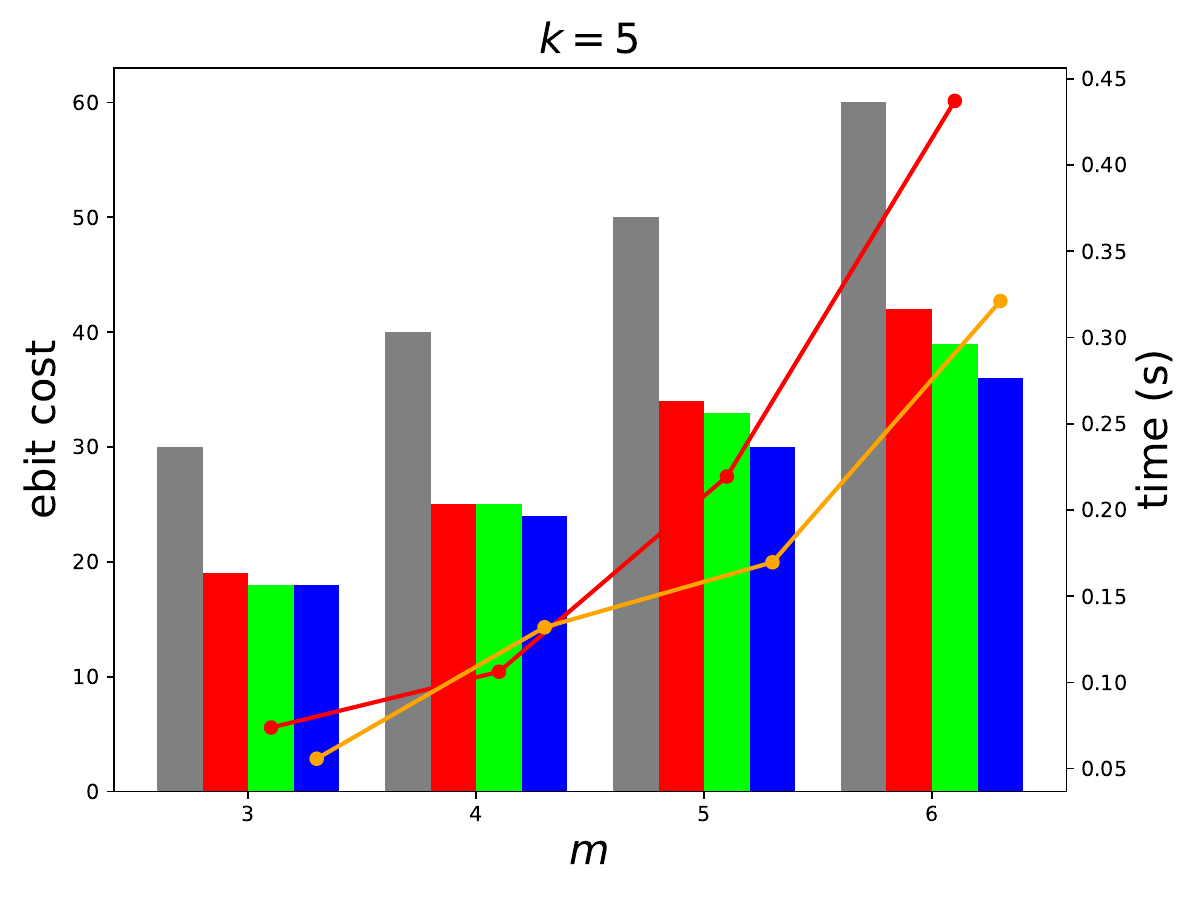}
    \end{subfigure}
    
    \begin{subfigure}{0.25\textwidth}
        \includegraphics[width=\linewidth]{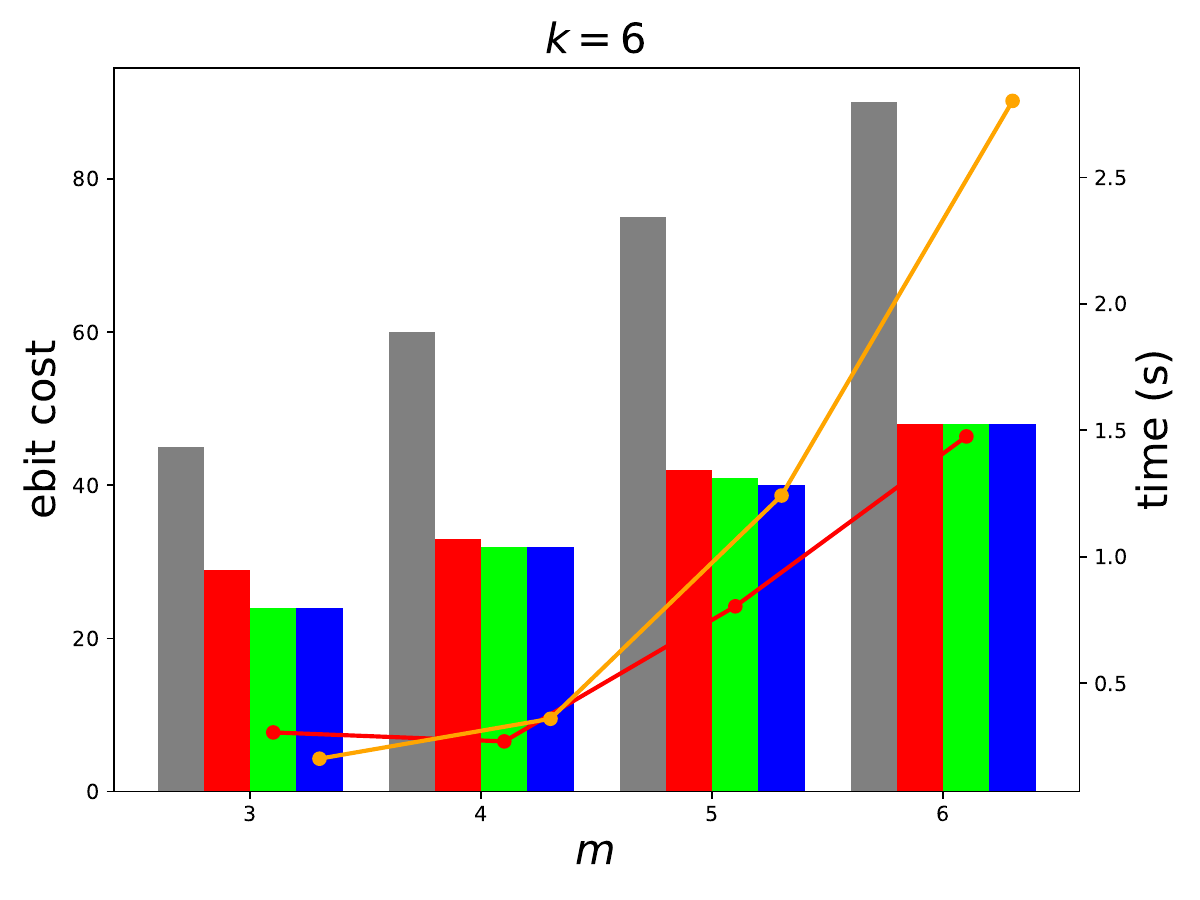}
    \end{subfigure}
    \begin{subfigure}{0.25\textwidth}
        \includegraphics[width=\linewidth]{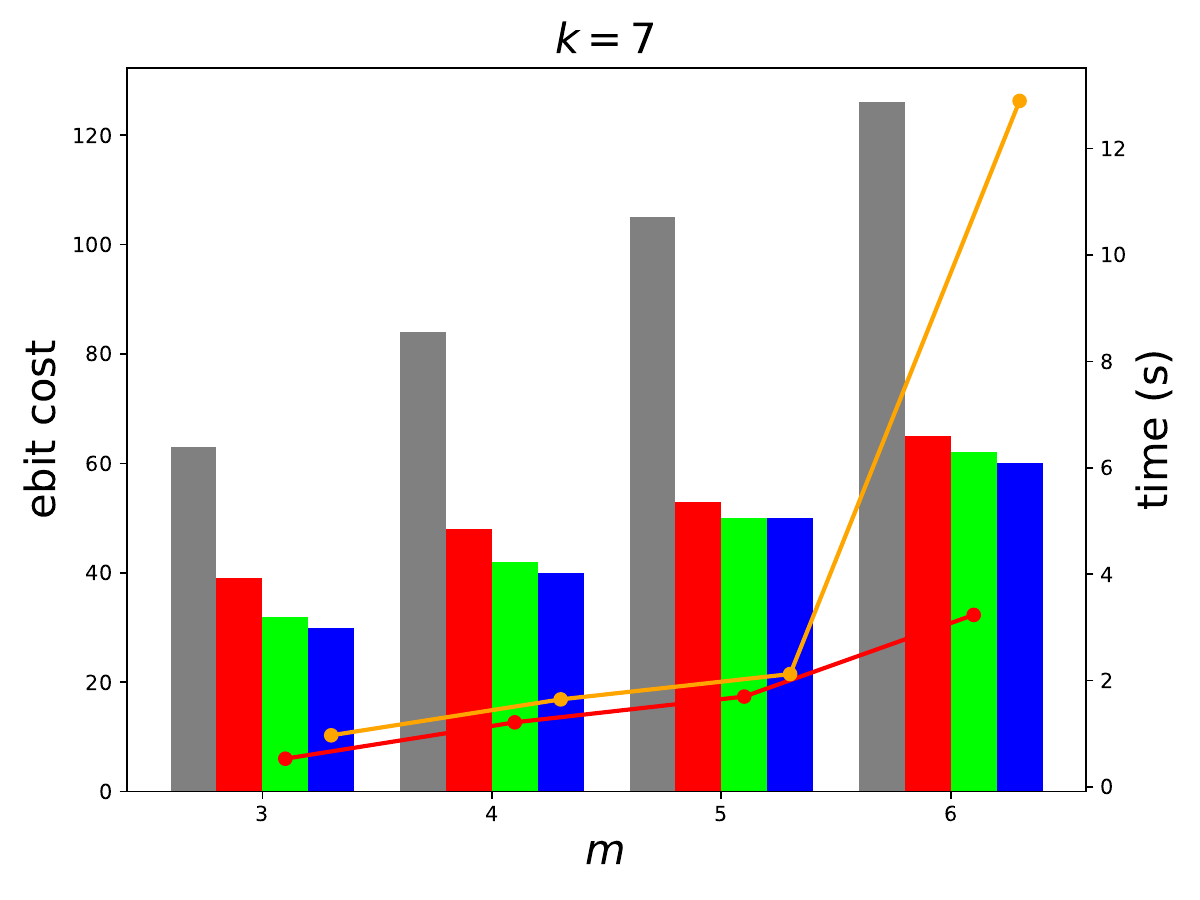}
    \end{subfigure}
    \begin{subfigure}{0.25\textwidth}
        \includegraphics[width=\linewidth]{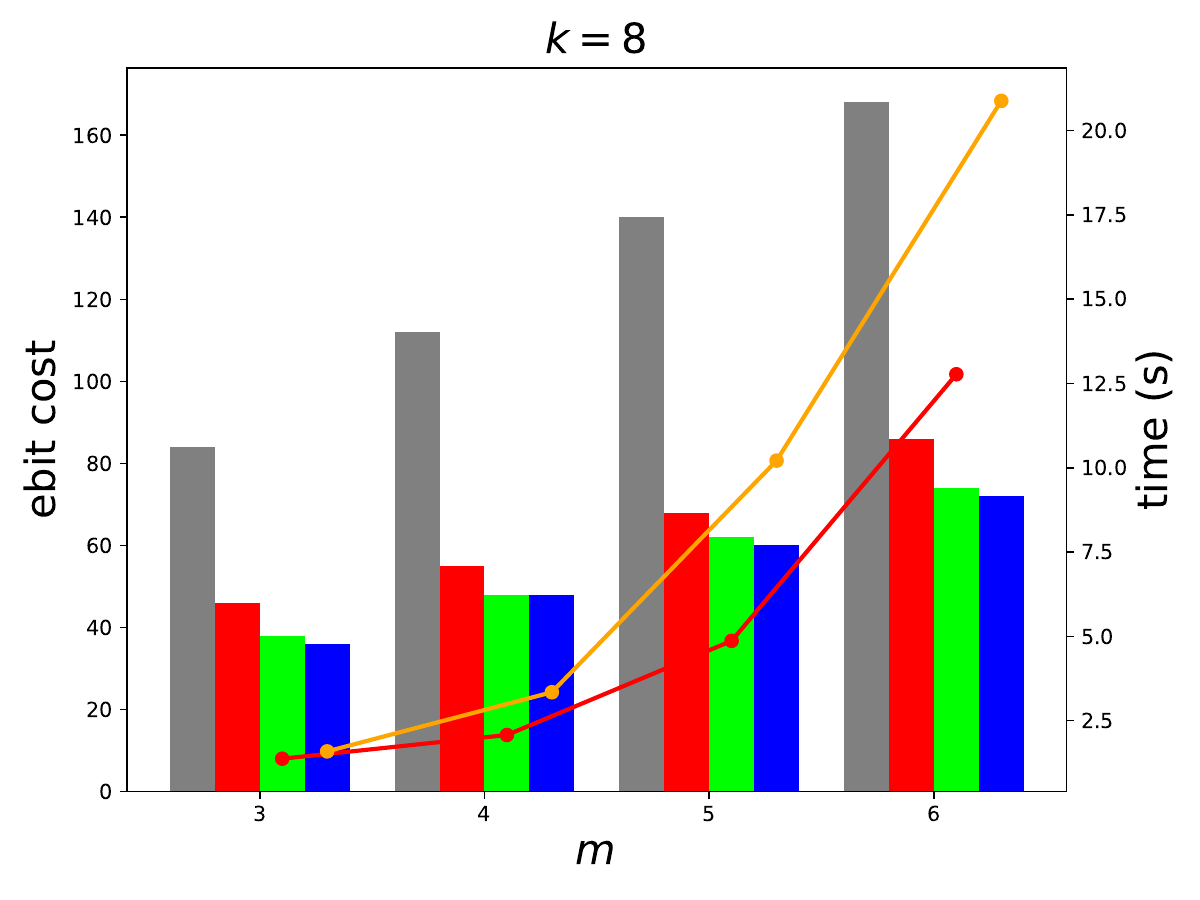}
    \end{subfigure}

    \caption{BIP and/or $\pi^*$ improvements observed in QFT circuits. \hl{Here, $k$ is the number of modules and $m$ is the number of qubits per module. Applying BIP to $\pi^*$ consistently yields better results than applying it to an HP-based allocation. This demonstrates that the current HP is suboptimal not only in how it selects migrations but also in the module allocations it produces.}}
    \label{fig:qft_result_figure}
\end{figure}

\clearpage

\subsection{DraperQFTAdder circuits}
\label{section:draper_appendix}
We first derive Eq.~\eqref{equation:swap_identity}. In matrix representations,
\newlength{\defaultarraycolsep}
\setlength{\defaultarraycolsep}{\arraycolsep}
\begin{align*}
\text{CP}_{12}(\theta) & =
\setlength{\arraycolsep}{1pt}
\begin{pmatrix}
1&&&&&&&\\
&1&&&&&&\\
&&1&&&&&\\
&&&1&&&&\\
&&&&1&&&\\
&&&&&1&&\\
&&&&&&e^{i\theta}&\\
&&&&&&&e^{i\theta}
\end{pmatrix}\\
& =
\setlength{\arraycolsep}{1pt}
\begin{pmatrix}
1&&&&&&&\\
&&1&&&&&\\
&1&&&&&&\\
&&&1&&&&\\
&&&&1&&&\\
&&&&&&1&\\
&&&&&1&&\\
&&&&&&&1
\end{pmatrix}
\setlength{\arraycolsep}{1pt}
\begin{pmatrix}
1&&&&&&&\\
&1&&&&&&\\
&&1&&&&&\\
&&&1&&&&\\
&&&&1&&&\\
&&&&&e^{i\theta}&&\\
&&&&&&1&\\
&&&&&&&e^{i\theta}
\end{pmatrix}
\setlength{\arraycolsep}{1pt}
\begin{pmatrix}
1&&&&&&&\\
&&1&&&&&\\
&1&&&&&&\\
&&&1&&&&\\
&&&&1&&&\\
&&&&&&1&\\
&&&&&1&&\\
&&&&&&&1
\end{pmatrix}
\setlength{\arraycolsep}{\defaultarraycolsep}\\
& = \text{SWAP}_{23}\text{CP}_{13}(\theta)\text{SWAP}_{23}.
\end{align*}

\begin{figure}[h]
    \centering
    \begin{subfigure}{0.15\textwidth}
        \includegraphics[width=\linewidth]{plots/legend_basic.pdf}
    \end{subfigure}
    \\
    \begin{subfigure}{0.25\textwidth}
        \includegraphics[width=\linewidth]{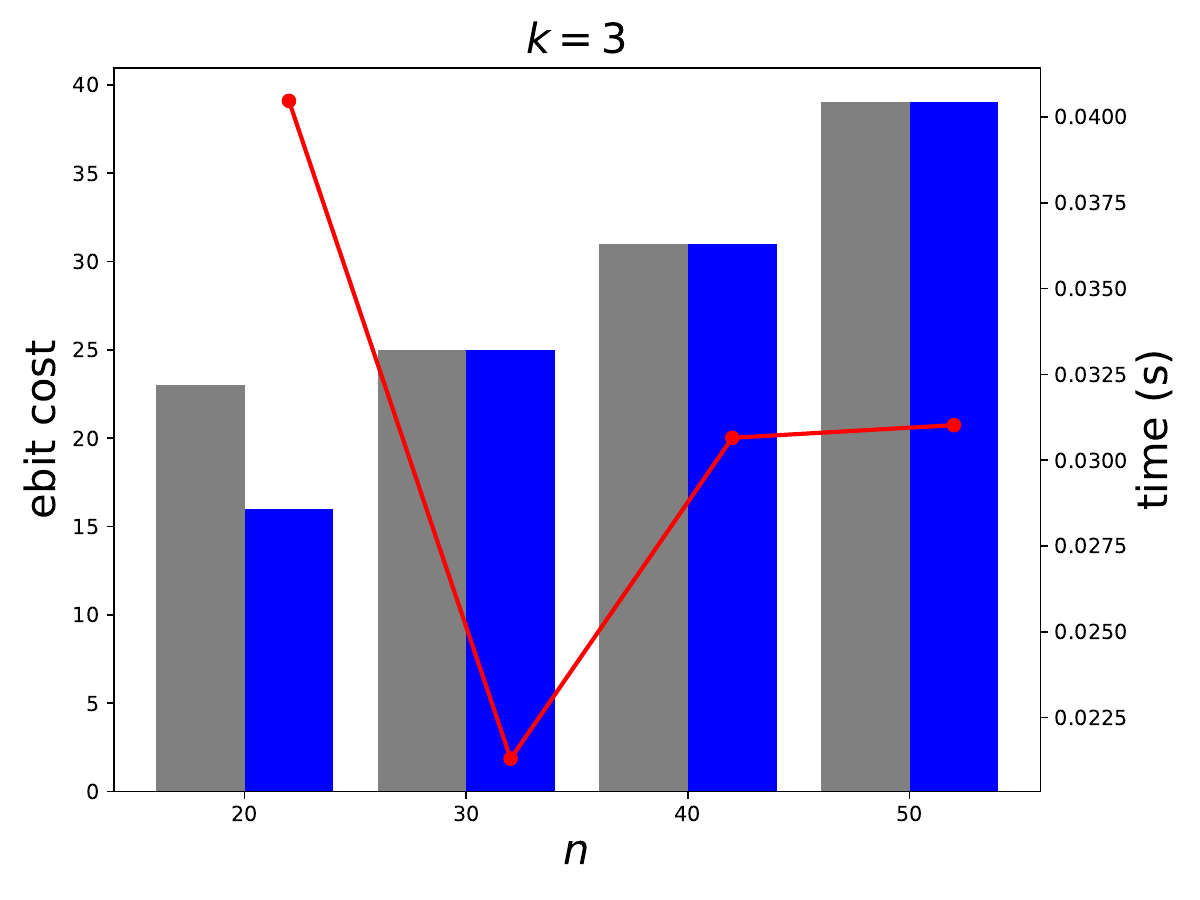}
    \end{subfigure}
    \begin{subfigure}{0.25\textwidth}
        \includegraphics[width=\linewidth]{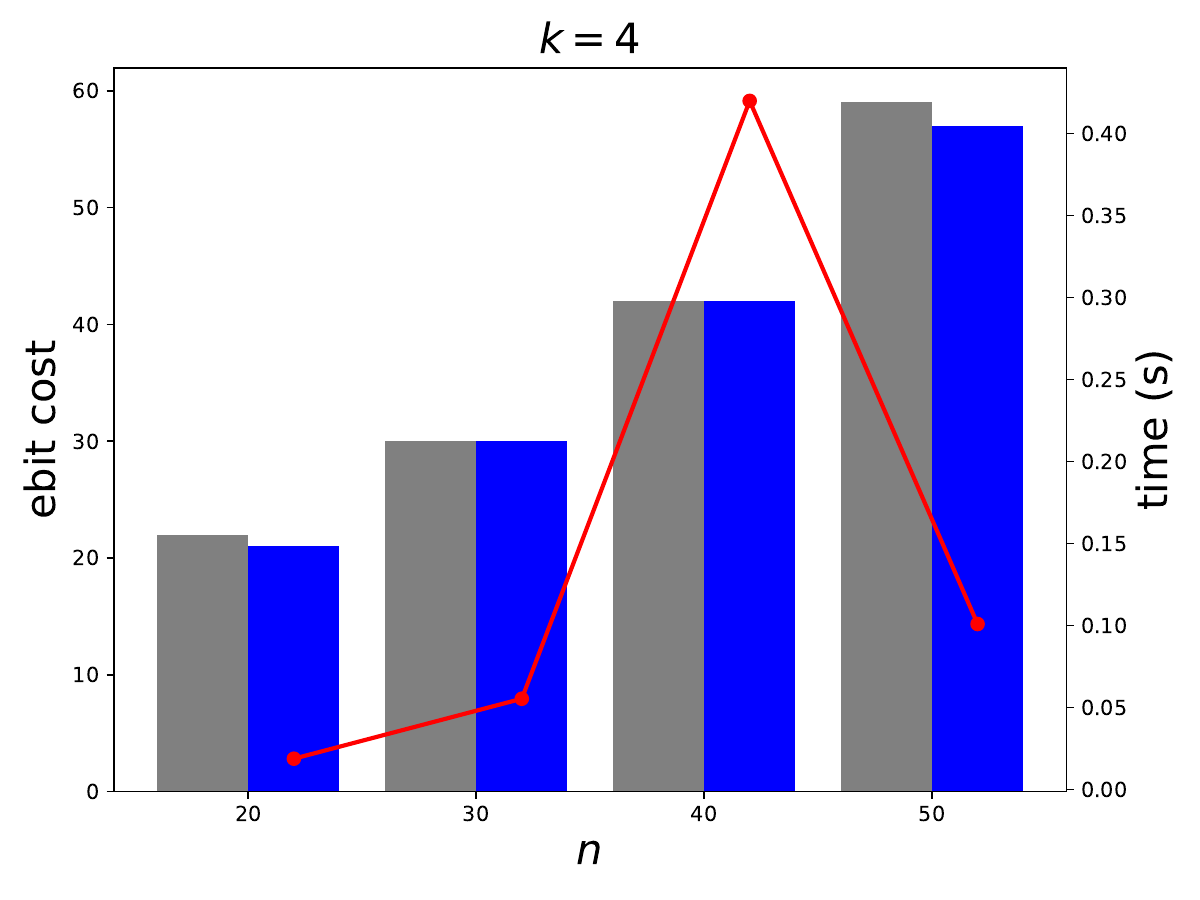}
    \end{subfigure}
    \begin{subfigure}{0.25\textwidth}
        \includegraphics[width=\linewidth]{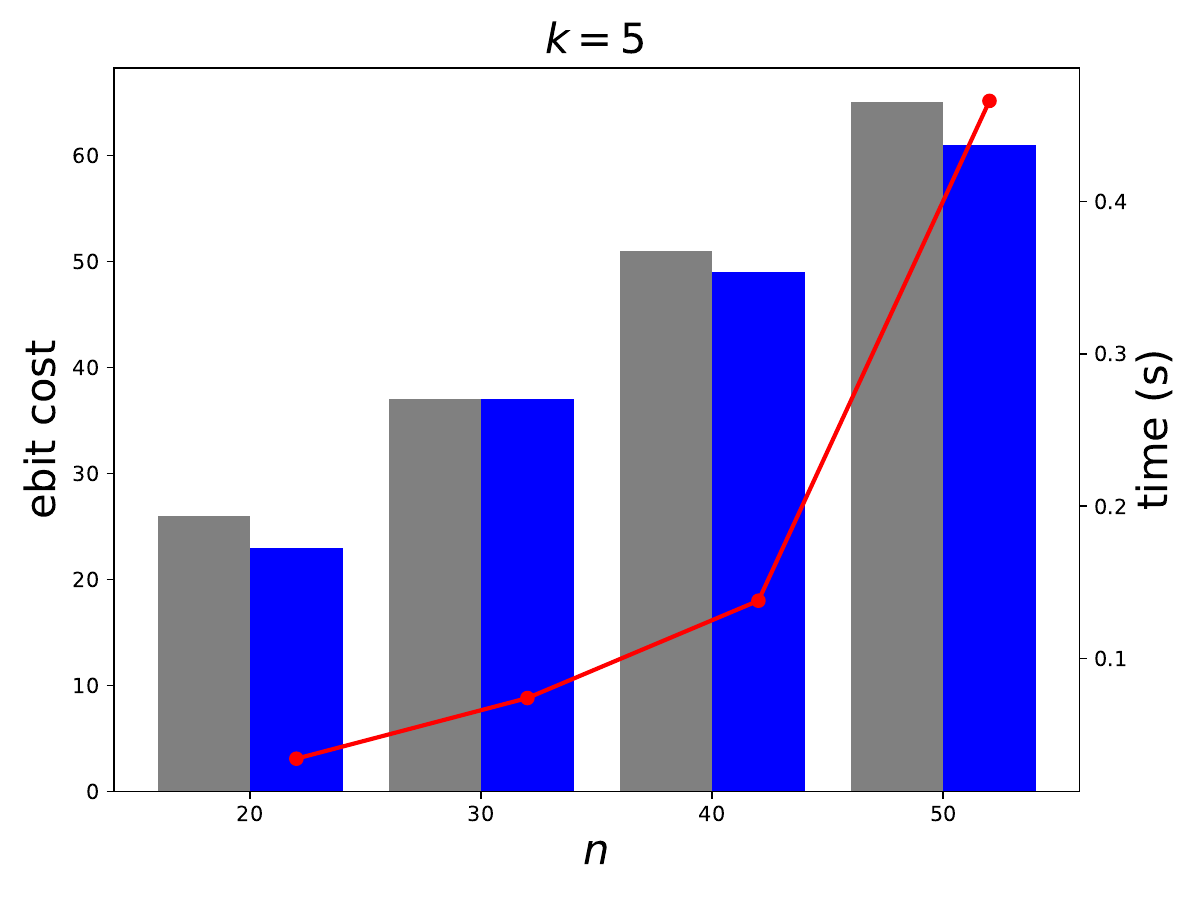}
    \end{subfigure}

    \begin{subfigure}{0.25\textwidth}
        \includegraphics[width=\linewidth]{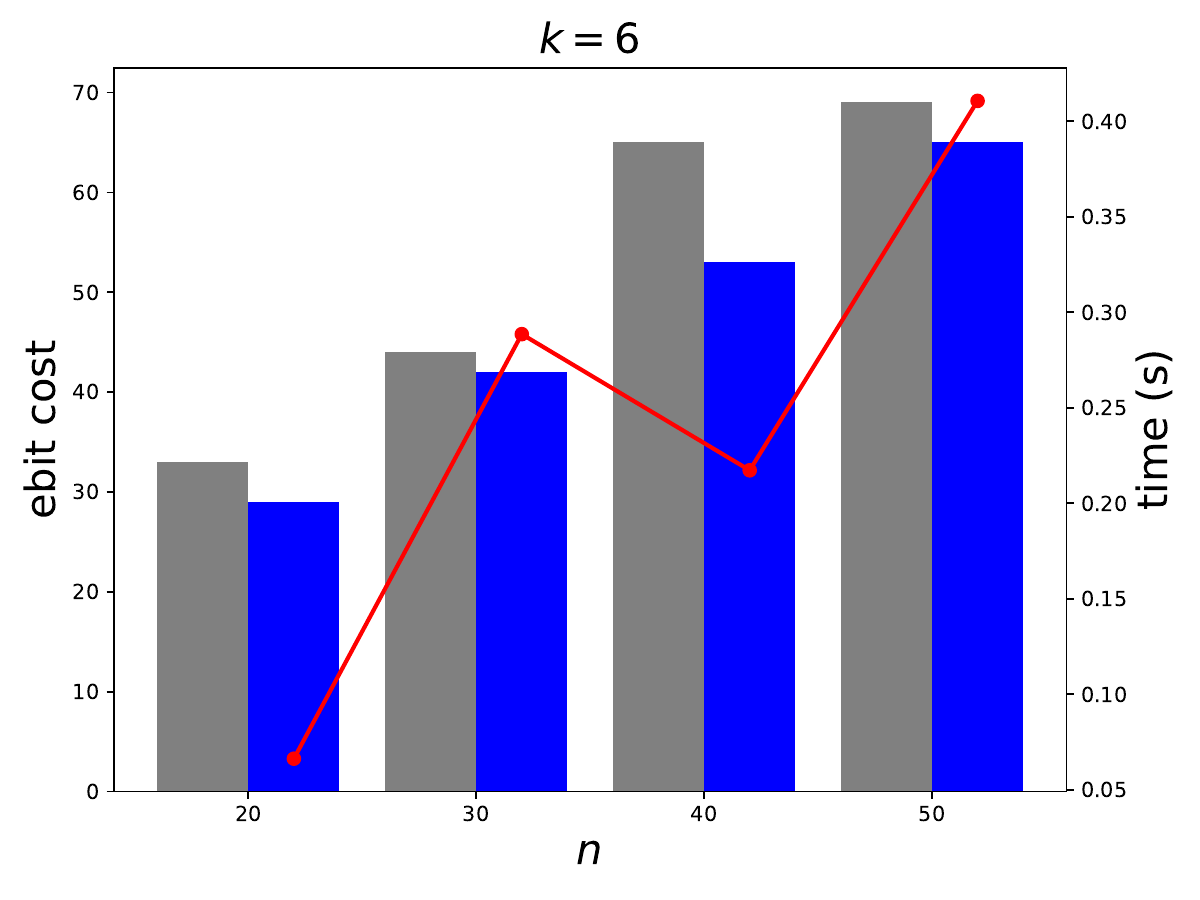}
    \end{subfigure}
    \begin{subfigure}{0.25\textwidth}
        \includegraphics[width=\linewidth]{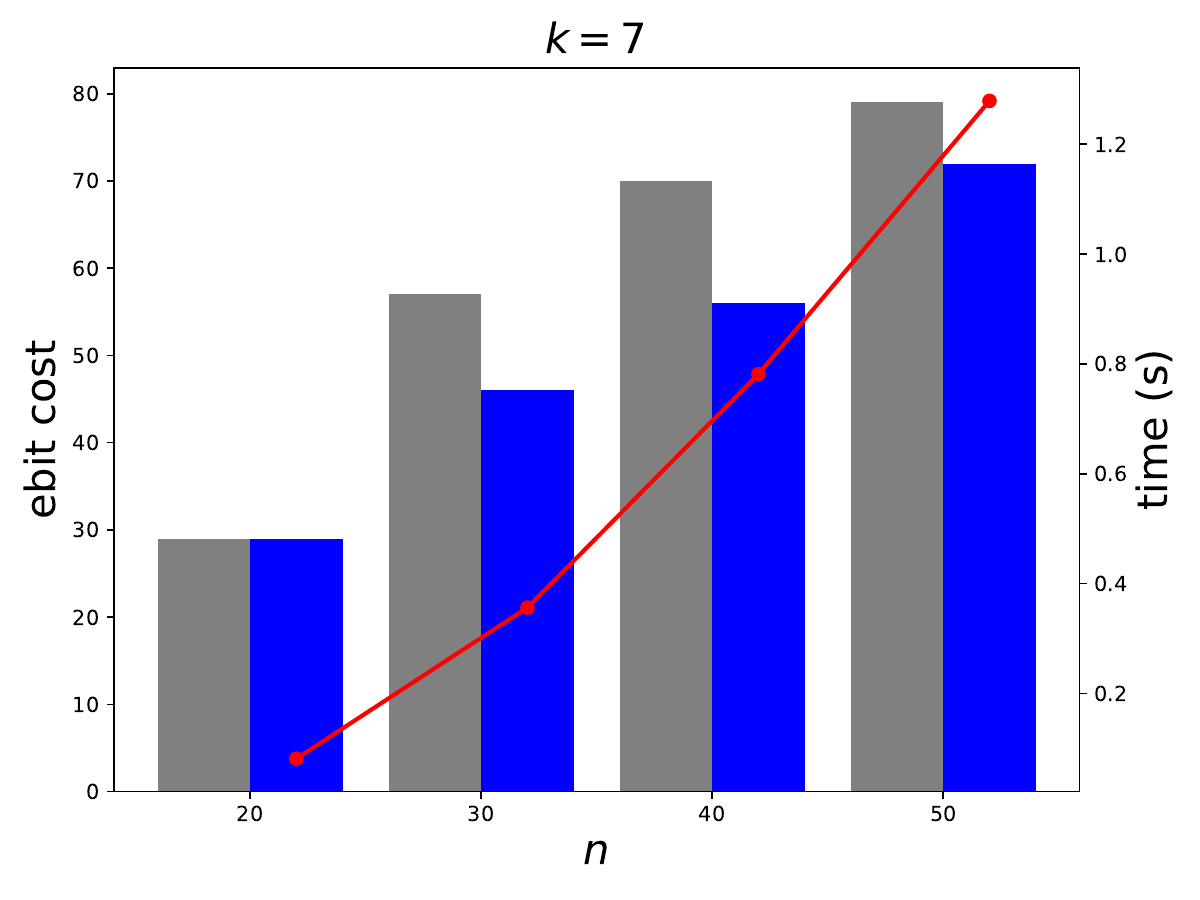}
    \end{subfigure}
    \begin{subfigure}{0.25\textwidth}
        \includegraphics[width=\linewidth]{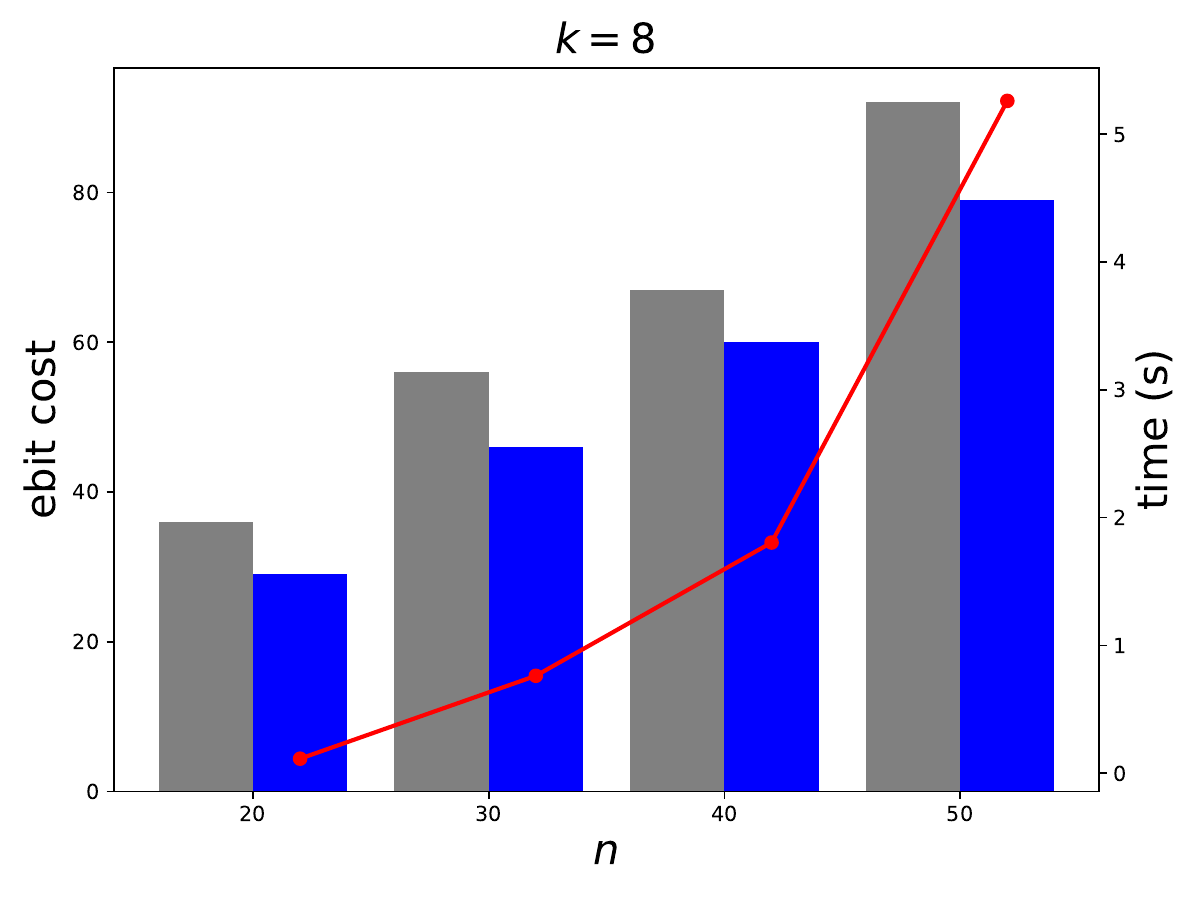}
    \end{subfigure}

    \caption{BIP improvements observed in DraperQFTAdder circuits. \hl{Here, $n$ is the number of qubits and $k$ is the number of modules.}}
    \label{fig:many_figures_draper}
\end{figure}

\clearpage

\subsection{RGQFTMultiplier circuits}
\label{section:rgqftmultiplier_appendix}

\begin{figure}[h]
    \centering
    \begin{subfigure}{0.15\textwidth}
        \includegraphics[width=\linewidth]{plots/legend_basic.pdf}
    \end{subfigure}
    \\
    \begin{subfigure}{0.25\textwidth}
        \includegraphics[width=\linewidth]{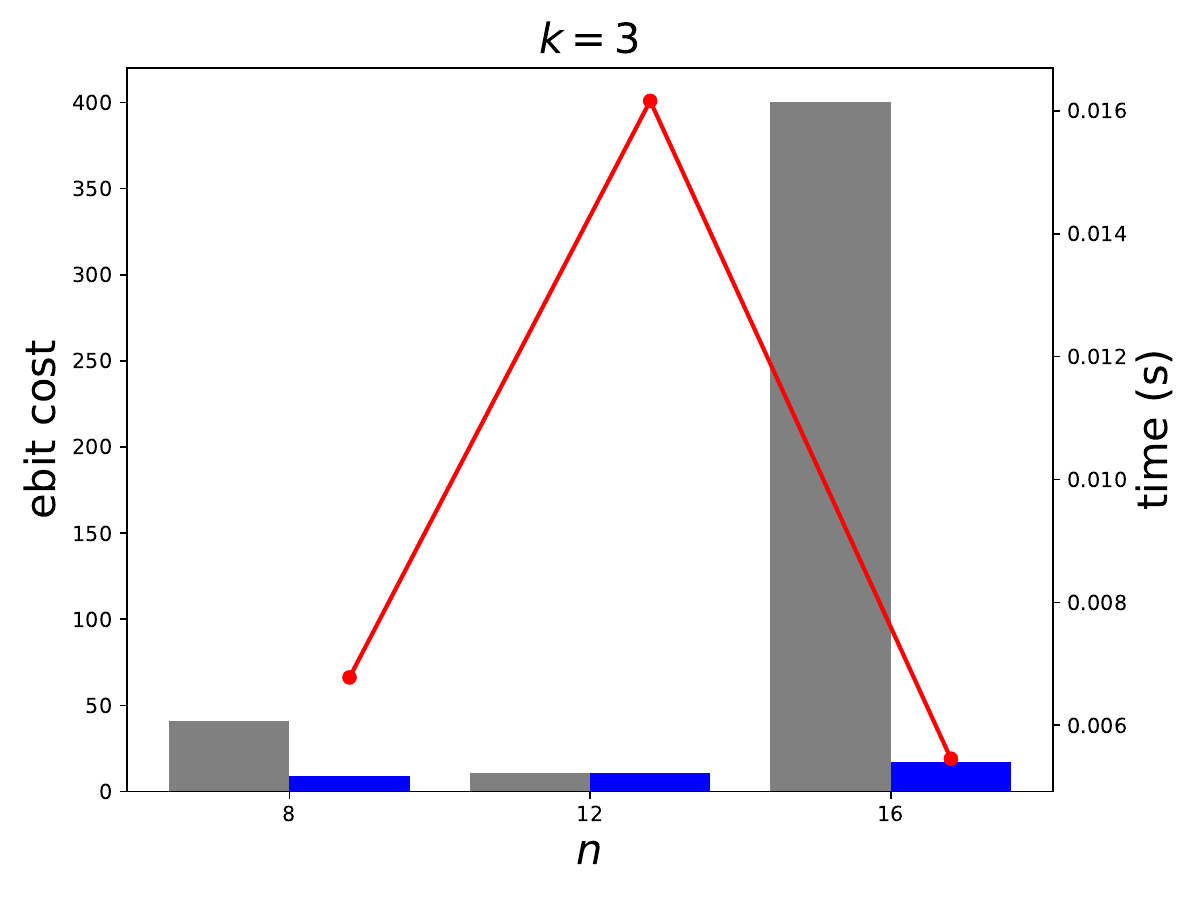}
    \end{subfigure}
    \begin{subfigure}{0.25\textwidth}
        \includegraphics[width=\linewidth]{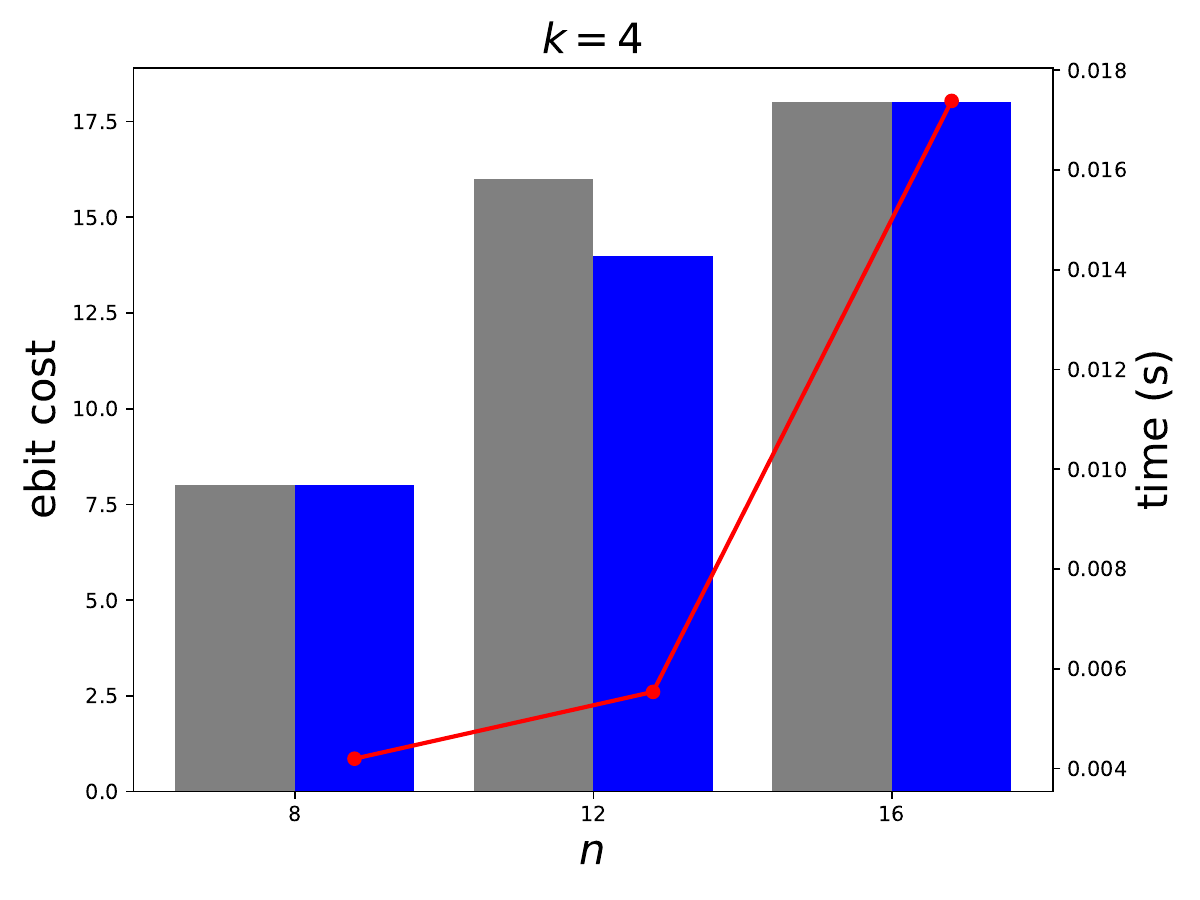}
    \end{subfigure}
    \begin{subfigure}{0.25\textwidth}
        \includegraphics[width=\linewidth]{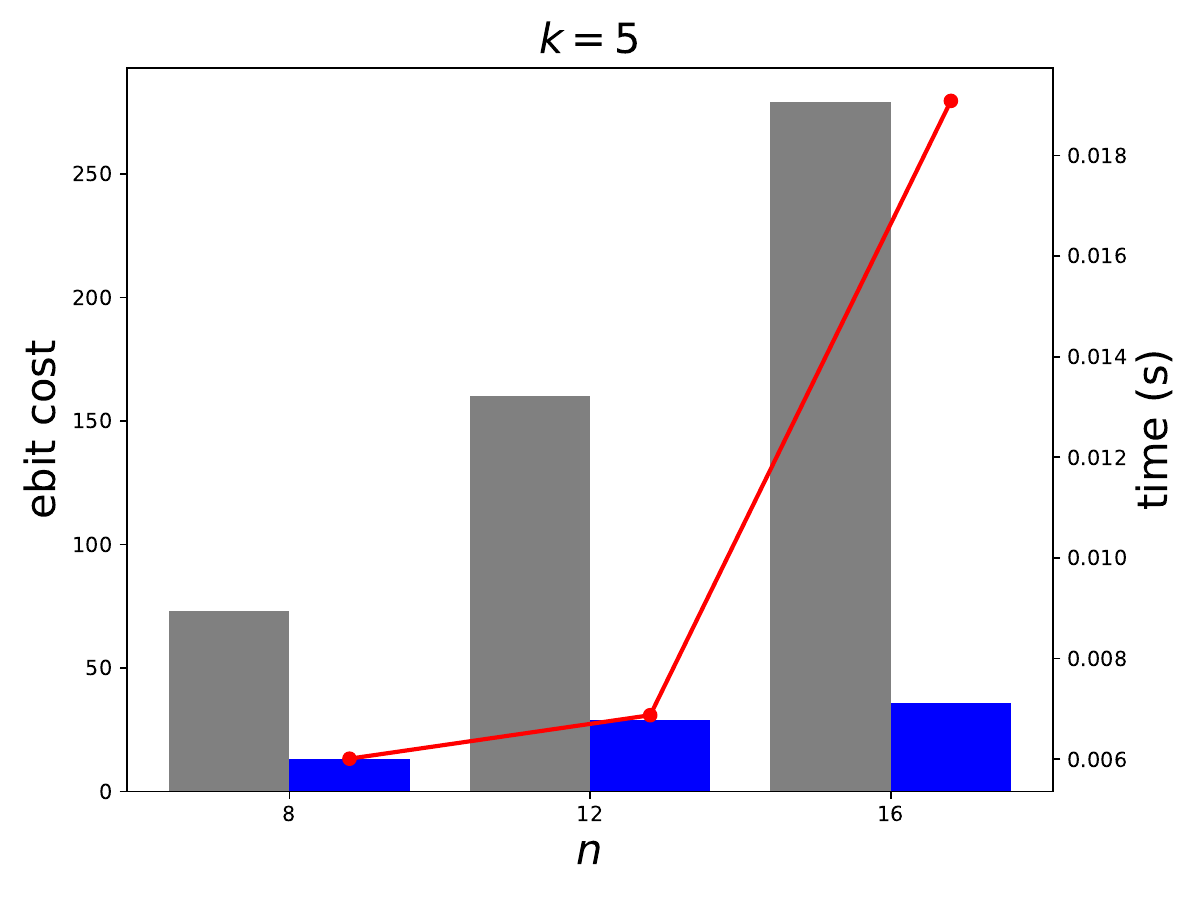}
    \end{subfigure}
    \begin{subfigure}{0.25\textwidth}
        \includegraphics[width=\linewidth]{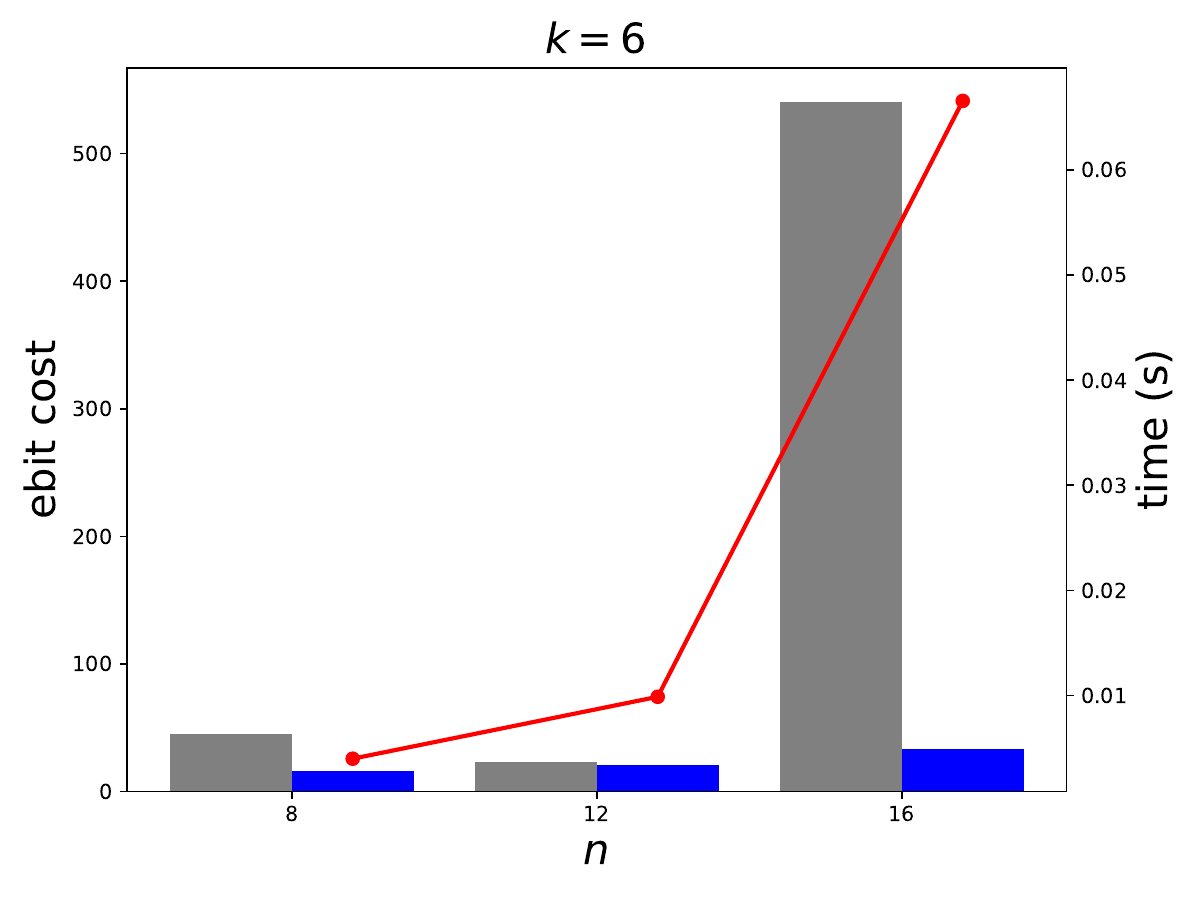}
    \end{subfigure}
    \begin{subfigure}{0.25\textwidth}
        \includegraphics[width=\linewidth]{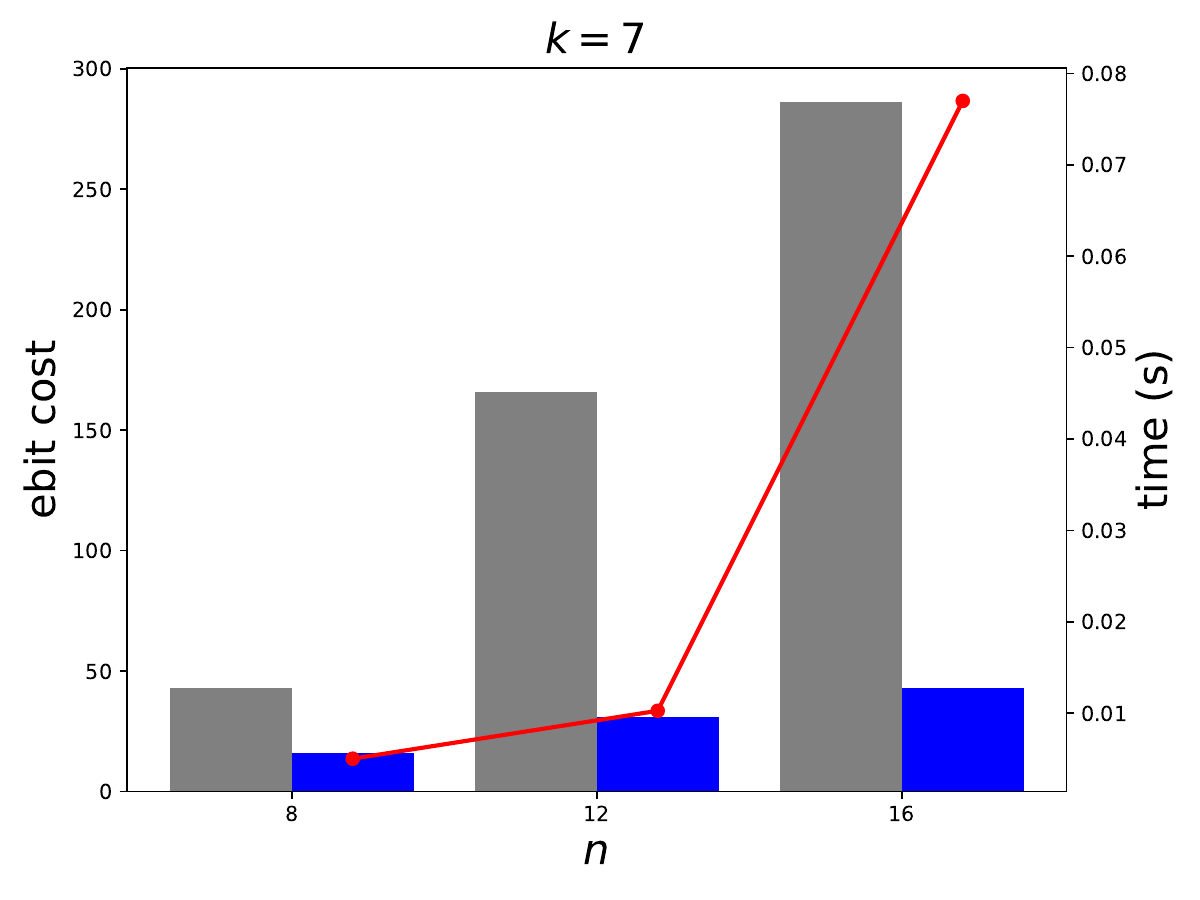}
    \end{subfigure}
    \begin{subfigure}{0.25\textwidth}
        \includegraphics[width=\linewidth]{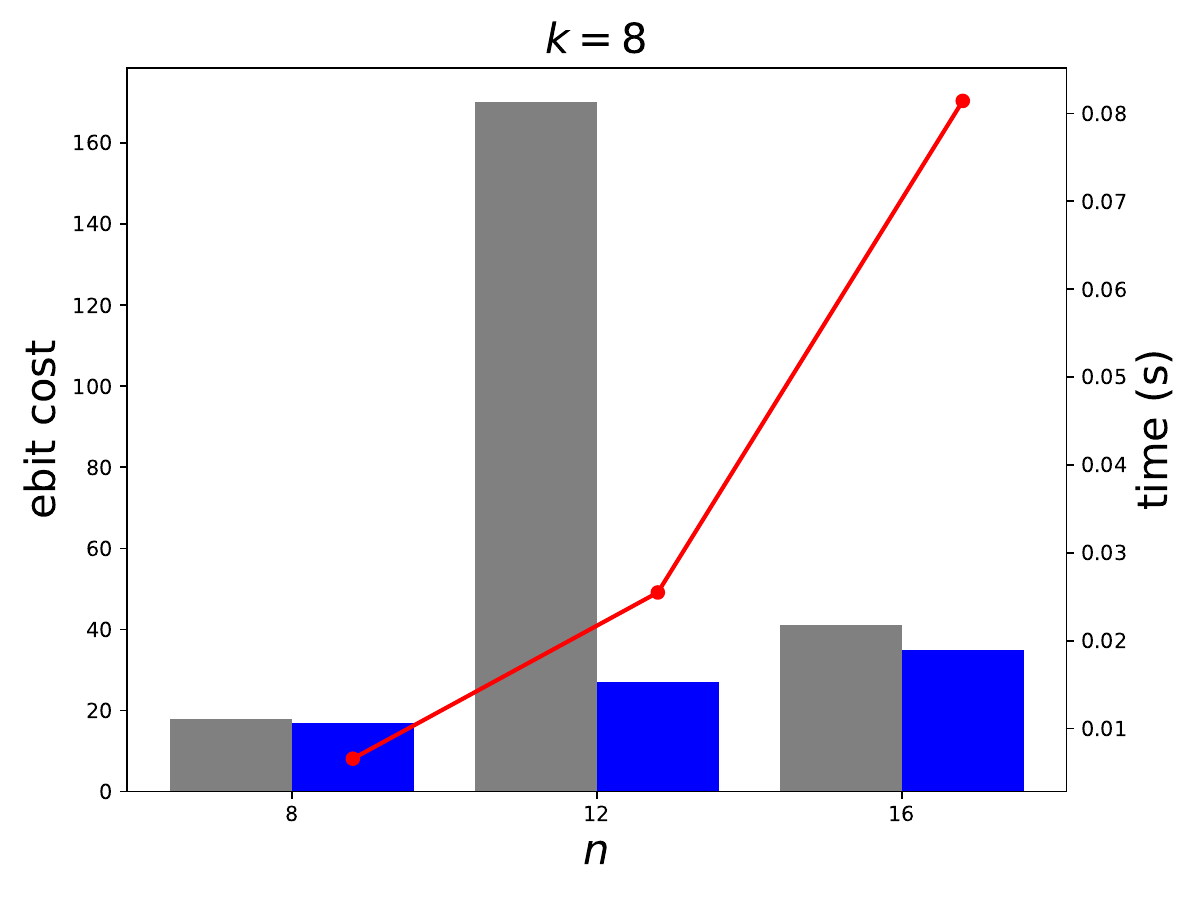}
    \end{subfigure}
    \caption{BIP improvements observed in RGQFTMultiplier circuits. \hl{Here, $n$ is the number of qubits and $k$ is the number of modules. In many configurations, our BIP post-processing achieves significant ebit savings compared to HP.}}
    \label{fig:rgqft_results}
\end{figure}

\subsection{AND circuits}
\label{section:and_appendix}

\begin{figure}[h]
    \centering
    \begin{subfigure}{0.15\textwidth}
        \includegraphics[width=\linewidth]{plots/legend_basic.pdf}
    \end{subfigure}
    \\
    \begin{subfigure}{0.25\textwidth}
        \includegraphics[width=\linewidth]{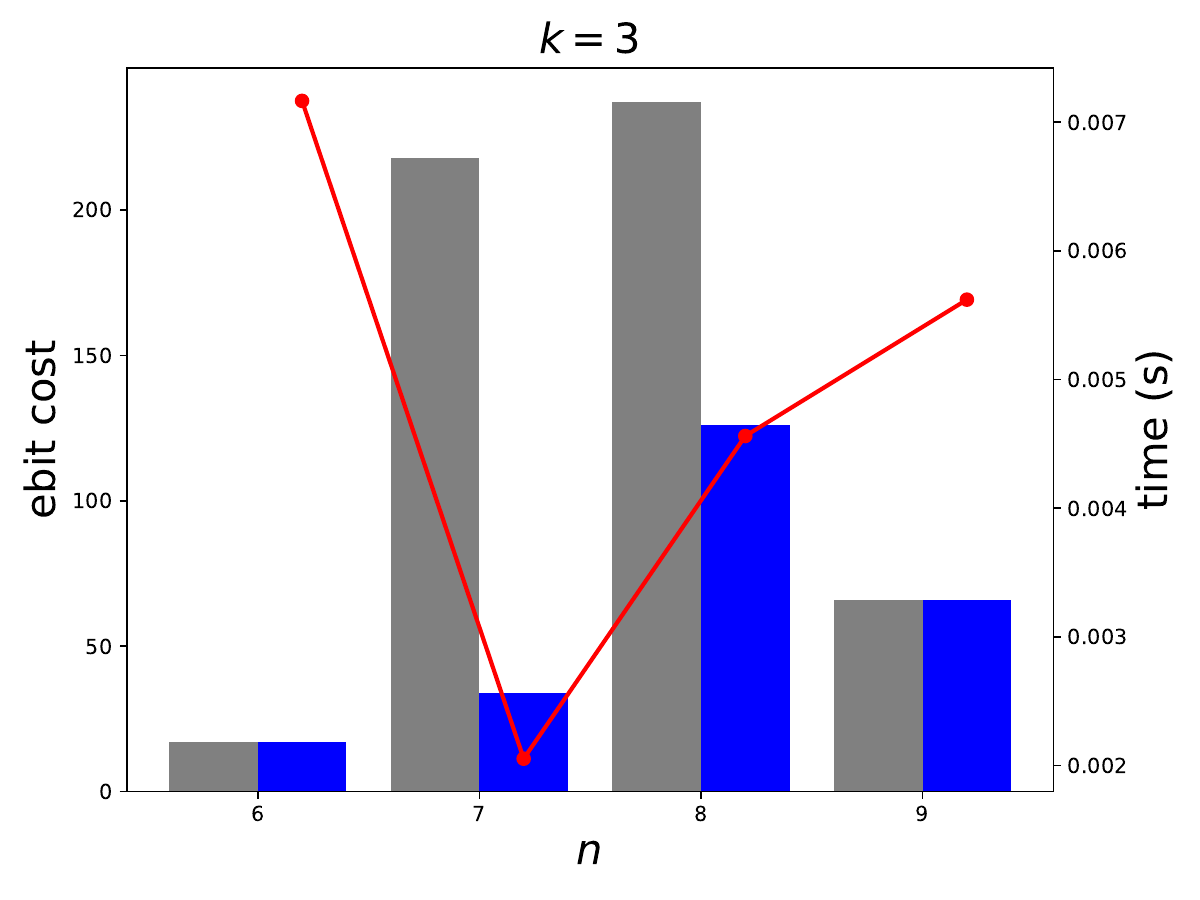}
    \end{subfigure}
    \begin{subfigure}{0.25\textwidth}
        \includegraphics[width=\linewidth]{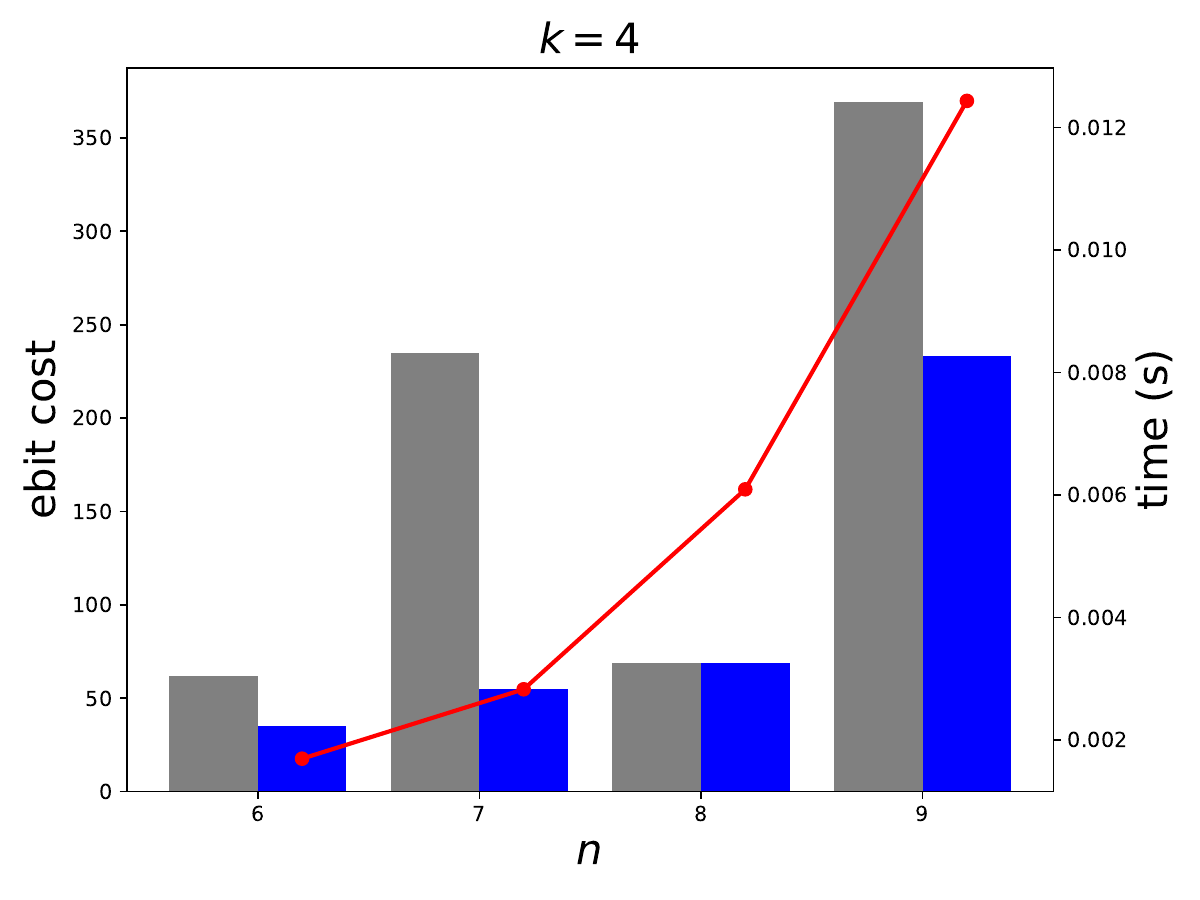}
    \end{subfigure}
    \begin{subfigure}{0.25\textwidth}
        \includegraphics[width=\linewidth]{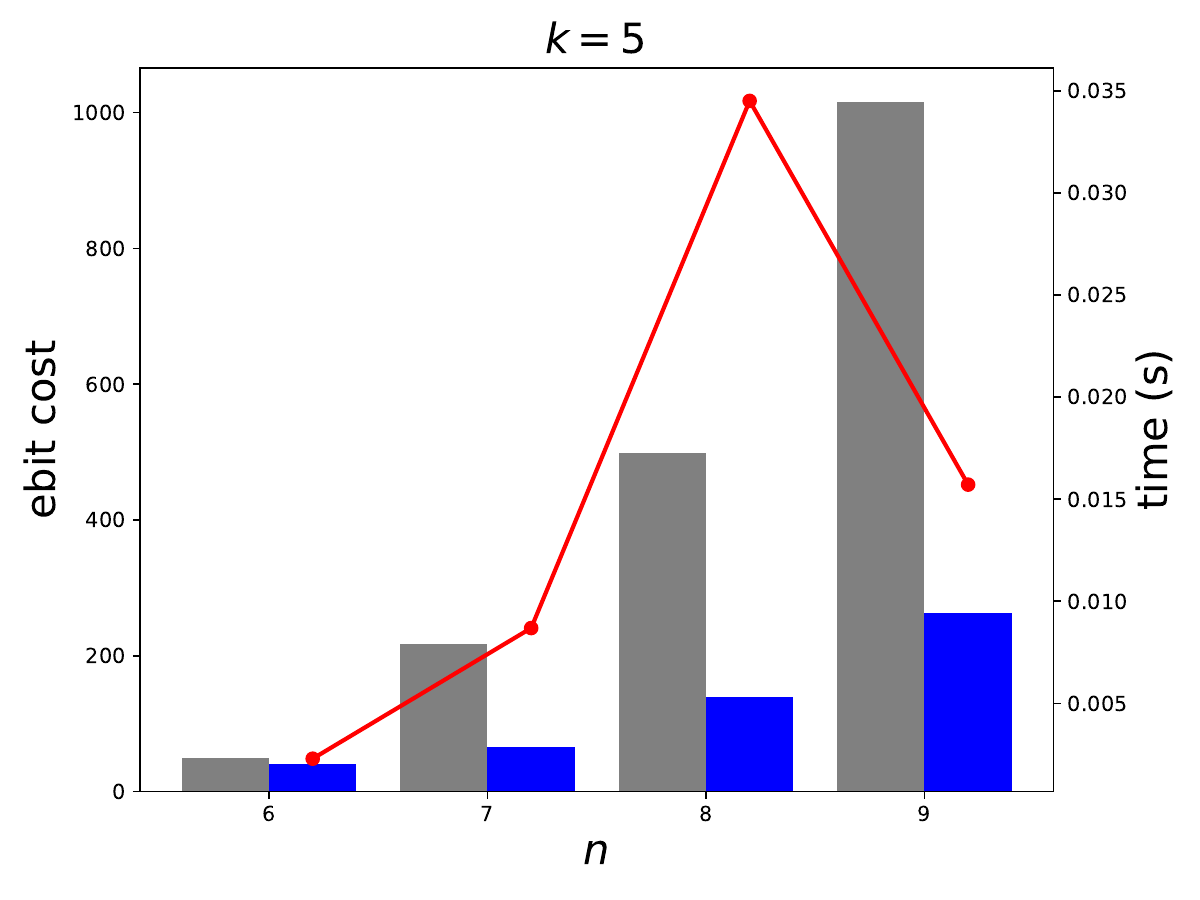}
    \end{subfigure}
    \caption{BIP improvements observed in AND circuits. \hl{Here, $n$ is the number of qubits and $k$ is the number of modules. In many configurations, our BIP post-processing achieves significant ebit savings compared to HP.}}
    \label{fig:and_plots}
\end{figure}

\clearpage

\subsection{InnerProduct circuits}
\label{section:innerproduct_appendix}

\begin{figure}[h]
    \centering
    \begin{subfigure}{0.15\textwidth}
        \includegraphics[width=\linewidth]{plots/legend_basic.pdf}
    \end{subfigure}
    \\
    \begin{subfigure}{0.25\textwidth}
        \includegraphics[width=\linewidth]{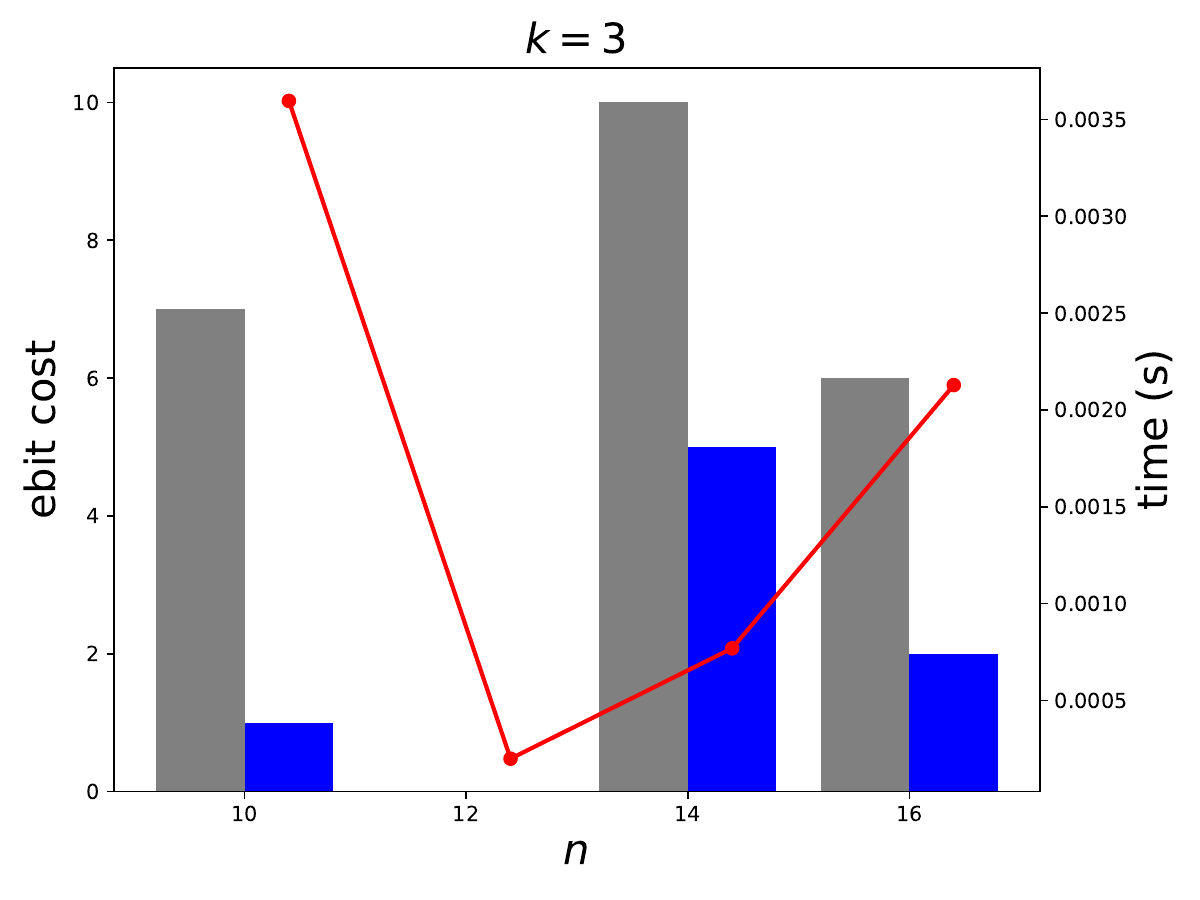}
    \end{subfigure}
    \begin{subfigure}{0.25\textwidth}
        \includegraphics[width=\linewidth]{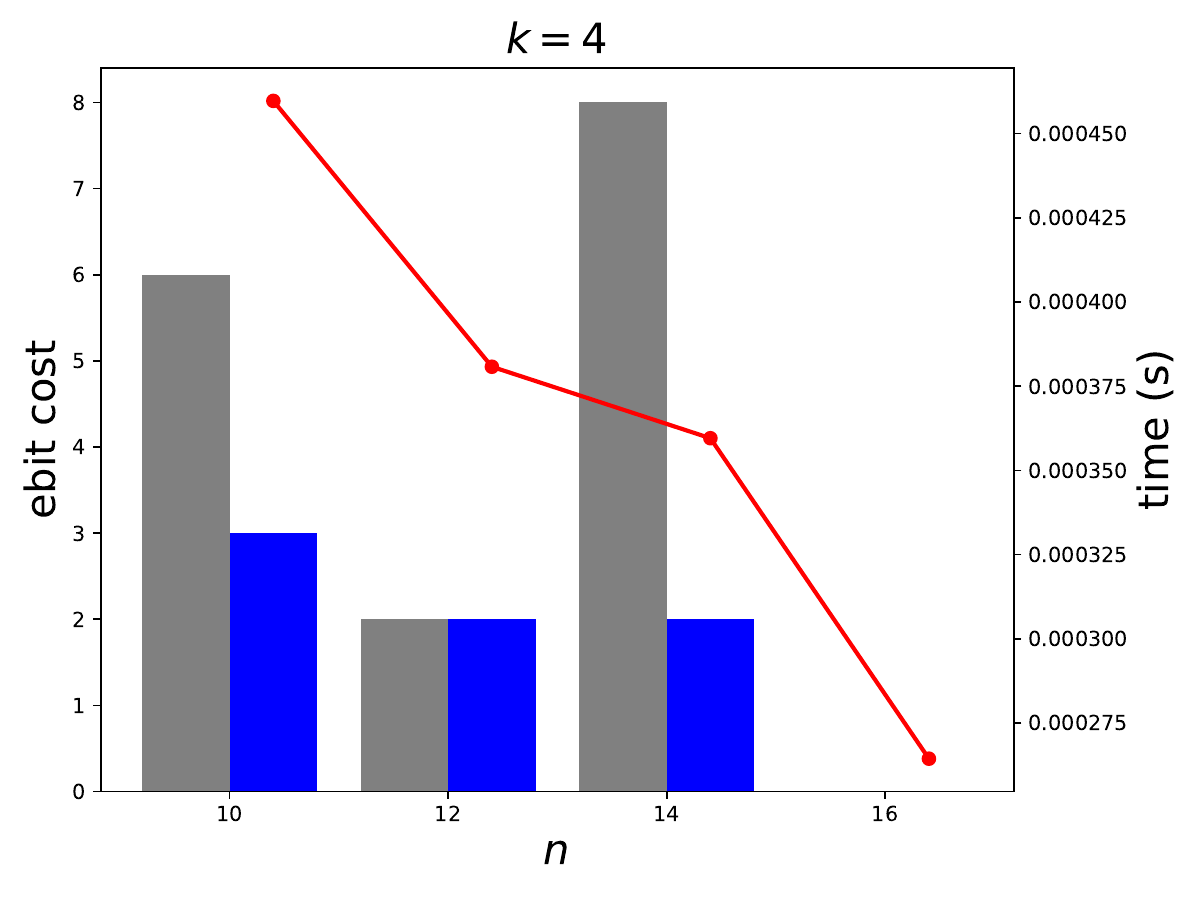}
    \end{subfigure}
    \begin{subfigure}{0.25\textwidth}
        \includegraphics[width=\linewidth]{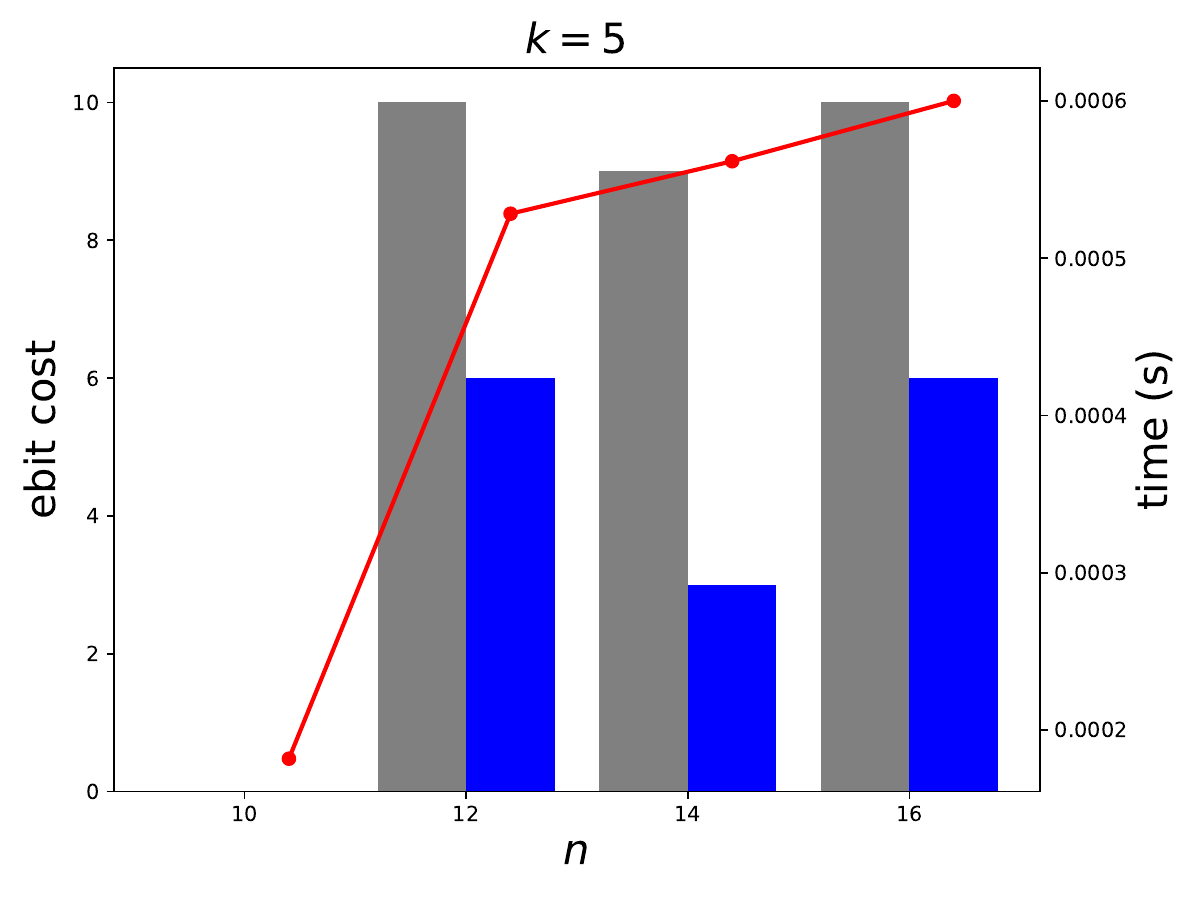}
    \end{subfigure}
    \caption{BIP improvements observed in InnerProduct circuits. \hl{Here, $n$ is the number of qubits and $k$ is the number of modules.} Empty bars indicate zero ebit cost. In many configurations, our BIP post-processing achieves significant ebit savings compared to HP.}
    \label{fig:inner_plots}
\end{figure}





\end{appendices}

\subsection*{Author contributions}
H.C. provided the main ideas, conducted the experiments, and wrote the manuscript. J.L. analyzed the results and revised the manuscript.

\subsection*{Data availability}
No datasets were generated or analyzed during the current study.

\section*{Declarations}
\subsection*{Competing interests}
The authors declare no competing interests.

\bibliography{sn-article}

@inproceedings{g2021efficient,
  title={Efficient distribution of quantum circuits},
  author={G Sundaram, Ranjani and Gupta, Himanshu and Ramakrishnan, CR},
  booktitle={35th International Symposium on Distributed Computing (DISC 2021)},
  pages={41--1},
  year={2021},
  organization={Schloss Dagstuhl--Leibniz-Zentrum f{\"u}r Informatik}
}

@article{andres2019automated,
  title={Automated distribution of quantum circuits via hypergraph partitioning},
  author={Andres-Martinez, Pablo and Heunen, Chris},
  journal={Physical Review A},
  volume={100},
  number={3},
  pages={032308},
  year={2019},
  publisher={APS}
}

@article{wu2023entanglement,
  title={Entanglement-efficient bipartite-distributed quantum computing},
  author={Wu, Jun-Yi and Matsui, Kosuke and Forrer, Tim and Soeda, Akihito and Andr{\'e}s-Mart{\'\i}nez, Pablo and Mills, Daniel and Henaut, Luciana and Murao, Mio},
  journal={Quantum},
  volume={7},
  pages={1196},
  year={2023},
  publisher={Verein zur F{\"o}rderung des Open Access Publizierens in den Quantenwissenschaften}
}

@article{barenco1995elementary,
  title={Elementary gates for quantum computation},
  author={Barenco, Adriano and Bennett, Charles H and Cleve, Richard and DiVincenzo, David P and Margolus, Norman and Shor, Peter and Sleator, Tycho and Smolin, John A and Weinfurter, Harald},
  journal={Physical review A},
  volume={52},
  number={5},
  pages={3457},
  year={1995},
  publisher={APS}
}

@article{arute2019quantum,
  title={Quantum supremacy using a programmable superconducting processor},
  author={Arute, Frank and Arya, Kunal and Babbush, Ryan and Bacon, Dave and Bardin, Joseph C and Barends, Rami and Biswas, Rupak and Boixo, Sergio and Brandao, Fernando GSL and Buell, David A and others},
  journal={Nature},
  volume={574},
  number={7779},
  pages={505--510},
  year={2019},
  publisher={Nature Publishing Group UK London}
}

@article{bravyi2018quantum,
  title={Quantum advantage with shallow circuits},
  author={Bravyi, Sergey and Gosset, David and K{\"o}nig, Robert},
  journal={Science},
  volume={362},
  number={6412},
  pages={308--311},
  year={2018},
  publisher={American Association for the Advancement of Science}
}

@article{neill2018blueprint,
  title={A blueprint for demonstrating quantum supremacy with superconducting qubits},
  author={Neill, Charles and Roushan, Pedran and Kechedzhi, K and Boixo, Sergio and Isakov, Sergei V and Smelyanskiy, V and Megrant, A and Chiaro, B and Dunsworth, A and Arya, K and others},
  journal={Science},
  volume={360},
  number={6385},
  pages={195--199},
  year={2018},
  publisher={American Association for the Advancement of Science}
}

@article{kjaergaard2020superconducting,
  title={Superconducting qubits: Current state of play},
  author={Kjaergaard, Morten and Schwartz, Mollie E and Braum{\"u}ller, Jochen and Krantz, Philip and Wang, Joel I-J and Gustavsson, Simon and Oliver, William D},
  journal={Annual Review of Condensed Matter Physics},
  volume={11},
  number={1},
  pages={369--395},
  year={2020},
  publisher={Annual Reviews}
}

@article{terhal2015quantum,
  title={Quantum error correction for quantum memories},
  author={Terhal, Barbara M},
  journal={Reviews of Modern Physics},
  volume={87},
  number={2},
  pages={307--346},
  year={2015},
  publisher={APS}
}

@article{fowler2012surface,
  title={Surface codes: Towards practical large-scale quantum computation},
  author={Fowler, Austin G and Mariantoni, Matteo and Martinis, John M and Cleland, Andrew N},
  journal={Physical Review A—Atomic, Molecular, and Optical Physics},
  volume={86},
  number={3},
  pages={032324},
  year={2012},
  publisher={APS}
}

@article{o2016scalable,
  title={Scalable quantum simulation of molecular energies},
  author={O’Malley, Peter JJ and Babbush, Ryan and Kivlichan, Ian D and Romero, Jonathan and McClean, Jarrod R and Barends, Rami and Kelly, Julian and Roushan, Pedram and Tranter, Andrew and Ding, Nan and others},
  journal={Physical Review X},
  volume={6},
  number={3},
  pages={031007},
  year={2016},
  publisher={APS}
}

@article{bourassa2021blueprint,
  title={Blueprint for a scalable photonic fault-tolerant quantum computer},
  author={Bourassa, J Eli and Alexander, Rafael N and Vasmer, Michael and Patil, Ashlesha and Tzitrin, Ilan and Matsuura, Takaya and Su, Daiqin and Baragiola, Ben Q and Guha, Saikat and Dauphinais, Guillaume and others},
  journal={Quantum},
  volume={5},
  pages={392},
  year={2021},
  publisher={Verein zur F{\"o}rderung des Open Access Publizierens in den Quantenwissenschaften}
}

@article{cirac1999distributed,
  title={Distributed quantum computation over noisy channels},
  author={Cirac, J Ignacio and Ekert, AK and Huelga, Susana F and Macchiavello, Chiara},
  journal={Physical Review A},
  volume={59},
  number={6},
  pages={4249},
  year={1999},
  publisher={APS}
}

@article{cacciapuoti2019quantum,
  title={Quantum internet: Networking challenges in distributed quantum computing},
  author={Cacciapuoti, Angela Sara and Caleffi, Marcello and Tafuri, Francesco and Cataliotti, Francesco Saverio and Gherardini, Stefano and Bianchi, Giuseppe},
  journal={IEEE Network},
  volume={34},
  number={1},
  pages={137--143},
  year={2019},
  publisher={IEEE}
}

@article{cuomo2023optimized,
  title={Optimized compiler for distributed quantum computing},
  author={Cuomo, Daniele and Caleffi, Marcello and Krsulich, Kevin and Tramonto, Filippo and Agliardi, Gabriele and Prati, Enrico and Cacciapuoti, Angela Sara},
  journal={ACM Transactions on Quantum Computing},
  volume={4},
  number={2},
  pages={1--29},
  year={2023},
  publisher={ACM New York, NY}
}

@article{yimsiriwattana2004generalized,
  title={Generalized GHZ states and distributed quantum computing},
  author={Yimsiriwattana, Anocha and Lomonaco Jr, Samuel J},
  journal={arXiv preprint quant-ph/0402148},
  year={2004}
}

@article{eisert2000optimal,
  title={Optimal local implementation of nonlocal quantum gates},
  author={Eisert, Jens and Jacobs, Kurt and Papadopoulos, Polykarpos and Plenio, Martin B},
  journal={Physical Review A},
  volume={62},
  number={5},
  pages={052317},
  year={2000},
  publisher={APS}
}

@book{nielsen2010quantum,
  title={Quantum computation and quantum information},
  author={Nielsen, Michael A and Chuang, Isaac L},
  year={2010},
  publisher={Cambridge university press}
}

@article{sych2009complete,
  title={A complete basis of generalized Bell states},
  author={Sych, Denis and Leuchs, Gerd},
  journal={New Journal of Physics},
  volume={11},
  number={1},
  pages={013006},
  year={2009},
  publisher={IOP Publishing}
}

@article{zaman2018counterfactual,
  title={Counterfactual Bell-state analysis},
  author={Zaman, Fakhar and Jeong, Youngmin and Shin, Hyundong},
  journal={Scientific reports},
  volume={8},
  number={1},
  pages={14641},
  year={2018},
  publisher={Nature Publishing Group UK London}
}

@article{andres2024distributing,
  title={Distributing circuits over heterogeneous, modular quantum computing network architectures},
  author={Andres-Martinez, Pablo and Forrer, Tim and Mills, Daniel and Wu, Jun-Yi and Henaut, Luciana and Yamamoto, Kentaro and Murao, Mio and Duncan, Ross},
  journal={Quantum Science and Technology},
  volume={9},
  number={4},
  pages={045021},
  year={2024},
  publisher={IOP Publishing}
}

@incollection{karp2009reducibility,
  title={Reducibility among combinatorial problems},
  author={Karp, Richard M},
  booktitle={50 Years of Integer Programming 1958-2008: from the Early Years to the State-of-the-Art},
  pages={219--241},
  year={2009},
  publisher={Springer}
}

@book{papadimitriou1998combinatorial,
  title={Combinatorial optimization: algorithms and complexity},
  author={Papadimitriou, Christos H and Steiglitz, Kenneth},
  year={1998},
  publisher={Courier Corporation}
}

@book{wolsey2020integer,
  title={Integer programming},
  author={Wolsey, Laurence A},
  year={2020},
  publisher={John Wiley \& Sons}
}

@article{coppersmith2002approximate,
  title={An approximate Fourier transform useful in quantum factoring},
  author={Coppersmith, Don},
  journal={arXiv preprint quant-ph/0201067},
  year={2002}
}

@article{ruiz2017quantum,
  title={Quantum arithmetic with the quantum Fourier transform},
  author={Ruiz-Perez, Lidia and Garcia-Escartin, Juan Carlos},
  journal={Quantum Information Processing},
  volume={16},
  pages={1--14},
  year={2017},
  publisher={Springer}
}

@article{weinstein2001implementation,
  title={Implementation of the quantum Fourier transform},
  author={Weinstein, Yaakov S and Pravia, MA and Fortunato, EM and Lloyd, Seth and Cory, David G},
  journal={Physical review letters},
  volume={86},
  number={9},
  pages={1889},
  year={2001},
  publisher={APS}
}

@article{cross2019validating,
  title={Validating quantum computers using randomized model circuits},
  author={Cross, Andrew W and Bishop, Lev S and Sheldon, Sarah and Nation, Paul D and Gambetta, Jay M},
  journal={Physical Review A},
  volume={100},
  number={3},
  pages={032328},
  year={2019},
  publisher={APS}
}

@article{draper2000addition,
  title={Addition on a quantum computer},
  author={Draper, Thomas G},
  journal={arXiv preprint quant-ph/0008033},
  year={2000}
}

@misc{gurobi,
  author = {{Gurobi Optimization, LLC}},
  title = {{Gurobi Optimizer Reference Manual}},
  year = 2024,
  url = "https://www.gurobi.com"
}

@incollection{land2009automatic,
  title={An automatic method for solving discrete programming problems},
  author={Land, Ailsa H and Doig, Alison G},
  booktitle={50 Years of Integer Programming 1958-2008: From the Early Years to the State-of-the-Art},
  pages={105--132},
  year={2009},
  publisher={Springer}
}

@article{clausen1999branch,
  title={Branch and bound algorithms-principles and examples},
  author={Clausen, Jens},
  journal={Department of computer science, University of Copenhagen},
  pages={1--30},
  year={1999}
}

@article{ferrari2023modular,
  title={A modular quantum compilation framework for distributed quantum computing},
  author={Ferrari, Davide and Carretta, Stefano and Amoretti, Michele},
  journal={IEEE Transactions on Quantum Engineering},
  volume={4},
  pages={1--13},
  year={2023},
  publisher={IEEE}
}

@article{kaur2025optimized,
  title={Optimized quantum circuit partitioning across multiple quantum processors},
  author={Kaur, Eneet and Pouryousef, Shahrooz and Shapourian, Hassan and Zhao, Jiapeng and Kilzer, Michael and Kompella, Ramana and Nejabati, Reza},
  journal={IEEE Transactions on Quantum Engineering},
  year={2025},
  publisher={IEEE}
}

@article{carrera2024combining,
  title={Combining quantum processors with real-time classical communication},
  author={Carrera Vazquez, Almudena and Tornow, Caroline and Riste, Diego and Woerner, Stefan and Takita, Maika and Egger, Daniel J},
  journal={Nature},
  volume={636},
  number={8041},
  pages={75--79},
  year={2024},
  publisher={Nature Publishing Group UK London}
}

@article{baumer2025measurement,
  title={Measurement-based long-range entangling gates in constant depth},
  author={B{\"a}umer, Elisa and Woerner, Stefan},
  journal={Physical Review Research},
  volume={7},
  number={2},
  pages={023120},
  year={2025},
  publisher={APS}
}

@inproceedings{neumann2020imperfect,
  title={Imperfect distributed quantum phase estimation},
  author={Neumann, Niels MP and van Houte, Roy and Attema, Thomas},
  booktitle={International Conference on Computational Science},
  pages={605--615},
  year={2020},
  organization={Springer}
}

\end{document}